\documentclass{aa} 

\usepackage{lscape}
\usepackage{afterpage}
\usepackage{subcaption} 
\usepackage{graphicx}
\usepackage{adjustbox}
\usepackage{xcolor}
\usepackage{ulem}
\usepackage{multirow}
\usepackage[colorlinks=true,citecolor=blue,linkcolor=blue]{hyperref}
\usepackage{mathptmx}
\usepackage{lscape} 
\usepackage{txfonts}
\usepackage{appendix}
\usepackage{float}

\usepackage{ragged2e}
\usepackage{blindtext}

\begin{document} 
\newcommand{\gc}[1]{\textcolor{magenta}{#1}}

   \title{Astrochemical modelling of infrared dark clouds}
\titlerunning{Astrochemical Modelling of IRDCs}
\authorrunning{Entekhabi et al.}
   \subtitle{}
   \author{N. Entekhabi\inst{1}, J. C. Tan\inst{1,2}
   G. Cosentino\inst{1}, C-J. Hsu\inst{1}, P. Caselli\inst{3}, C. Walsh\inst{4}, W. Lim\inst{5}, J. D. Henshaw\inst{6}, A. T. Barnes\inst{7}, F. Fontani\inst{8}, I. Jim\'enez-Serra\inst{9}
%          \and
%          G. Cosentino\inst{2}\fnmsep\thanks{Just to show the usage
%          of the elements in the author field}
          }

   \institute{\inst{1}Dept. of Space, Earth \& Environment, Chalmers University of Technology, Gothenburg, Sweden\\
   \inst{2}Dept. of Astronomy, University of Virginia, Charlottesville, VA, USA\\
   %\inst{3}Max Planck Institute for Extraterrestrial Physics, Giessenbachstrasse 1, D-85748 Garching bei M\"unchen, Germany\\
   \inst{3}Max-Planck-Institut f\"ur extraterrestrische Physik, Gie{\ss}enbachstrasse 1, 85748 Garching bei M\"unchen, Germany\\
   \inst{4}School of Physics and Astronomy, University of Leeds, Leeds LS2 9JT, UK\\
   \inst{5}SOFIA Science Center, Mountain View, CA, USA\\
   \inst{6}Max Planck Institute for Astronomy, K\"onigstuhl 17, D-69117 Heidelberg, Germany\\
   \inst{7}Argelander-Institut f\"ur Astronomie, Universit\"at Bonn, Auf dem H\"ugel 71, D-53121 Bonn, Germany\\
   \inst{8}INAF Osservatorio Astronomico di Arcetri, Largo E. Fermi 5, I-50125 Florence, Italy\\
   \inst{9}Centro de Astrobiolog\'ia (CSIC/INTA), Ctra. de Torrej\'on a Ajalvir km 4, E-28850 Madrid, Spain
             % T\"urkenschanzstrasse 17, A-1180 Vienna\\
              %\email{jonathan.tan@chalmers.se}
             %\thanks{The university of heaven temporarily does notaccept e-mails}
             }

  % \date{Received September 15, 1996; accepted March 16, 1997}

% \abstract{}{}{}{}{} 
% 5 {} token are mandatory
 
\abstract
   {Infrared dark clouds (IRDCs) are cold, dense regions of the interstellar medium (ISM) that are likely to represent the initial conditions for massive star and star cluster formation. It is thus important to study the physical and chemical conditions of IRDCs to provide constraints and inputs for theoretical models of these processes.
   }
   {We aim to determine the astrochemical conditions, especially the cosmic ray ionisation rate (CRIR) and chemical age, in different regions of the massive IRDC G28.37+00.07 by comparing observed abundances of multiple molecules and molecular ions with the predictions of astrochemical models.}
   {We have computed a series of single-zone,  
   time-dependent, astrochemical models with a gas-grain network that
   systematically explores the parameter space of the density, temperature, CRIR, and visual extinction. We have also investigated the effects of choices of CO ice binding energy and temperatures achieved in the transient heating of grains when struck by cosmic rays.
   We selected ten positions across the IRDC that are known to have a variety of star formation activity. We utilised mid-infrared (MIR) extinction maps and sub-millimetre (sub-mm) emission maps to measure the mass surface densities of these regions needed for abundance and volume density estimates. The sub-mm emission maps were also used to measure temperatures. We then used Instituto de Radioastromía Milimétrica (IRAM) 30m observations of various tracers, especially C$^{18}$O(1-0), H$^{13}$CO$^+$(1-0), HC$^{18}$O$^+$(1-0), and N$_2$H$^+$(1-0), to estimate column densities and thus abundances. Finally, we investigated the range of astrochemical conditions that are consistent with the observed abundances.
   }
   {
   The typical physical conditions of the IRDC regions are $n_{\rm H}\sim 3\times 10^4$ to $10^5\:{\rm cm}^{-3}$ and $T\simeq 10$ to $15$~K. Strong emission of H$^{13}$CO$^+$(1-0) and N$_2$H$^+$(1-0) is detected towards all the positions and these species are used to define relatively narrow velocity ranges of the IRDC regions, which are used for estimates of CO abundances, via C$^{18}$O(1-0). We would like to note that CO depletion factors are estimated to be in the range $f_D \sim 3$ to 10. Using estimates of the abundances of CO, HCO$^+$, and N$_2$H$^+$, we find consistency with astrochemical models that have relatively low CRIRs of $\zeta \sim10^{-18}$ to $\sim10^{-17}\:{\rm s}^{-1}$, with no evidence for systematic variation with the level of star formation activity. Astrochemical ages, which are defined with a reference to an initial condition of all H in $\rm H_2$, all C in CO, and all other species in atomic form, are found to be $<1$~Myr. We also explore the effects of using other detected species, that is HCN, HNC, HNCO, $\rm CH_3OH$, and $\rm H_2CO$, to constrain the models. These generally lead to implied conditions with higher levels of CRIRs and older chemical ages. Considering the observed $f_D$ versus $n_{\rm H}$ relation of the ten positions, which we find to have relatively little scatter, we discuss potential ways in which the astrochemical models can match such a relation as a quasi-equilibrium limit valid at ages of at least a few free-fall times, that is $\gtrsim 0.3$~Myr, including the effect of CO envelope contamination, small variations in temperature history near 15~K, CO-ice binding energy uncertainties, and CR-induced desorption. We find general consistency with the data of $\sim 0.5$~Myr-old models that have $\zeta \sim 2-5\times 10^{-18}\:{\rm s}^{-1}$ and CO abundances set by a balance of freeze-out with CR-induced desorption.}
  %Using abundances of CO broad velocity range, HCO$^+$ obtained from H$^{13}$CO$^+$ together with N$_2$H$^+$ and performing the fitting in the restricted region of the parameters, leads to obtain estimated $\zeta$ 1.0e-19-4.6e-19 s$^{-1}$ at p1,p2,p3,p4,p7,p8,p9, 2.2e-19-1.0e-18 s$^{-1}$ at p5,p10 and 4.6e-19-1.0e-18 s$^{-1}$ at p6, whereas the age of the cloud has been approximated older than 1 Myrs at all positions, except p1 and p5 for which the age has been estimated before 1 Myrs. Repeating the same fitting in ultra restricted range of the parameters, and using CO narrow and HCO$^+$ obtained from HC$^{18}$O$^+$, estimated $\zeta$ 2.2e-18-2.2e-17 $^{-1}$ at p1, 2.2e-18-1.0e-17 $^{-1}$ at p2,p4,p7,p9, 4.6e-19-4.6e-18 $^{-1}$ at p3, 1.0e-18-1.0e-17 $^{-1}$ at p5, 1.0e-18-2.2e-18 $^{-1}$ at p6, 2.2e-18-4.6e-18 $^{-1}$ at p8 and 1.0e-18-4.6e-18 $^{-1}$ at p10. Using this method, the age of the cloud has been estimated younger than 1.0e6 years for all positions, and even before 1.0e5 years for p1. The range of CRIR has been picked up based on the $\chi^2$<3} 
   {We have constrained the astrochemical conditions in ten regions in a massive IRDC, finding evidence for relatively low values of CRIR compared to diffuse ISM levels. 
   %based on comparison to fiducial astrochemical networks. 
   We have not seen clear evidence for variation in the CRIR with the level of star formation activity. We favour models that involve relatively low CRIRs ($\lesssim 10^{-17}\:{\rm s}^{-1}$) and relatively old chemical ages ($\gtrsim 0.3\:{\rm Myr}$, i.e. $\gtrsim 3 t_{\rm ff}$). We discuss potential sources of systematic uncertainties in these results and the overall implications for IRDC evolutionary history and astrochemical models.
   %Although we expected to see differences between quiescent and active star forming regions, the results showed very similar distributions of $\zeta$ at all positions. In general, CRIR has been estimated approximately 1.0e-18-1.0e-17 s$^{-1}$, whereas the Cloud is younger than 1.0e6 years.
   }


   \maketitle
%
%--------------------------------------------------------------

%\input{chapters/main_file}

%%%%%%%%%%%%%%%%%%%%%%%%%%%%%%%%%%%%%%%
\section{Introduction}\label{S:introduction}

%\nocite{*}
\nocite{bron2021tracers}
Infrared dark clouds (IRDCs) are cold ($T<25\:$K) \citep[e.g.][]{2006A&A...450..569P,peretto2010mapping,lim2016distribution} and dense ($n_{\rm H} \gtrsim 10^3\:{\rm cm}^{-3}$) \citep[e.g.][]{2013A&A...549A..53K,lim2016distribution} regions of the interstellar medium (ISM) that have the potential to host the formation of massive stars and star clusters \citep[for a review, see, e.g.][]{2014prpl.conf..149T}. Thus, it is important to study the physical and chemical properties of IRDCs to understand the initial conditions for the formation of these massive stellar systems.

One important astrochemical process that is expected to occur in the cold, dense conditions of IRDCs is the freeze-out of certain molecules to form ice mantles around dust grains, thus depleting their abundance in the gas phase. The CO depletion factor ($f_{\rm D,CO}$) is defined as the ratio of the CO gas phase abundance if all C was present as gas phase CO compared to the actual abundance. For example, in the filamentary IRDC G035.39-00.33 (Cloud H), this depletion factor has been measured to be $\sim 3$ on parsec scales by \cite{hernandez2012virialized} and up to $\sim10$ in some positions by \cite{jimenez2014gas}.

High CO depletion factors are generally expected in cold conditions, that is when $T\lesssim 20\:$K. The rate of CO depletion scales inversely with density, and the timescale to achieve a high depletion factor can be relatively short, that is shorter than the free-fall time for densities $n_{\rm H}\gtrsim 10^3\:{\rm cm}^{-3}$. In this cold temperature regime, the equilibrium abundance of gas phase CO is then expected to be set by a balance between the freeze-out rate and rates of non-thermal desorption processes, that is photo-desorption (including by UV photons induced by cosmic rays [CRs]) and by direct CR desorption, that is due to localised transitory heating of dust grains when impacted by a CR \citep{hasegawa1993new}. While CO and other species, such as $\rm H_2O$, are expected to be heavily depleted in cold, dense conditions, from studies of low-mass cores it has been inferred that N-bearing species suffer lower amounts of depletion \citep[e.g.][]{caselli1999co,crapsi2007observing}.

%(e.g., Ceccarelli et al. 2014).({\bf is that this paper: \citep{ceccarelli2014herschel}) I added these references too: \citep{caselli1999co,crapsi2007observing}.}

In regions where CO is highly depleted and the ortho-to-para ratio of $\rm H_2$ has declined to small values, another astrochemical process that is expected to occur is the build-up of relatively large abundances of $\rm H_2D^+$, which in turn can lead to the deuteration of species such as $\rm N_2H^+$. Relatively high levels of deuteration of $\rm N_2H^+$ have been observed in IRDCs on both parsec scales \citep{barnes2016widespread} and in more localised, sub-parsec scale cores \citep[e.g.][]{kong2016deuterium}. These studies have concluded that the conditions in these IRDCs are likely to be relatively chemically evolved, that is with the gas having been in a cold and dense state for at least several local free-fall times in order to achieve the high levels of deuteration that are observed. However, this is not a unique solution: models of relatively rapid collapse of cores can achieve the high observed levels of deuteration if the CR ionisation rate (CRIR) is relatively high. For example, \cite{hsu2021deuterium} have presented models of a massive ($60\:M_\odot$) core that can achieve abundance ratios of [$\rm N_2D^+$] / [$\rm N_2H^+$] $\sim 0.1$ within one free-fall time of $\sim 80,000\:$yr if the CRIR is $\sim 10^{-16}\:{\rm s}^{-1}$.

%CO depletion factor, the Deuterium fractionation \citep{2015ApJ...804...98K} and consequently Ionization fraction X(e)=$\frac{\rm n(e)}{\rm n(H_2)}$ \citep{2002P&SS...50.1133C}, may be another parameters leading to obtain information about the age and the evolution of IRDCs. The studies by \cite{1999ApJ...523L.165C} and \cite{ceccarelli2014deuterium} shows a very high degree of CO ($\sim 10$) depletion factor and deuteration as well, at a dense region of IRDCs where the formation of massive stars begins.\\

%Another study by \cite{2003A&A...403L..37C} indicates that at the earliest stage prior to the formation of protostellar core, the major molecular ion dominating the chemistry evolution is H$_2$D$^+$ (the deuterated forms of the main ion H$_3$ $^+$). On the other side, the formation of ions are produced by Cosmic rays (CRs) and the molecules in ISM are formed mainly by ion-molecule gas phase reactions \citep{2002P&SS...50.1133C}.\\

In addition to the deuteration process, much of interstellar chemistry is driven by CR ionisation, and the CRIR, $\zeta$, (defined as the rate of ionisations per H nucleus) is an important parameter of astrochemical models. Examples of inferred values of CRIR include the study of the $\rm H^{13}CO^+$ emission and $\rm H_3^+$ absorption towards seven massive protostellar cores by \cite{2000A&A...358L..79V}, who estimated $\zeta \sim (2.6\pm 1.8) \times 10^{-17}\:{\rm s}^{-1}$. More recent studies of hot cores with more sophisticated chemical networks include the work of \cite{barger2020constraining}, who derived more elevated levels of CRIRs.
%jct - we need to give some more quantitative info here; and make sure definition of \zeta is the same.
Studies of the diffuse ISM have found values of $\zeta\simeq 2\times 10^{-16}\:{\rm s}^{-1}$ \citep[e.g.][]{neufeld2017cosmic}. In regions of active star formation, such as OMC-2 FIR 4, much higher values of CRIR, 
$\gtrsim 10^{-15}\:{\rm s}^{-1}$, 
%jct - we should check this number
have been reported \citep{ceccarelli2014herschel, fontani2017seeds,favre2018solis}, which may be explained by local production in protostellar accretion and/or outflow shocks \citep{padovani2016protostars, gaches2018exploration}.
Another region where the CRIR is inferred to be very high is the Galactic Centre, with values of $\zeta\gtrsim10^{-15}\:{\rm s}^{-1}$ \citep{carlson2016cosmic}. 
%Carlson et al. 2016, PRD, 94, 3504

Lower energy CRs, which dominate the CRIR, are expected to be absorbed by high column densities of gas and potentially more easily shielded by magnetic fields, whose strength increases in denser regions \citep{crutcher2012magnetic} \citep[however, see][]{2018ApJ...863..188S}. Thus, it is possible that the CRIR in IRDCs may be smaller than that of the diffuse ISM in which they reside. To date, there have been relatively few studies of the CRIR in IRDCs. 

IRDCs have the advantage of being probed by mid-infrared (MIR) extinction (MIREX) mapping methods \citep[e.g.][]{butler2009mid,butler2012mid, 2013A&A...549A..53K}, which yield measurements of cloud mass surface density, $\Sigma$. {\it Herschel} sub-mm emission maps also constrain $\Sigma$ and additionally provide a measure of the dust temperature, although careful subtraction of contributions from the diffuse Galactic plane are important \citep{lim2016distribution}. IRDCs are expected to be at a relatively early evolutionary stage, which should simplify their modelling, especially with respect to hot cores \citep[e.g.][]{2000A&A...358L..79V,2019A&A...631A.142G}. However, studies with ALMA and {\it Herschel} do indicate that active star formation can be proceeding even in MIR-dark regions of the clouds \citep[e.g.][]{2016ApJ...821L...3T,kong2019widespread,moser2020high}.

In this paper, we aim to constrain the astrochemical conditions, including CRIR and chemical age, in ten different positions in the massive IRDC G28.37+00.07. We first present a grids of astrochemical models, based on the work of \cite{2015A&A...582A..88W}, that cover a wide range of physical conditions, including those relevant to IRDCs (\S\ref{astrochemical_model}). We next, in \S\ref{observations}, measure the physical properties of mass surface density ($\Sigma$), number density ($n_{\rm H}$) and temperature ($T$) of the IRDC regions, followed by measurements of the column densities and thus abundances of various molecules and molecular ions from IRAM-30m telescope observations. Then, in \S\ref{S:results}, we first discuss the derived abundances, then determine which parts of the multi-dimensional astrochemical model parameter space, including CRIR and time, are consistent with the observations. We also discuss potential effects of systematic uncertainties in the measurements of abundances and astrochemical modelling.
%the implications of our results in \S\ref{discussion}. 
We present our conclusions in \S\ref{S:conclusions}.

%For this purpose, we present the comparison of molecular abundances observed by IRAM 30m telescope, with the astrochemical model provided by \cite{2015A&A...582A..88W}, to be able to find the best models with associated cosmic ray ionization rate. The paper is structured as follows. In sections \ref{astrochemical_model} we describe the model and its properties and interpret its results. In section \ref{observations}, the properties of the observed cloud, detected lines and their obtained abundances is presented. In section \ref{comparison}, the comparison between observation and the model is made. Finally, a summery of our work and obtained results comprise section \ref{final}.

%%%%%%%%%%%%%%%%%%%%%%%%%%%%%%%%%%%%%%%%%%%%%%%%%%%%%%%%%
\section{Astrochemical model}\label{astrochemical_model}

\subsection{Overview of the model}

We use the astrochemical model developed by \cite{2015A&A...582A..88W} to predict the evolution of abundances of different species in the gas and grain ice mantle phases. 
%The model, initially used to reproduce molecular abundances in T-tauri type protoplanetary disks, is used here to reproduce chemical compositions of IRDCs in physical property parameter space \citep{2015A&A...582A..88W}.
The model uses a gas-phase network extracted from the UMIST 2012 Database for Astrochemistry \citep{2013A&A...550A..36M}, complemented by both thermal and non-thermal gas-grain processes, including CR-induced thermal desorption (CRID) (with standard value for CR-induced temperature of $T_{\rm CRID}=70\:$K), photodesorption, X-ray desorption, and grain-surface reactions. The reaction types included in the astrochemical code are described in \cite{walsh2010chemical} and the subsequent works \citep{walsh2012chemical,walsh2013molecular,2014A&A...563A..33W,2015A&A...582A..88W}.

For the first model grid (`Grid 1') we have adopted the default CO ice binding energy of 855~K and 
%However, of these extended reactions, 
we have switched off CR-induced thermal desorption reactions \citep{hasegawa1993new}, since their rates are highly uncertain \citep{cuppen2017grain}. Still, CR-induced photoreactions are included, including those that lead to desorption of species from grain surfaces. Grid 1 is the fiducial model we have used for most of our investigation. However, as we discuss later, after comparison with the observational results we find the need to investigate the following changes in the astrochemical modelling, that is `Grid 2': an increase in the CO ice binding energy to 1100~K, relevant in more realistic situations in which CO is mixed with water ice \citep{oberg2005competition,cuppen2017grain}; CRID reactions turned on with various values of $T_{\rm CRID}$ explored up to about $100\:$K. In all of our modelling we do not include any X-ray background, so the reactions involving X-rays do not play any role.
%In the main code of the model, the reactions of direct X-ray ionization has not been considered as well. 

Overall the network includes a total of 8763 reactions following the time evolution of the abundances of 669 species. The elements that are followed in the network and their assumed initial abundances are shown in Table~\ref{tab:initial_abundances}. All elements are initially in atomic form, except for: H, which is almost entirely in the form of $\rm H_2$; C, which is entirely in the form of CO; and 44\% of O that is in the form of CO. This choice is made given that IRDCs are expected to typically be structures within larger-scale molecular clouds.

%%%%%%%%%%%%%%%%%%%%%%%%%%%%%%%%%%%%
%\input{chapters/tables/initial_abundances}
\begingroup
\setlength{\tabcolsep}{10pt} % horizontal space
\renewcommand{\arraystretch}{1.3} % vertical space

\begin{table}[H]
\caption{Fiducial initial abundances}
\centering
\begin{tabular}{c c}
      \hline
      \hline
      Species & Abundance ($n_{\rm X}/n_{\rm H}$)\\
      \hline
      H$_2$ & 5.00e-01\\
      H & 5.00e-05 \\
      He & 9.75e-02 \\
      CO & 1.40e-04 \\
      O & 1.80e-04 \\
      N & 7.50e-05 \\
      S &  8.00e-08 \\
      F & 2.00e-08 \\
      Si & 8.00e-09 \\
      Cl & 4.00e-09 \\
      P & 3.00e-09 \\
      GRAIN & 1.30e-12\\
    \hline 
    \end{tabular}
    %\vspace{2mm}
    %\caption{Fiducial initial elemental abundances.}
    \label{tab:initial_abundances}
\end{table}
%%%%%%%%%%%%%%%%%%%%%%%%%%%%%%%%%%%%
%In order to produce the grid of models 
We have adopted the standard parameter values of a Draine FUV radiation field equal to $6.4\times 10^{-3}\:{\rm erg\: cm^{-2}\:s^{-1}}$ and a Habing FUV field equal to $4.0\times 10^8\:{\rm cm^{-2}\:s^{-1}}$. Other standard parameters are the assumed dust particle radius of $a_0=0.1\:{\rm \mu m}$, so that the grain surface area per H nucleus is $1.63\times 10^{-19}\:{\rm cm}^2$.
%jct - this needs to be checked.

%In addition, we have set the initial abundances as reported in Table \ref{tab:initial_abundances}. We note that, for the adopted initial abundances, all the carbon is the form of CO, and corrected to Oxygen abundance accordingly. This assumptions indicates that the chemical evolution in our models occurs after the cloud is formed i.e., when the material is mainly in the molecular form.

Each astrochemical model in a grid is run for a fixed set of values of number density of H nuclei ($n_{\rm H}$), gas temperature ($T$), CRIR ($\zeta$), visual extinction ($A_V$), and FUV radiation field intensity ($G_0$). 
%The models are evolved in time for $10^8$~years. 
The range of the parameter space of the grid of the models for this work is presented in Table \ref{tab:initial_parameters}. This parameter space is sampled at a resolution of 3 values per decade for density and CRIR. For temperatures, with a focus on lower values, we run models at 5, 10, 15, 20, 25, 30, 40, 50, ..., 100, 120, ..., 200, 250, 300, ..., 500, 600, ..., 1000 ~K. For $A_V$, we run models at 1, 2, 3, 4, 5, 7, 10, 20, 50, 100~mag.
%focusing on lower temperatures and reducing the resolution as temperature raises, higher resolution of A$_V$s below 10 mag and 
We only run models with one value of external FUV radiation field, that is $G_0 = 4$, which is the value expected in the inner Galaxy relevant to most IRDCs \citep{wolfire2010dark}.
%(Wolfire et al. 2010). 
In any case, the typical values of extinctions in IRDCs are very high, so that the external FUV radiation field is expected to play a negligible role in the chemical evolution.

With the above choices, we have a total of 137,750 different physical models in each grid. Each of these models is evolved in time for $10^8$~years, with a sampling of 100 outputs per decade in time (after $10^3$~years). We note that while we have developed here a relatively wide range of models for a first general exploration, when modelling IRDCs, we focus on a much narrower range of conditions that are constrained by observational data on the clouds.

%%%%%%%%%%%%%%%%%%%%%%%%%%%%%%%%%%%%%%%
%\input{chapters/tables/initial_parameters}
\begingroup
\setlength{\tabcolsep}{10pt} % horizontal space
\renewcommand{\arraystretch}{1.3} % vertical space

\begin{table}[H]
\caption{Model grid parameter space}
\centering
\begin{tabular}{c c c}
      \hline
      \hline
      Parameter & Explored Range \\
      \hline
      $n_{\rm H}$ (cm$^{-3}$) & 10 - $10^9$\\
      $T$ (K) & 5 - 1000\\
      $\zeta$ (s$^{-1}$) & $10^{-19}$ - $10^{-13}$\\
      $A_V$ (mag) & 1 - 100\\
%      t & time & 1.0e3-1.0e8~~year \\
      $G_0$ (Habings) & 4\\
%      a$_0$ & dust particle radius & 0.1 $\mu$m \\
    \hline 
    \end{tabular}
    \label{tab:initial_parameters}
\end{table}
%%%%%%%%%%%%%%%%%%%%%%%%%%%%%%%%%%%%%%%%%%%

\subsection{\large CO depletion factor}
\label{co depletion factor}

In this section we study how the CO depletion factor, $f_{D}$, varies with astrochemical conditions. CO is typically the second most abundant molecule in molecular clouds after $\rm H_2$ and so it is important to understand its behaviour.
%The physical and chemical properties of different regions in the IRDCs may be reflected in the abundance of CO, causing variation in the CO abundance and consequently its depletion factor across the cloud \citep{2011ApJ...738...11H}. 
One of the main expectations is that CO freezes out rapidly from the gas phase at temperatures $\lesssim 20$~K, but then with eventual abundances set by a balance of the non-thermal desorption processes from the grains.

\begin{figure*}[!htb]
    \centering
     \includegraphics[width=0.9\textwidth]{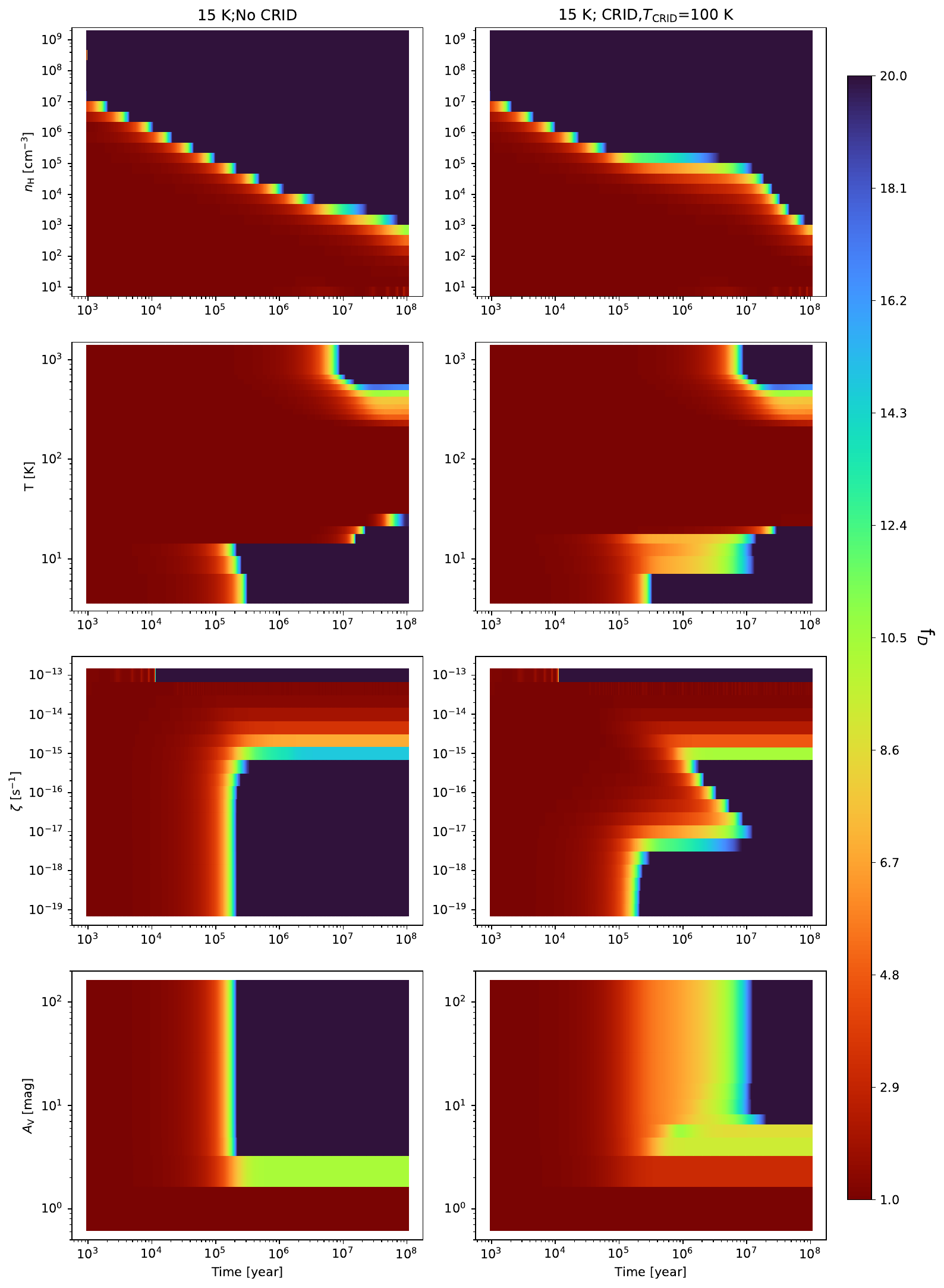}
     \vspace*{-1mm}
     \caption{Time evolution of CO depletion factor, $f_D$, evaluated in astrochemical model Grid 1 (left column) and Grid 2 (right column) as a function of the environmental variables of: (a) top row - number density of H nuclei, $n_{\rm H}$; (b) second row - temperature, $T$; (c) third row - cosmic ray ionisation rate, $\zeta$; (d) fourth row - visual extinction, $A_V$. In each panel the three remaining variables are held fixed at their fiducial level (i.e. $n_{\rm H}=10^5\:{\rm cm}^{-3}$, $T=15\:$K, $\zeta=10^{-17}\:{\rm s}^{-1}$, $A_V=100$~mag). Each time-evolving model is shown by a horizontal line with $f_D$ indicated by the colour scale.
     }
     \label{fig:fD_15K-1.0e-17}
 \end{figure*}

  \begin{figure*}[!htb]
    \centering
     \includegraphics[width=0.9\textwidth]{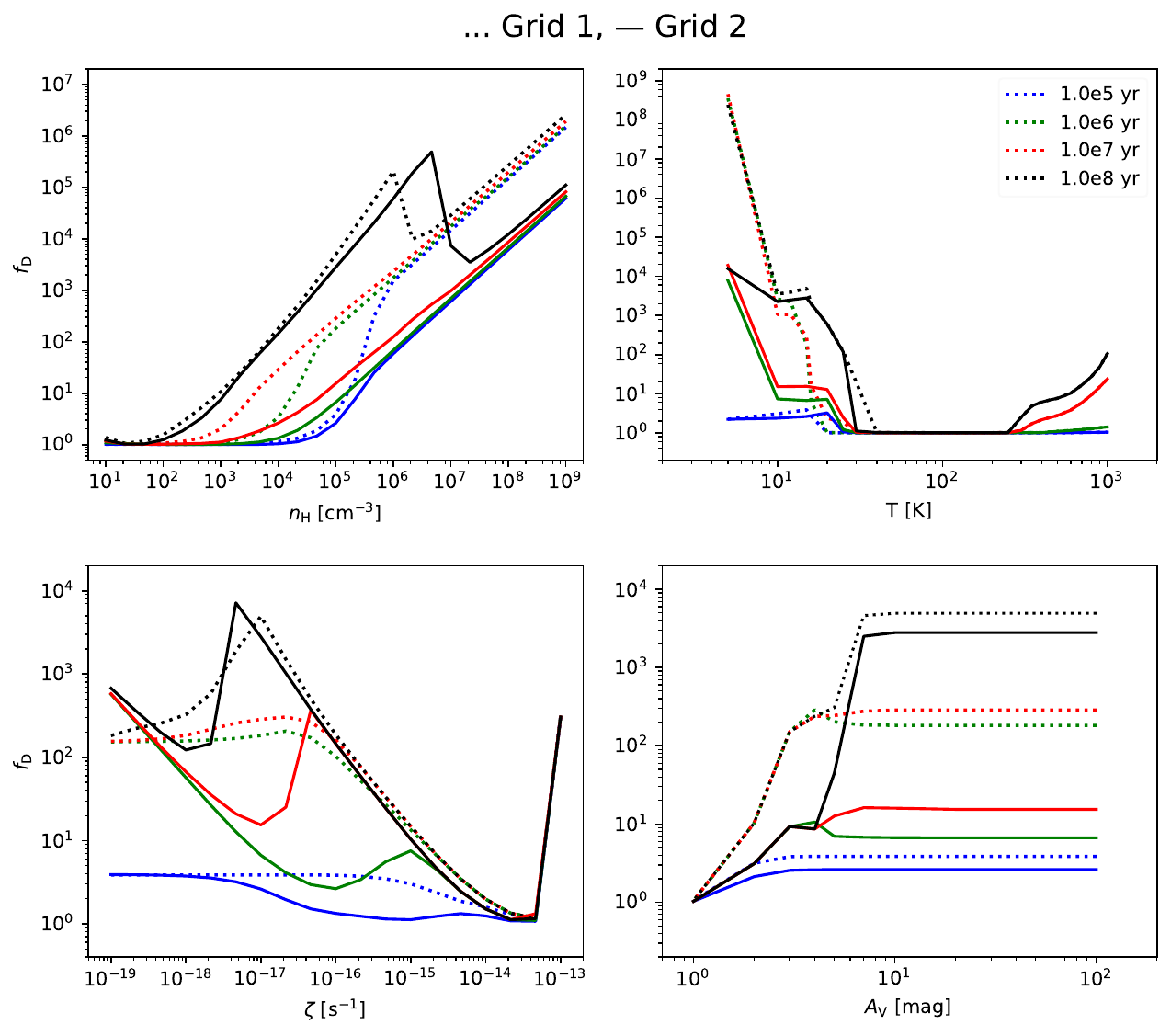}
     \vspace*{-2mm}
     \caption{CO depletion factor, $f_D$, evaluated in astrochemical model Grid 1 (dotted lines) and Grid 2 (solid lines), at fixed times of $10^5$ (blue), $10^6$ (green), $10^7$ (red) and $10^8$~yr (black) as a function of the environmental variables of: (a) top left - number density of H nuclei, $n_{\rm H}$; (b) top right - temperature, $T$; (c) bottom left - cosmic ray ionisation rate, $\zeta$; (d) bottom right - visual extinction, $A_V$. In each panel the three remaining variables are held fixed at their fiducial level, that is $n_{\rm H}=10^5\:{\rm cm}^{-3}$, $T=15\:$K, $\zeta=10^{-17}\:{\rm s}^{-1}$, $A_V=100$~mag.
     }
     \label{fig:fD_values_15K-1.0e-17}
 \end{figure*}
 
 %%%%%%%%%%%%%%%%%%%%%%%%%%%%%%%%%%%%%%%

In the following, we define $f_{D}$ via: %(\ref{rel:fd}) and will discuss its variation as a function of the model parameters.
\begin{equation}
    f_{D}(t) = \frac{1.4 \times 10^{-4}}{X[{\rm CO}](t)},
    \label{rel:fd}
\end{equation}
where $1.4 \times 10^{-4}$ is the initial gas phase abundance of CO with respect to H nuclei in our models (i.e. assuming all C is in this form) and $X[{\rm CO}](t)$ is the actual gas phase abundance of CO with respect to H nuclei, which evolves in time.

To explore the time evolution of $f_D$ and its dependence on density, temperature, CRIR and $A_V$, we choose a fiducial reference case with $n_{\rm H}=10^5\:{\rm cm}^{-3}$, $T=15\:$K, $\zeta=10^{-17}\:{\rm s}^{-1}$ and $A_V=100\:$mag. We then vary each of these variables systematically, while holding the others constant. The results are shown in %Figures~\ref{fig:fD}a-d ({\bf and 
Fig.~\ref{fig:fD_15K-1.0e-17}a-d). Simple profiles of $f_D$ versus these environmental variables at several fixed times are shown in %Figure~\ref{fig:fD_values}a-d ({\bf and
Fig.~\ref{fig:fD_values_15K-1.0e-17}a-d). The time evolution of the abundances of CO, $\rm HCO^+$ and $\rm N_2H^+$ are shown in %Figures~\ref{fig:CO_evolution_fD}a-d and 
Fig.~\ref{fig:CO_evolution_fD_15K_1.0e-17}a-d. We now discuss these results in the following four sub-sections.

\begin{figure*}[!htb]
\centering
\includegraphics[width=\textwidth]{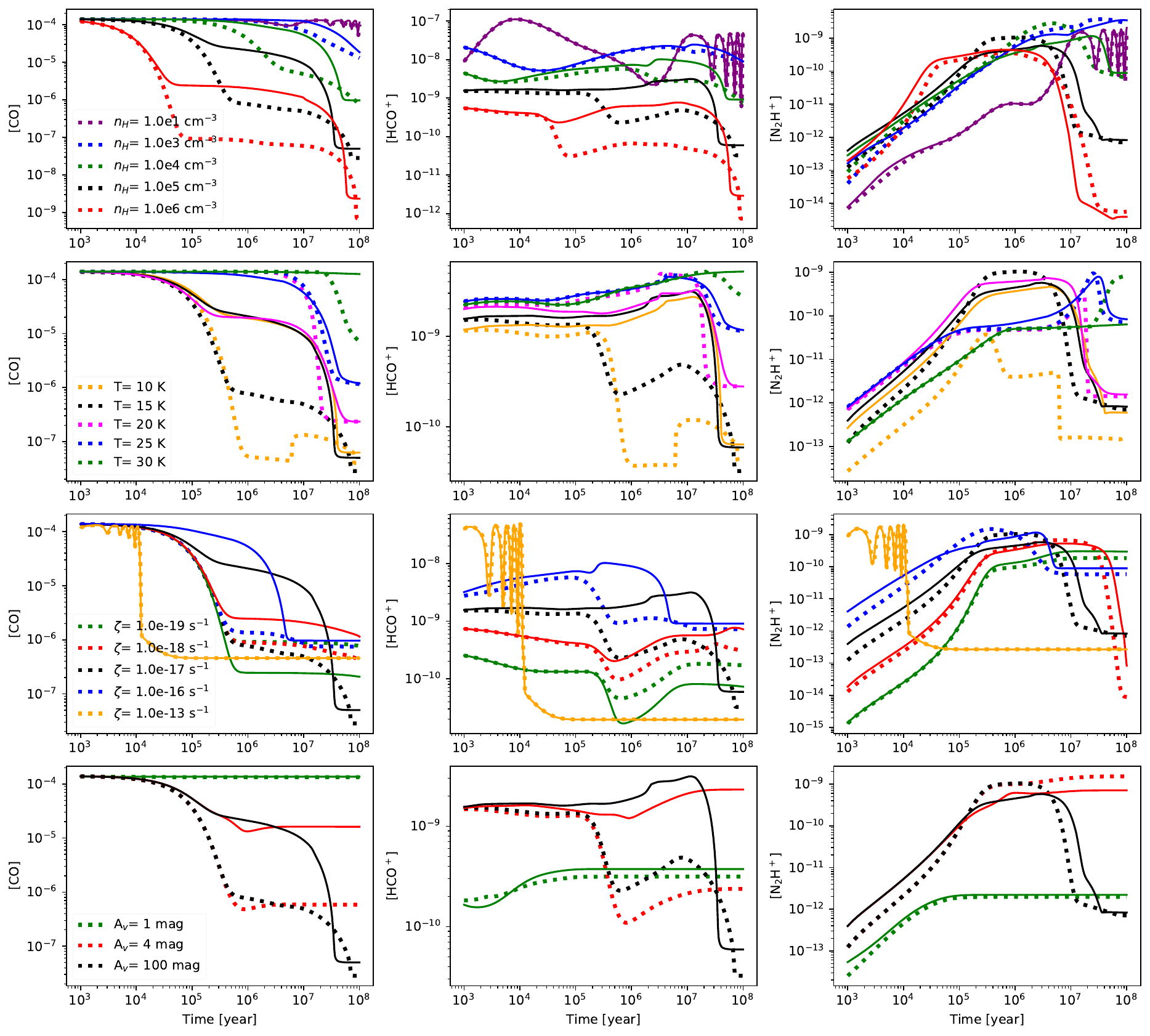}
     \vspace*{-5mm}
     \caption{Time evolutions of the abundances of CO (left column), HCO$^+$ (middle column) and N$_2$H$^+$ (right column) evaluated in astrochemical model Grid 1 (dotted lines) and Grid 2 (solid lines), exploring the effects of the environmental variables of: (a) top row - number density of H nuclei, $n_{\rm H}$; (b) second row - temperature, $T$; (c) third row - cosmic ray ionisation rate, $\zeta$; (d) bottom row - visual extinction, $A_V$. In each row the three remaining variables are held fixed at their fiducial level (i.e. $n_{\rm H}=10^5\:{\rm cm}^{-3}$, $T=15\:$K, $\zeta=10^{-17}\:{\rm s}^{-1}$, $A_V=100$~mag).
     }
     \label{fig:CO_evolution_fD_15K_1.0e-17}
\end{figure*}

%%%%%%%%%%%%%%%%%%%%%%%%%%%%%%%%%%%%%%%%%%%
%\begin{figure*}[!ht]
%    \centering
%    \addtolength{\leftskip} {-2.3cm}
%    \addtolength{\rightskip}{-1cm}
%     \includegraphics[width=23cm, height=4cm]
%     {chapters/images/n_rates.pdf}
%     \caption{\tiny Evolution of the rates of multiple reactions with the highest rates, in the models with different n(H). The rest of the parameters have been set to the value of fiducial model (Fig~\ref{tab:fiducial_parameters})}
%     \label{fig:n_rates}
% \end{figure*}
 %%%%%%%%%%%%%%%%%%%%%%%%%%%%
% \begin{figure*}[!ht]
%    \centering
%    \addtolength{\leftskip} {-2.3cm}
%    \addtolength{\rightskip}{-1cm}
%     \includegraphics[width=23cm, height=4cm]
%     {chapters/images/T_rates.pdf}
%     \caption{\tiny Evolution of the rates of multiple reactions with the highest rates, in the models with different T. The rest of the parameters have been set to the value of fiducial model (Fig~\ref{tab:fiducial_parameters}).}
%     \label{fig:T_rates}
% \end{figure*}
 %%%%%%%%%%%%%%%%%%%%%%%%%%%%%%%%%%%%%%%

%\begin{figure*}[!ht]
%    \centering
%    \addtolength{\leftskip} {-2.3cm}
%    \addtolength{\rightskip}{-1cm}
%     \includegraphics[width=23cm, height=4cm]
%     {chapters/images/CRIR_rates}
%     \caption{Evolution of the rates of multiple reactions with the highest rates, in the models with different CRIRs. The rest of the parameters have been set to the value of fiducial model (Fig~\ref{tab:fiducial_parameters}).}
%     \label{fig:CRIR_rates}
% \end{figure*}
%%%%%%%%%%%%%%%%%%%%%%%%%%%%%%%%%%%%%%%%%

\subsubsection{Dependence on density}

In the high density regime, $n_{\rm H}\gtrsim 10^5\:{\rm cm}^{-3}$, the general trend shown in Fig.~\ref{fig:fD_15K-1.0e-17}a is that the CO depletion factor increases with time and can reach very high values, $f_D\gtrsim20$, within a few Myr. As expected, the evolution proceeds more quickly at higher densities, which lead to higher rates of CO molecules hitting and sticking to the dust grains.

At low densities, $n_{\rm H}\lesssim 10^2\:{\rm cm}^{-3}$, the freeze-out rates become very low and the desorption rates are high enough to keep most CO in the gas phase. We note that at these densities and at relatively higher CRIRs ($\gtrsim 10^{-16}\:{\rm s}^{-1}$) (not shown in Fig.~\ref{fig:fD_15K-1.0e-17}a) competing processes that destroy CO, that is via interaction with CR-produced He$^+$ or CR-induced FUV photons, become more important (it is important to note that when $A_V=100\:$mag, the destruction of CO via external FUV photons is unimportant). The system can then evolve to having most of the C in atomic or ionised form, and hence the CO abundance is lowered and $f_D$ appears large.

Focusing on the density range that is relevant to IRDCs (at least as averaged on $\sim$parsec scales), that is $n_{\rm H}\sim 10^4$ to $10^5\:{\rm cm}^{-3}$, we see from Fig.~\ref{fig:fD_values_15K-1.0e-17}a that after 100,000~yr $f_D$ is still quite small ($\lesssim 2$) at the low density end of this range, but is rising steeply with density at the high-density end. By 1~Myr, it is still relatively low at the low density end, while at the high-density end of the range it is approaching quasi-equilibrium values of $f_D\gtrsim 100$ for grid 1 and $f_D\gtrsim 10$ for grid 2, which would hold even at $\sim 10\:$Myr. However, as we see later, this evolution of $f_D$ with time and its dependence on density can be very sensitive to the temperature in the range from $\sim15$ to $20\:$K.

 \subsubsection{Dependence on temperature}
 
Time dependant evolutions of $f_D$ at different temperatures are shown in Figs.~\ref{fig:fD_15K-1.0e-17}b and \ref{fig:fD_values_15K-1.0e-17}b. The actual gas phase abundances of CO versus time for these models are shown in Fig.~\ref{fig:CO_evolution_fD_15K_1.0e-17}, along with similar evolutions of $\rm HCO^+$ and $\rm N_2H^+$. The strong dependence of $f_D$ evolution with $T$ in the range from $\sim10-50$~K is clearly illustrated, that is above this temperature range, gas phase CO remains the main reservoir of C. At temperatures $\gtrsim500\:$K, C is still mostly in the gas phase, but mostly in the form of HCN, rather than CO.

\subsubsection{Dependence on cosmic ray ionisation rate}

Time dependant evolutions of $f_D$ at different CRIRs are shown in Figs.~\ref{fig:fD_15K-1.0e-17}c and \ref{fig:fD_values_15K-1.0e-17}c. The actual gas phase abundances of CO versus time for these models are shown in Fig.~\ref{fig:CO_evolution_fD_15K_1.0e-17}c, along with similar evolutions of $\rm HCO^+$ and $\rm N_2H^+$. In this figure, we sometimes see oscillations in the abundances of CO, HCO$^+$ and N$_2$H$^+$, that is in the models with high CRIR and where the condition $10^{18}\:{\rm cm^{-3}\:s}< n_{\rm H}/\zeta <2.2\times 10^{18}\:{\rm cm^{-3}\:s}$.
%at which the CRIR is in highest level and the value of $\frac{n(H)}{\zeta}$ is 1.0e18 cm$^{-3}$s. 
%Figures~\ref{fig:os_n} and \ref{fig:os_cr} indicates that the oscillatory feature appears in the models at which this ratio is between 1.0e18 and 2.2e18 cm$^{-3}$ s$^{-1}$, the models with lower densities and higher CRIR consistenting with 
Such behaviour has been noted in the previous study of \citet{maffucci2018astrochemical}, corresponding to the high ionisation phase (HIP) case discussed by \citet{roueff2020sustained}. 
%In upper left panel in figure \ref{fig:fD} and the upper left panel in figure ~\ref{fig:CO_evolution_fD} the variation of f$_D$ with respect to $\zeta$ is shown. 
More generally, the results of Figures \ref{fig:fD_15K-1.0e-17}c and \ref{fig:fD_values_15K-1.0e-17}c indicate that as $\zeta$ increases, CO depletion begins later in time, but reaches a lower equilibrium value. A different mode of evolution occurs at the highest value of CRIR ($\zeta = 10^{-13}\:{\rm  s}^{-1}$), in which C is mostly in the form of $\rm C^+$.

\subsubsection{Dependence on visual extinction}

%\begin{figure*}[!ht]
%    \centering
%    \addtolength{\leftskip} {-2.7cm}
%    \addtolength{\rightskip}{-1.8cm}
%     \includegraphics[width=17cm, height=4cm]
%     {chapters/images/Av_rates.pdf}
%     \caption{\tiny Evolution of the rates of multiple reactions with the highest rates, in the models with different A$_v$. The rest of the parameters have been set to the value of fiducial model (Fig~\ref{tab:fiducial_parameters}).}
%     \label{fig:Av_rates}
% \end{figure*}

Time dependant evolutions of $f_D$ at different levels of $A_V$ are shown in Figs.~\ref{fig:fD_15K-1.0e-17}d and \ref{fig:fD_values_15K-1.0e-17}d. The actual gas phase abundances of CO versus time for these models are shown in Fig.~\ref{fig:CO_evolution_fD_15K_1.0e-17}d, along with similar evolutions of $\rm HCO^+$ and $\rm N_2H^+$.
These figures show that in the models with $A_V>5$~mag, CO evolution has little dependency on the level of visual extinction so that CO depletion becomes large after about 1 Myr for grid 1 and after about 10 Myr for grid 2, no matter the value of $A_V$.

%It is demonstrated in Figure~\ref{fig:Av_rates} (panel a) that at very low A$_v$, CO remains in gas phase and f$_D$ stays below 2, which is the result of equilibrium between thermal desorption and freeze-out.

%As A$_v$ increases (panel b and c in Figure~\ref{fig:Av_rates}), the rates of thermal desorption and freeze-out decreases, but instead the rates of the reactions~\ref{rel:top_rates_zeta}, which are responsible of CO formation and destruction, increases Figure~\ref{fig:Av_rates} (panel b and c) and Figure~\ref{fig:fD} shows that in the model with the highest A$_v$ (100 mag), CO starts to be depleted since $\approx$ 1.0e6 years and reaches equilibrium before $\approx$ 1.0e7 years.

%%%%%%%%%%%%%%%%%%%%%%%%%%%%%%%%%%%%%%%%%%%

\section{Observational data and abundance estimates}\label{observations}

%%%%%%%%%%%%%%%%%%%%%%%%%%%%%%%%%%%%%%%%%
%\input{chapters/tables/table2}
\setlength{\tabcolsep}{7pt} % horizontal space
\renewcommand{\arraystretch}{1.4} % vertical space

\begin{table*}[t]\centering
\caption{Observed molecular species, rest frequencies, rotational quantum numbers of the transitions, rotational constants, rotational partition function, Einstein $A$ coefficients, energy of lower state, degeneracy of upper state, velocity resolution, $B_{\rm eff}$ and HPBWs.}
  \begin{tabular}{c c c c c c c c c c c}
  \hline
  \hline
  molecule & $\nu$ (GHz) & $J$ & $B$~(Hz) & $Q_{\rm rot}$($T_{\rm ex}$=7.5K) & $A$~(s$^{-1}$) & $E_{L}$~(K) & $g_{u}$ & $d\rm{v}~(km s$$^{-1}$) & $B_{\rm eff}$ & $\theta(\arcsec)$ \\
  \hline
   HC$^{18}$O$^+$ & 85.1621570 & 1-0 & 4.26e10 & 3.90  & 3.6450E-05 & 0.00 & 3 &  0.6876 & 0.81 & 28.89\\
    \hline
    H$^{13}$CO$^+$ & 86.7542884 & 1-0 & 4.34e10 & 3.83 & 3.8539E-05 & 0.00 & 3 &  0.6749 & 0.81 & 28.36 \\
    \hline
    %SiO & 86846.96 & Q($T_{ex}$) = 1.010600$\times T^{0.990020}$ & 2.9275E-05 & 2.08 & 5 & 0.6742 & 83.39\\
    %\hline
     HN$^{13}$C & 87.0908500 & 1-0 & 4.35e10 &  3.82 & 2.3840E-05 & 0.00 & 3 & 0.6723 & 0.81 & 28.25\\
    \hline
     %CCH & 87316.925 & Q($T_{ex}$) = 2.106200$\times T^{0.981580}$ & 1.5314E-06 & 0.00216 & 3 & 0.6706 & 83.30\\
    %\hline
    HNCO & 87.9250443 & 4-3 & 1.11e10 & 36.90 & 8.7801E-06 & 6.33 & 9 & 0.6659 & 0.81 & 27.98\\
    \hline
    HCN & 88.6304160 & 1-0 & 4.43e10 & 3.88 & 8.0884E-06 & 0.00 & 3 & 0.6606 & 0.81 & 27.76\\
    \hline
    N$_2$H$^+$ & 93.1761300 & 1-0 & 4.66e10 & 3.71 & 3.6300E-05 & 0.00 & 3 & 0.0628 & 0.80 & 26.40\\
    \hline
    C$^{18}$O & 109.7821734 & 1-0 & 5.49e10 & 3.11& 6.2661E-08 & 0.00 & 3 & 0.0533 & 0.78 & 22.41\\
    \hline
     %DCO+ & 144077.2144 & Q($T_{ex}$) = 1.842760$\times T^{0.991545}$ & 2.1184E-04 & 3.46 & 5 & 0.4064 & 72.82 \\
    %\hline
     CH$_3$OH & 145.1031850 & 3-2 & 2.47e10 & 13.25 & 1.2323E-05 & 6.96 & 7 & 0.4035 & 0.73 & 16.95\\
    \hline
    H$_2$CO & 145.6029490 & 2-1 & 3.88e10 & 10.52 & 7.8127E-05 & 3.49 & 5 & 0.4021 & 0.73 & 16.90\\
    \hline
    %\dots & \dots
  \end{tabular}
  \vspace*{1mm}
  
  \label{tab:modelcules_parameters_Ntot}
\end{table*}
%%%%%%%%%%%%%%%%%%%%%%%%%%%%%%%%%%%%%%%

\subsection{Molecular line data}

The rotational transitions listed in Table~\ref{tab:modelcules_parameters_Ntot} were mapped towards the IRDC G28.37+00.07 (hereafter Cloud C, Butler \& Tan \citeyear{butler2009mid,butler2012mid}) in August 2008, May 2013 and September 2013 using the IRAM (Instituto de Radioastrom\'ia Milim\'etrica) 30m telescope in Pico Velata, Spain. Observations were performed in On-The-Fly mode with map size $264\arcsec\times252\arcsec$; central coordinates $\alpha = 18^{h}42^{m}52.3^{s}$, $\delta =-4^{\circ}02\arcmin26.2\arcsec$; and a relative off position of ($-370\arcsec$, 30$\arcsec$). 
An angular separation in the direction perpendicular to the scanning direction of 6$\arcsec$ was adopted.

The VESPA receiver was utilised for observations of $\rm C^{18}O(1-0)$ and $\rm N_2H^+(1-0)$, which provided a spectral resolution of 20 kHz, corresponding to a velocity resolution of 0.053 and 0.063~$\rm km\:s^{-1}$, respectively. The other species were observed using the FTS spectrometer with 200 kHz resolution, corresponding to velocity resolutions of $\sim0.4$ to $0.7\:{\rm km\:s}^{-1}$. 
%For the other transitions, the data have velocity resolutions of $\sim0.4$ to $0.7\:{\rm km\:s}^{-1}$.
%jct - more information is needed about this observing mode
In the worst case, that is at the lowest frequency of 81.5~GHz, the angular resolution of the molecular line data has a half power beam width (HPBW) of $32\arcsec$ (corresponding to 0.78~pc at the 5~kpc distance of the IRDC).
Peak intensities were measured in units of antenna temperature, $T^{\star}_{A}$, and converted into main-beam temperature, $T_{\rm mb}$. Beam and forward efficiencies of 0.73 and 0.95, respectively, for $\nu\sim145\:$GHz, and 0.86 and 0.95 for $\nu\sim86\:$GHz were used. The final data cubes were produced using the {\sc CLASS} and {\sc MAPPING} software within the {\sc GILDAS} package\footnote{https://www.iram.fr/IRAMFR/GILDAS/}.

%%%%%%%%%%%%%%%%%%%%%%%%%%%%%%%%%%%%%%%%%%%%%%%%%%
\subsection{Mass surface density and temperature data}

The mid-infrared (MIR) extinction (MIREX) map of \cite{butler2012mid}, derived from 8~$\rm \mu m$ {\it Spitzer}-IRAC imaging data, with the near-infrared (NIR) correction of \cite{2013A&A...549A..53K}, was used as a first method to estimate the mass surface density, $\Sigma$, of selected regions in the IRDC. The advantage of this map is that it has a high angular resolution of 2\arcsec\ and the derivation of $\Sigma$ does not depend on the temperature of the cloud. The disadvantages of the map are that it cannot make a measurement in regions that are MIR bright. Furthermore, the map saturates at values of $\Sigma\sim 0.5\:{\rm g\:cm}^{-2}$, that is in regions with $\Sigma$ approaching this value the map underestimates the true value of $\Sigma$.

An independent measure of $\Sigma$ for the IRDC has been derived by \cite{lim2016distribution} based on {\it Herschel}-PACS and SPIRE sub-millimetre (sub-mm) emission imaging data, with this analysis also simultaneously yielding an estimate of the dust temperature. These maps have an angular resolution of $18\arcsec$ for the $\Sigma$ map and $36\arcsec$ for the $T$ map, that is much lower than that of the MIREX map. As discussed by \cite{lim2016distribution}, there are significant effects due to galactic plane foreground and background emission that contaminate the emission arising from the IRDC itself. In our analysis, we make use of the map derived from the galactic gaussian (GG) foreground-background subtraction method.

%Finally, we have made use, in our analysis, of the dust temperature and H$_2$ mass surface density ($\Sigma$) maps obtained by \cite{lim2016distribution} and \cite{2013A&A...549A..53K}, from Herschel and a combination of NIR and MIR data, respectively.
%of cloud C (see Figure\ref{fig:sigma_T_map}), obtained by \citep{lim2016distribution}, from a combination of Herschel images at 70 $\mu$m and Spitzer images at 8 $\mu$m and 24 $\mu$m. We also extracted $\Sigma$ values from the map obtained by \citep{2013A&A...549A..53K} (see table\ref{tab:physical_parameters_IRDC}), although we will continue with the data obtained by \citep{lim2016distribution}.

\subsection{Ten selected IRDC regions}

From the final data cubes, we extracted molecular line spectra from ten regions in the IRDC (see Fig.~\ref{fig:sigma_T_map}). Each region has a circular aperture of radius $16\arcsec$, which was set so that the angular resolution of the HPBW of the lowest frequency transition, that is from $\rm HC^{18}O^+(1-0)$, fits within the region. At the 5~kpc distance of the IRDC, the radius of each region corresponds to 0.39~pc.

The locations of the ten positions were selected for having relatively high values of $\Sigma$ and for having a range of star formation activities. In particular, positions P1 to P6 (shown with red circles in Fig.~\ref{fig:sigma_T_map}) are known to be sites of active star formation based on the presence of CO outflows \citep{kong2019widespread} and containing {\it Herschel}-Hi-GAL 70~$\rm \mu m$ point sources \citep{moser2020high}. However, these positions are still dark structures at $8~{\rm \mu m}$: indeed the P1 to P6 positions correspond approximately to the MIREX $\Sigma$ peaks of C1, C2, C4, C5, C6 and C8, respectively, identified and characterised by \cite{butler2012mid}.

\begin{figure*}[htb!]
    \centering
    %\addtolength{\leftskip} {-1.9cm}
    %\addtolength{\rightskip}{-1.5cm}
     \includegraphics[width=\textwidth]{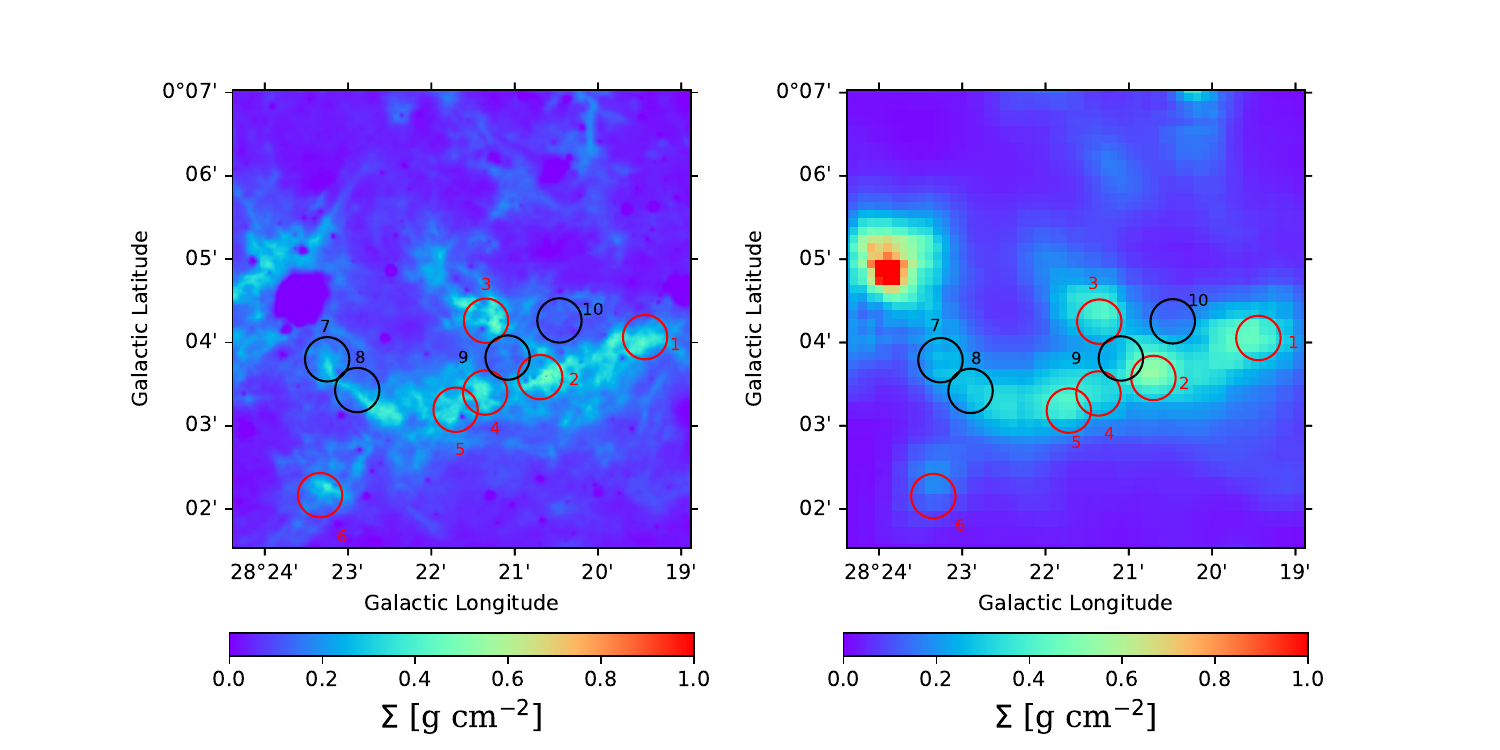}
    \vspace*{-7mm}
    \caption{(a) Left: MIREX derived mass surface density map of IRDC G28.37+00.07 from \cite{2013A&A...549A..53K} (scalebar is in $\rm g\:cm^{-2}$). The ten regions selected for our astrochemical study are marked with circles, labelled 1 to 10. Regions 1 to 6 (red) are known to harbour some star formation activity, while regions 7-10 have been specifically selected to avoid known protostellar sources. (b) Right: Sub-mm emission derived mass surface density map of the IRDC from \cite{lim2016distribution} using the Galactic Gaussian method of foreground-background subtraction (scalebar is in $\rm g\:cm^{-2}$). We note that it is of significantly lower angular resolution than the MIREX map (see text). The analysed positions are marked as in (a).}
    \label{fig:sigma_T_map}
\end{figure*}
%%%%%%%%%%%%%%%%%%%%%%%%%%%%%%%%%%%%%%%%%%%%5
\begin{figure}[htb!]
    \centering
    \addtolength{\leftskip} {-0.8cm}
    %\addtolength{\rightskip}{-8.5cm}
     \includegraphics[width=0.45\textwidth]{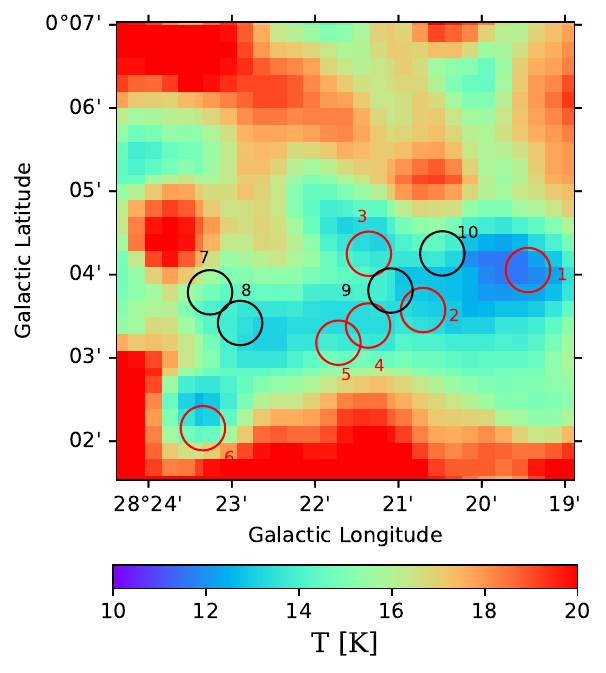}
    \vspace*{-4mm}
    \caption{Sub-mm emission derived dust temperature map of IRDC G28.37+00.07 from \cite{lim2016distribution} using the galactic gaussian method of foreground-background subtraction (scalebar in K). The ten regions selected for our astrochemical study are marked with circles, labelled 1 to 10, as in Fig.~\ref{fig:sigma_T_map}.
    }
    \label{fig:T_map}
\end{figure}

%%%%%%%%%%%%%%%%%%%%%%%%%%%%%%%%%%%%%%%%%%%%%%%%5

The positions from P7 to P10 (shown with black circles in Figs.~\ref{fig:sigma_T_map} and \ref{fig:T_map}) are not directly associated with massive condensations previously identified within the cloud \citep{rathborne2006infrared,butler2009mid,butler2012mid} and have been selected for being relatively quiescent in terms of their star formation activity \citep{kong2019widespread,moser2020high}.

%%%%%%%%%%%%%%%%%%%%%%%%%%%%%%%%%%%%%%%%%%%
%\input{chapters/tables/table1}
%\input{chapters/tables/table3_positions_abundances}

%\input{chapters/tables/landscape}
%\begin{landscape}
\setlength{\tabcolsep}{5pt} % horizontal space (was15pt)(Guiliana put 10pt)

\begin{table*}[htpb!]\centering
\Large
\caption{Estimated physical properties and molecular abundances of the P1 to P10 positions in IRDC G28.37+00.07.}
 \begin{adjustbox}{width=\textwidth} %(was width=\textwidth )(it was 1.3 textwidth)
  \begin{tabular}{c c c c c c c c c c c}
  \hline
  \hline
  Parameter & P1 & P2 & P3 & P4 & P5 & P6 & P7 & P8 & P9 & P10 \\
   \hline
   R.A. (J2000) & 18:42:46.687 & 18:42:50.709 & 18:42:49.490 & 18:42:52.583 & 18:42:53.970 & 18:43:00.610 & 18:42:54.636 & 18:42:55.289 & 18:42:50.591 & 18:42:47.860\\
   [0.5cm]
   Dec. (J2000) &  -4:04:08.830 & -4:03:14.763 & -4:02:21.514 & -4:02:44.647 & -4:02:31.320 & -4:01:32.639 & -4:00:52.451 & -4:01:21.835 & -4:02:47.469 & -4:03:08.458\\
   [0.3cm]
   \hline
   $\Sigma$ [g~cm$^{-2}$] & 0.394 & 0.438 & 0.330 & 0.322 & 0.336 & 0.130 & 0.239 & 0.270 & 0.315 & 0.171\\
   & 0.270 & 0.296 & 0.265 & 0.267 & 0.259 & 0.187 & 0.193 & 0.195 & 0.164 & 0.116\\
   [0.3cm]
   \hline
   $N_{\rm H}$ [10$^{23}$ cm$^{-2}$] & 1.68 & 1.87 & 1.41 & 1.38 & 1.44 & 0.56 & 1.02 & 1.15 & 1.35 & 0.73 \\
   & 1.15 & 1.26 & 1.13 & 1.14 & 1.11 & 0.80 & 0.83 & 0.83 & 0.70 & 0.50\\
   [0.3cm]
   \hline
   $n_{\rm H}$ [10$^{4}$ cm$^{-3}$] & 10.50 & 11.70 & 8.84 & 8.65 & 9.02 & 3.48 & 6.39 & 7.21 & 8.46 & 4.58 \\
   & 7.19 & 7.88 & 7.06 & 7.12 & 6.94 & 4.99 & 5.16 & 5.21 & 4.38 & 3.11 \\
   [0.3cm]
   \hline
   $t_{\rm ff}$ [yr] & 1.3$\times$10$^5$ & 1.3$\times$10$^5$ & 1.5$\times$10$^5$ & 1.5$\times$10$^5$ & 1.5$\times$10$^5$ & 2.3$\times$10$^5$ & 1.7$\times$10$^5$ & 1.6$\times$10$^5$ & 1.5$\times$10$^5$ & 2.0$\times$10$^5$\\
   [0.3cm]
   \hline
   $A_{V,{\rm tot}}$ [mag] & 88.4 & 98.4 & 74.2 & 72.6 & 75.8 & 29.2 & 53.7 & 60.5 & 71.1 & 38.5 \\
   & 60.7 & 66.6 & 59.6 & 60.1 & 58.3 & 42.1 & 43.4 & 43.9 & 36.9 & 26.1 \\
   [0.3cm]
   \hline
   $\mathrm{}{\bar{A}}_{V}$ [mag] & 22.1 & 24.6 & 18.6 & 18.2 & 19.0 & 7.3 & 13.4 & 15.1 & 17.8 & 9.6\\
   & 15.2 & 16.6 & 14.9 & 15.0 & 14.6 & 10.5 & 10.9 & 11.0 & 9.2 & 6.5 \\
   [0.3cm]
   \hline
   $T$ [K] & 11.8 & 13.1 & 13.5 & 13.4 & 13.8 & 14.1 & 14.9 & 13.6 & 13.3 & 13.6\\
   [0.3cm]
   \hline
   $\mathrm{[C^{18}O]}$ [10$^{-8}$] & 4.77$\pm$1.47 & 4.31$\pm$1.33 & 4.86$\pm$1.50 & 5.19$\pm$1.60 & 5.58$\pm$1.72 & 11.00$\pm$3.40 & 6.68$\pm$2.06 & 5.79$\pm$1.79 & 5.13$\pm$1.58 & 6.43$\pm$1.98\\
   [0.3cm]
   \hline
   $\mathrm{[CO]}$ [10$^{-5}$] & 1.52$\pm$0.56 & 1.37$\pm$0.51 & 1.54$\pm$0.57 & 1.65$\pm$0.61 & 1.77$\pm$0.65 & 3.51$\pm$1.29 & 2.12$\pm$0.78 & 1.84$\pm$0.68 & 1.63$\pm$0.60 & 2.05$\pm$0.75\\
   [0.3cm]
   \hline
   $\mathrm{[H^{13}CO^{+}]}$ [10$^{-11}$] & 1.02$\pm$0.31 & 1.11$\pm$0.34 & 0.87$\pm$0.27 & 0.97$\pm$0.30 & 1.02$\pm$0.32 & 1.52$\pm$0.48 & 0.84$\pm$0.26 & 0.69$\pm$0.22 & 0.96$\pm$0.30 & 1.36$\pm$0.42 \\
   [0.3cm]
   \hline
   $\mathrm{[HCO^+]}$ [10$^{-10}$] & 4.11$\pm$1.51 & 4.48$\pm$1.64 & 3.49$\pm$1.28 & 3.89$\pm$1.43 & 4.10$\pm$1.52 & 6.11$\pm$2.27 & 3.37$\pm$1.25 & 2.77$\pm$1.04 & 3.84$\pm$1.41 & 5.45$\pm$2.01\\
   [0.3cm]
   \hline
   $\mathrm{[HC^{18}O^{+}]}$ [10$^{-12}$] & 1.96$\pm$0.69 & 1.84$\pm$0.67 & 1.88$\pm$0.74 & 1.64$\pm$0.65 & 2.19$\pm$0.78 & & & & & 1.40$\pm$0.36\\
   [0.3cm]
   \hline
   $\mathrm{[HCO^+]}$ [10$^{-10}$] & 6.23$\pm$2.54 & 5.84$\pm$2.43 & 5.96$\pm$2.64 & 5.23$\pm$2.31 & 6.98$\pm$2.86 & & & & & 4.45$\pm$1.45\\
   [0.3cm]
   \hline
  $\mathrm{[N_2H^+]}$ [10$^{-10}$] & 0.55$\pm$0.17 & 1.79$\pm$0.56 & 1.53$\pm$0.47 & 1.52$\pm$0.47 & 1.10$\pm$0.34 & 2.62$\pm$0.81 & 0.91$\pm$0.28 & 0.80$\pm$0.25 & 1.14$\pm$0.35 & 1.47$\pm$0.46\\
  [0.3cm]
   \hline
   $\mathrm{[HN^{13}C]}$ [10$^{-11}$] & 1.41$\pm$0.43 & 1.61$\pm$0.50 & 1.85$\pm$0.57 & 1.64$\pm$0.51 & 1.24$\pm$0.39 & 2.85$\pm$0.89 & 0.94$\pm$0.30 & 0.91$\pm$0.29 & 1.41$\pm$0.44 & 1.88$\pm$0.59\\
   [0.3cm]
   \hline
  $\mathrm{[HNC]}$ [10$^{-10}$] & 5.65$\pm$2.08 & 6.47$\pm$2.37 & 7.43$\pm$2.73 & 6.59$\pm$2.43 & 4.97$\pm$1.84 & 11.40$\pm$4.24 & 3.79$\pm$1.43 & 3.67$\pm$1.38 & 5.66$\pm$2.09 & 7.54$\pm$2.80\\
  [0.3cm]
   \hline
   $\mathrm{[HCN]}$ [10$^{-11}$] & 11.10$\pm$3.44 & 10.30$\pm$3.22 & 5.86$\pm$1.88 & 8.16$\pm$2.57 & 8.53$\pm$2.69 & 26.60$\pm$8.29 & 10.10$\pm$3.28 & 10.50$\pm$3.33 & 7.66$\pm$2.40 & 10.00$\pm$3.32\\
   [0.3cm]
   \hline
   $\mathrm{[HNCO]}$ [10$^{-10}$] & 2.17$\pm$0.74 & 4.29$\pm$1.45 & 4.15$\pm$1.41 & 3.45$\pm$1.18 & 2.74$\pm$0.94 & 1.91$\pm$0.72 & 1.78$\pm$0.64 & 1.40$\pm$0.51 & 3.47$\pm$1.18 & 4.32$\pm$1.49\\
   [0.3cm]
   \hline
   $\mathrm{[H_2CO]}$ [10$^{-11}$] & 3.89$\pm$1.26 & 10.40$\pm$3.33 & 7.33$\pm$2.37 & 7.66$\pm$2.47 & 6.98$\pm$2.25 & 3.31$\pm$1.19 & 2.76$\pm$0.94 & 2.19$\pm$0.73 & 9.16$\pm$2.95 & 9.94$\pm$3.21\\
   [0.3cm]
   \hline
   $\mathrm{[CH_3OH]}$ [10$^{-9}$] & 0.52$\pm$0.18 & 1.73$\pm$0.60 & 1.74$\pm$0.61 & 1.33$\pm$0.46 & 1.26$\pm$0.44 & 0.49$\pm$0.18 & 0.45$\pm$0.16 & 0.33$\pm$0.12 & 1.54$\pm$0.54 & 1.93$\pm$0.67\\
   [0.3cm]
   \hline
       \hline
       %%%\hline
  \end{tabular}
 \end{adjustbox}
  \vspace*{-1mm}
  
  \label{tab:landscape-tab}
\end{table*}
%\end{landscape}

%%%%%%%%%%%%%%%%%%%%%%%%%%%%%%%%%%%%%%%%%%%
%\begin{landscape}
\setlength{\tabcolsep}{8pt} % horizontal space
\renewcommand{\arraystretch}{2} % vertical space

%\Large
\begin{table*}[htpb!]
 \caption{Extracted astrochemical parameters of the best models in R, UR and XR regions, using observed abundances of [CO],[HCO$^+$] and [N$_2$H$^+$] as Case 1, and all observed abundances as Case 2. The extracted parameters are $\zeta$ [s$^{-1}$], Time [year], $\mathrm{n}_H$ [cm$^{-3}$], T [K], $\mathrm{A}_{V}$ [mag] and $\chi^2$, respectively from left to right. For the Case 2 in the XR region, we have excluded the models with higher $\zeta$ (>1.0e-14 s$^{-1}$). Hence the models which could give us early time solution (<1.0e4 years) were eliminated in consequence.}\centering
 \begin{adjustbox}{width=\textwidth}
  \begin{tabular}{c c c c c}
  \hline
  \hline
  & & XR & UR & R\\
  & & [$\zeta/t/n_{\rm H}/T/A_V/\chi^2$]   &  [$\zeta/t/n_{\rm H}/T/A_V/\chi^2$]  &  [$\zeta/t/n_{\rm H}/T/A_V/\chi^2$]\\
  \hline
 P1 & Case 1 & 1.0e-18/1.70e5/1.0e5/15/20/0.11 & 1.0e-18/1.70e5/1.0e5/15/50/0.11 & 1.0e-19/2.00e7/2.2e4/20/5/0.10 \\
  & Case 2 & 4.6e-18/1.26e7/1.0e5/15/20/56.95 & 1.0e-18/1.00e8/2.2e5/20/10/37.75 & 4.6e-18/1.00e8/1.0e6/30/50/8.56\\
  %1.0e-13/9.33e3/1.0e5/15/20/43.015\\
  \hline
  P2 & Case 1 & 2.2e-18/2.00e5/1.0e5/15/20/0.69 & 1.0e18/4.17e5/4.6e4/15/50/0.17 & 2.2e-19/2.24e6/1.0e4/15/5/0.02\\
 & Case 2 & 2.2e-17/4.47e6/1.0e5/15/20/60.82  & 1.0e-16/1.45e5/2.2e5/10/10/42.64 & 1.0e-17/5.62e7/1.0e6/30/50/16.26\\
 %1.0e-13/2.88e3/1.0e5/15/20/52.752\\
  \hline
P3 & Case 1 & 2.2e-18/1.95e5/1.0e5/15/20/0.97 & 1.0e-18/3.98e5/4.6e4/15/10/0.49 & 2.2e-19/2.09e6/1.0e4/15/5/0.17\\
 & Case 2 & 1.0e-17/6.92e6/1.0e5/15/20/66.17 & 2.2e-18/5.25e7/2.2e5/20/10/47.71 & 4.6e-18/1.00e8/1.0e6/30/50/18.56\\
 %1.0e-13/9.33e3/1.0e5/15/20/55.295 \\
  \hline
  P4 & Case 1 & 2.2e-18/1.91e5/1.0e5/15/20/0.86 & 1.0e-18/3.89e5/4.6e4/15/10/0.31 & 2.2e-19/2.00e6/1.0e4/15/5/0.03\\
 & Case 2 & 1.0e-17/6.92e-17/1.0e5/15/20/64.20 & 2.2e-18/5.13e7/2.2e5/20/10/43.41 & 4.6e-18/1.00e8/1.0e6/30/50/15.68\\
 %1.0e-13/7.24e3/1.0e5/15/20/50.513\\
  \hline
  P5 & Case 1 & 2.2e-18/1.74e5/1.0e5/15/20/0.48 & 1.0e-18/3.55e5/4.6e4/15/10/0.21 & 4.6e-19/7.59e5/2.2e4/15/5/0.13\\
 & Case 2 & 1.0e-17/7.24e6/1.0e5/15/20/64.38 & 2.2e-18/5.13e7/2.2e5/20/10/43.08 & 4.6e-18/1.00e8/1.0e6/30/50/12.58\\
 %1.0e-13/9.33e3/1.0e5/15/20/46.611\\
  \hline
  P6 & Case 1 & 4.6e-18/2.88e5/4.6e4/15/7/2.47 & 1.0e-18/7.24e5/2.2e4/15/5/1.73 & 4.6e-19/1.66e6/1.0e4/15/5/1.33\\
 & Case 2 & 4.6e-17/2.51e5/4.6e4/15/7/53.73 & 4.6e-19/1.48e7/2.2e4/20/5/13.31 & 4.6e-19/1.48e7/2.2e4/20/5/13.31\\
 %4.6e-14/6.17e3/4.6e4/15/7/34.657\\
  \hline
  P7 & Case 1 & 1.0e-18/3.31e5/4.6e4/15/10/0.72 & 2.2e-19/8.32e5/2.2e4/15/20/0.39 & 1.0e-19/2.00e6/1.0e4/15/7/0.21 \\
 & Case 2 & 2.2e-18/2.45e7/4.6e4/15/10/58.95 & 1.0e-18/1.00e8/1.0e5/20/7/39.32 & 4.6e-18/1.00e8/1.0e6/30/50/11.61 \\
 %4.6e-14/2.45e4/4.6e4/15/10/40.362\\
  \hline
  P8 & Case 1 & 4.6e-19/3.80e5/4.6e4/15/10/0.43 & 2.2e-19/8.13e5/2.2e4/15/7/0.27 & 1.0e-19/1.95e6/1.0e4/15/5/0.04\\
 & Case 2 & 2.2e-18/2.45e7/4.6e4/15/10/55.88 & 1.0e-19/7.76e6/2.2e4/15/7/37.03 & 4.6e-18/1.00e8/1.0e6/30/50/12.06 \\
 %4.6e-14/2.45e4/4.6e4/15/10/39.054\\
  \hline
  P9 & Case 1 & 2.2e-18/1.78e5/1.0e5/15/20/0.48 & 1.0e-18/3.63e5/4.6e4/15/10/0.27 & 1.0e-19/2.34e6/1.0e4/15/100/0.12\\
 & Case 2 & 1.0e-17/7.08e6/1.0e5/15/20/63.83 & 2.2e-18/5.13e7/2.2e5/20/10/44.47 & 4.6e-18/1.00e8/1.0e6/30/50/13.95\\
 %1.0e-13/9.33e3/1.0e5/15/20/49.929\\
  \hline
  P10 & Case 1 & 2.2e-18/3.16e5/4.6e4/15/10/0.51 & 4.6e-19/7.94e5/2.2e4/15/20/0.20 & 2.2e-19/1.91e6/1.0e4/15/100/0.06\\
 & Case 2 & 4.6e-18/1.38e7/4.6e4/15/10/60.40 & 2.2e-17/5.01e5/4.6e4/10/7/38.31 & 4.6e-18/1.00e8/1.0e6/30/50/14.36\\
 %4.6e-14/1.00e3/4.6e4/15/10/44.001\\

        \hline
        \hline
    \end{tabular}
  \end{adjustbox}
  \vspace*{-1mm}
 
  \label{tab:gen_table}
  
\end{table*}
%\end{landscape}

%%%%%%%%%%%%%%%%%%%%%%%%%%%%%%%%%%%%%%%%

From the two mass surface density maps shown in Fig.~\ref{fig:sigma_T_map}, we have measured the average values of $\Sigma_{\rm MIREX}$ and $\Sigma_{\rm sub-mm}$ in each of the ten regions P1 to P10 (see Table~\ref{tab:landscape-tab}). This also allows an estimate of $N_{\rm H}$, assuming a mass per H of $\mu_{\rm H}=1.4 m_{\rm H} = 2.34\times 10^{-24}\:$g (i.e. approximately due to $n_{\rm He}=0.1 n_{\rm H}$ and ignoring contributions from other species). We assume an uncertainty of $\sim$30\% in the measurements of $\Sigma$ and $N_{\rm H}$ \citep{2013A&A...549A..53K,lim2016distribution}.

We have then estimated the average total amount of visual extinction, $A_V$, through each region using the standard dust model assumptions of \cite{2013A&A...549A..53K}, that is:
%By using Equation 4 in \cite{2013A&A...549A..53K}, we have used the N(H) measurements to calculate the visual extinction level, A$_v$, at each position:
\begin{equation}
   N_{\rm H} = 1.9\times 10^{21} \: {\rm cm}^{-2} \:\left(\frac{A_{V,{\rm tot}}}{\rm mag}\right).
   \label{eq:N_av}
\end{equation}
However, we then divide this estimate of the total extinction through the region by a factor of 4 in order to obtain an estimate of the typical extinction from the nearest boundary of the IRDC, that is $\bar{A}_V=A_{V,{\rm tot}}/4$.

%we have measured the dust temperature T, Hydrogen column density, N(H), Hydrogen volume density n(H), and visual extinction, A$_v$, toward the ten positions across cloud C.

%For each position, we obtained the mass surface density, $\Sigma$, as the mean value within a beam aperture, calculated across all pixels within a beam. Hence, we have converted the obtained $\Sigma$ values into Hydrogen column density values, N(H), using Equation 5 in \cite{2013A&A...549A..53K}:
%\begin{equation}
 %   \Sigma_{\rm H} = 1.4\rm m_{H}\rm N_{H}
  %  \label{rel:Hydrogen density}
%\end{equation}
%where, m$_{H}$ is the Hydrogen mass ($1.6735\times 10^{-24} g$) and the 1.4 factor takes into account a 40$\%$ helium contribution to the column density (implying the H:He number ratio of 10:1).\\

We have estimated the average number density of H nuclei, $n_{\rm H}$, in each region by assuming the material is distributed in a spherical volume with radius equal to the circular aperture radius of the region projected on the sky. These values are also listed in Table \ref{tab:landscape-tab}.

%converted the Hydrogen column density N(H) into Hydrogen volume density, n(H), assuming that the Hydrogen nuclei are distributed in a sphere of the same radius of the beam aperture. The obtained values of $\Sigma$, N(H) and n(H) are listed in Table \ref{tab:physical_parameters_IRDC}.\\

Finally, by using the dust temperature map from \cite{lim2016distribution}, we have estimated the $\Sigma$-averaged $T$ values for the ten selected positions (Fig.~\ref{fig:T_map}).
These values are also listed in Table \ref{tab:landscape-tab}. The uncertainties in $T$ are assumed to be at the level of $30\%$ \citep{lim2016distribution}.

%Toward each position, we have measured the mean weighted T , averaged across the pixels within the 32$^{\prime\prime}$ beam aperture and using as weight the corresponding $\Sigma$ value at that pixel (we note that the two maps have the same pixel size and shape).\\

%The obtained values of $\Sigma_{\rm H}$, T, N(H), n(H) and A$_v$ are listed in Table \ref{tab:physical_parameters_IRDC}. For the $\Sigma_{\rm H}$ and T measurements, we have assumed an uncertainty of $\sim$30\% \citep{2013A&A...549A..53K,lim2016distribution} and calculated the uncertainty in the other quantities following statistical error propagation rules. 

%\input{chapters/molecular_abundances}

\subsection{Molecular abundances}

%All spectra have been extracted over a beam-aperture of 32$^{\prime\prime}$ HPBW, corresponding to the poorest angular resolution for the observed transitions (81.5 GHz). In some cases, spectra were smoothed to improve the signal-to-noise- ratio and the final velocity resolution of the lines are reported in Table~\ref{tab:modelcules_parameters_Ntot}.

We now estimate column densities of the nine observed species listed in Table~\ref{tab:modelcules_parameters_Ntot} in the 10 regions of the IRDC and use these results to calculate their abundances with respect to the total H nuclei, that is $N_X/N_H$. For the column density estimation we have used Equation A4 of \cite{caselli2002molecular}:
% \begin{equation}
 \begin{eqnarray}
 %\begin{split}
  N_{X} & = & \frac{8 \pi \nu^{2}}{A g_{u}c^{3}}\frac{k}{h} \frac{1}{J_{\nu}(T_{\rm ex})-J_{\nu}(T_{\rm bg})}\nonumber\\
     & \times & \frac{Q_{\rm rot}}{1-{\rm exp} (-h\nu/[kT_{\rm ex}])} \frac{1}{{\rm exp}(-E_{L}/[kT_{\rm ex}])} \int T_{\rm mb} d v,
     \label{rel:N_tot}
 %\end{split}
\end{eqnarray}
% \end{equation}
where $\nu$ is the frequency of the observed transition, $A$ is the Einstein coefficient obtained from the CDMS catalog\footnote{https://cdms.astro.uni-koeln.de/classic/}, $g_{u}=2J+1$ is the statistical weight of the upper level, $E_{L}$ is the energy of the lower level, and
%in unit of temperature (K), 
$J_{\nu}(T_{\rm ex})$ and $J_{\nu}(T_{\rm bg})$ are the equivalent rayleigh-jeans (RJ) excitation and background temperatures,
%Rayleigh-Jeans 
%({\bf aren't these equivalent temperature of a black body (RJ equivalent temperature)? }), 
%jctfinal - where does this comment come from?
respectively. For these we set $T_{\rm ex}$= 7.5 K based on the multi-transition CO study of an IRDC by \cite{hernandez2012virialized}, while $T_{\rm bg}$=2.73 K from the cosmic microwave background (CMB). Finally, $Q_{\rm rot}$ is the rotational partition function at the assumed $T_{\rm ex}$. For each molecule, $Q_{\rm rot}$ has been calculated by fitting its values from jet propulsion laboratory (JPL) catalogue\footnote{https://spec.jpl.nasa.gov/ftp/pub/catalog/catdir.html}, except in the case of N$_2$H$^+$ and HCN for which, since their main rotational transitions split into hyperfine components, the partition function reported in the JPL catalogue includes an additional factor and it does not coincide with the pure rotational partition function. Hence, for N$_2$H$^+$ and HCN, we have calculated Q$_{\rm rot}$ as:
\begin{equation}
 %\begin{split}
  Q_{\rm rot} = \sum_{J=0}^{\infty} (2J+1) {\rm exp} \left(\frac{-E_{J}}{kT_{\rm ex}}\right),
  \label{rel:Q_rot}
 %\end{split}
 \end{equation}
 %jctfinal - should T be T_{\rm ex} in the above equation?
where, $E_{J}=J(J+1)hB$ is the energy of the $J$ level and $B$ is the rotational constant of the molecule.
The various parameters used to derive the abundances of the species are listed in Table \ref{tab:modelcules_parameters_Ntot}.
Finally, we note that $N_{\rm N_2H^+}$ and $N_{\rm HCN}$ obtained from Eq.~(\ref{rel:N_tot}) have been corrected for the transition statistical weights of 0.11 and 0.3333, respectively, since only one of the hyperfine lines was used in each case to derive total abundances. %to take into account the presence of hyperfine structures, i.e., $\approx$ 0.11 for N$_2$H$^+$ \citep{henshaw2014dynamical} and $\approx$ 0.3333 for HCN. All used spectroscopic quantities are reported in Table \ref{tab:modelcules_parameters_Ntot}.

In deriving column densities from the observed spectra, which are shown in Figs.~\ref{fig:spectra_A} and \ref{fig:spectra_B}, we must make a choice of the velocity range over which to integrate the line intensities. From inspection of the spectra, this range is chosen to have a common extent of 77 to 82 km~s$^{-1}$, motivated by the goal of covering the dense gas tracers of $\rm H^{13}CO^+(1-0)$, $\rm HC^{18}O^+(1-0)$ and $\rm N_2H^+(1-0)$. Examining the spectra of Figs.~\ref{fig:spectra_A} and \ref{fig:spectra_B}, we see that C$^{18}$O(1-0) generally shows emission from several components that span a broader velocity range than the above dense gas tracers. In our analysis, we only measure the CO column density from the part of C$^{18}$O(1-0) emission that falls in the velocity range from 77 to 82 km~s$^{-1}$.

%Looking at the spectra of C$^{18}$O, H$^{13}$CO$^{+}$, HC$^{18}$O$^{+}$  and N$_2$H$^+$ (Figures~\ref{fig:spectra_A} and \ref{fig:spectra_B}) for all positions, the C$^{18}$O emission likely shows multiple velocity components and shows a broader velocity range compared to the other species. Due to the ubiquitousness of CO along the line of sight, the CO abundance may be overestimated. On the other hand, H$^{13}$CO$^{+}$, HC$^{18}$O$^{+}$ and N$_2$H$^+$ show much narrower line widths. Thus, comparing the velocities of H$^{13}$CO$^{+}$, HC$^{18}$O$^{+}$ and N$_2$H$^+$ transition and choosing the narrowest range of velocity (77-82 km.s$^{-1}$), mostly based on N$_2$H$^+$ spectra, extracted abundance are demonstrated in Table \ref{tab:abundances}.\\

%We estimated the area under the curve of all spectra, in the range of 77-82 km.s$^{-1}$. Since resolution of the spectra is different for each molecule, we defined a condition for the channels chosen in 77-82 km.s$^{-1}$. For lower boundary (77 km.s$^{-1}$), if the central velocity of the channel (V$_c$) is bigger than 77 (V$_c$ > 77), the channel is included to calculate the area of the spectrum, otherwise it is excluded. For upper boundary (82 km.s$^{-1}$), if the central velocity of the channel is smaller than 82 (V$_c$ < 82), the channel is included, otherwise is excluded.\\

To estimate the uncertainties in the total column densities, we consider the following factors. First is the uncertainty in the integrated intensity of each line, which we assess to be equal to the noise integrated in a line-free spectral range and calculated as:
\begin{equation}
    %\sigma_{area}=\sqrt{\rm N}\times rms \times dv,
    \int T_{\rm mb,noise} d v = \sqrt{n} \Delta T_{\rm mb,noise} d v
    \label{rel:Uncertainty}
\end{equation}
where $n$ is the number of channels, each of velocity width $dv$, that are summed in the integration and $\Delta T_{\rm mb,noise}$ is the noise in an individual channel.

Second is the systematic uncertainty due to the need to assume a value of $T_{\rm ex}$. We note that $T_{\rm ex}=7.5\:$K has been adopted for all the species. If $T_{\rm ex}$ is varied by $\pm30\%$ about this value, the estimated total column density changes by $\sim 7-28\%$. For most species, this systematic error dominates over that due to the noise in the integrated intensity. In our analysis, we estimate the overall uncertainty in total column densities by summing the above two sources of uncertainties in quadrature. However, we note that there are other potential sources of systematic error, such as the assumption that the emission in the particular chosen velocity range is a good representation of the physical component of the molecular cloud that is being modelled.

For some of the species, like CO, $\rm HCO^+$ and HNC, the column density is estimated by conversion from a rarer isotopologue. This involves another source of systematic uncertainty. For these conversions, we assume $\rm [^{12}C]/[^{13}C]=40.2\pm 8.0$ \citep{zeng201715n} and $\rm [^{16}O]/[^{18}O]=318\pm 64$ \citep{hezareh2008simultaneous} and these levels of uncertainty are also propagated in quadrature for the final column density estimate.

We note that, since no significant signal is detected for the HC$^{18}$O$^{+}$(1-0) transition towards the positions P6, P7, P8, P9, and P10, we have estimated their HC$^{18}$O$^{+}$ column density from the average of their spectra. Then, we have apportioned this column density among the five positions in proportion to their H column densities.

%divided obtained value by the average Hydrogen column density toward the five positions p6-p10. The obtained HC$^{18}$O$^{+}$ abundance has been used for the 5 positions, p6 to p10.

%The CO and HCO$^+$ molecular abundances have been obtained by converting the abundances of C$^{18}$O and H$^{13}$CO$^{+}$ (or HC$^{18}$O$^{+}$) respectively. 

The final step of the analysis is to estimate abundances of species with respect to H nuclei that are assumed to be present in the same molecular cloud component. For this we simply divide the above estimates of total species column density by the estimate of H nucleus column density obtained previously either via {\it Herschel}-observed sub-mm dust emission or via MIR extinction. Each of these methods is assumed to have a 30\% uncertainty, which is summed in quadrature for the final abundance estimate. The obtained results are summarised in Table \ref{tab:landscape-tab}. We note that the H column densities obtained by both methods are themselves quite similar (see above) and, for simplicity, in the remaining analysis we adopt those estimates based on the sub-mm dust continuum emission.

%Finally, we have obtained molecular abundances, as the ratio between molecular and hydrogen column density, at each position and for each molecule. 

%%%%%%%%%%%%%%%%%%%%%%%%%%%%%%%%%%%%%%%%%%
\begin{figure*}
    \centering
    %\includegraphics[width=\textwidth]{chapters/images/both_maps_16_arsec.pdf}
    %\addtolength{\leftskip} {-3.1cm}
    %\addtolength{\rightskip}{-1.5cm}
     \includegraphics[height=25cm,width=15cm,]{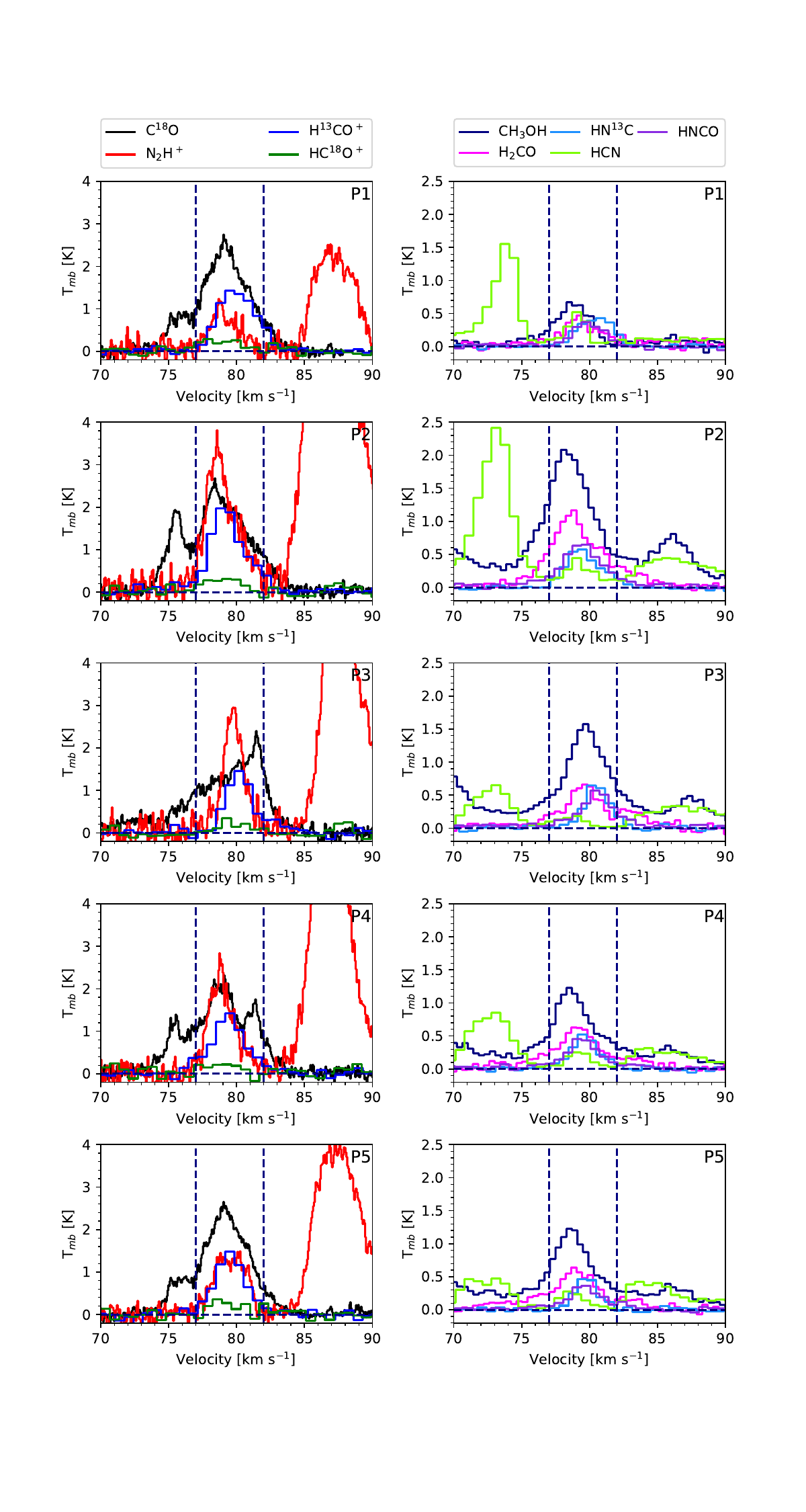}
    \vspace*{-21mm}
    \caption{Spectra of observed species at positions P1-P5. The dashed lines indicate the velocity range 77 to 82 km~s$^{-1}$, which has been selected to cover the dense gas as traced by N$_2$H$^+$ and H$^{13}$CO$^+$. In the left column panels, the amplitudes of the signals of $\rm H^{13}CO^+(1-0)$, $\rm HC^{18}O^+(1-0)$ and $\rm N_2H^+(1-0)$ have been multiplied by a factor of three for clarity of display.} 
    \label{fig:spectra_A}
\end{figure*}
%%%%%%%%%%%%%%%%%%%%%%%%%%%%%%%%%%%%%%%%%
\begin{figure*}
    \centering
    %\includegraphics[width=\textwidth]{chapters/images/both_maps_16_arsec.pdf}
    %\addtolength{\leftskip} {-3.1cm}
    %\addtolength{\rightskip}{-1.5cm}
     \includegraphics[height=25cm,width=15cm,]{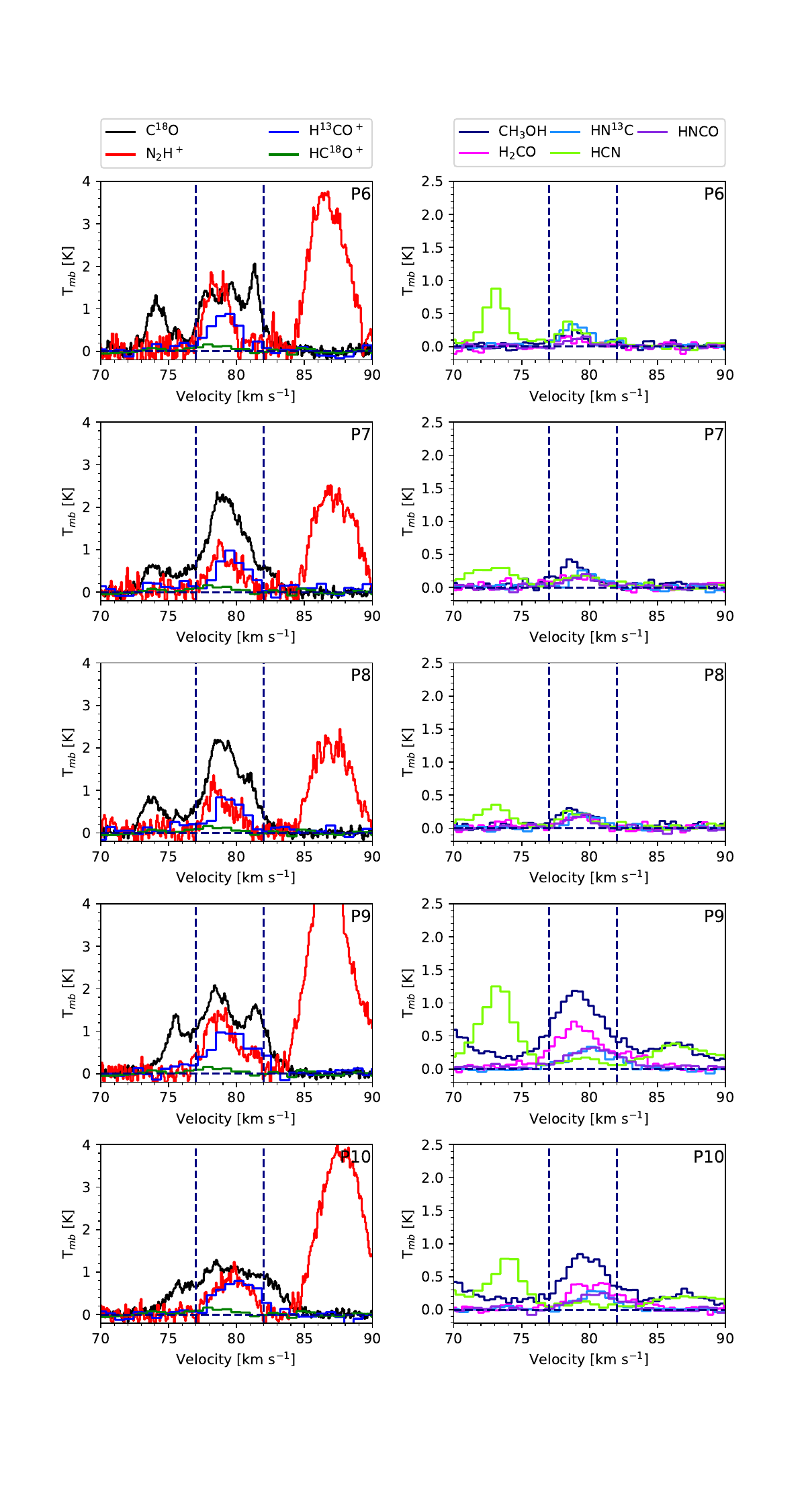}
    \vspace*{-21mm}
    \caption{As Fig.~\ref{fig:spectra_A}, but now for positions P6-P10.
    } 
    \label{fig:spectra_B}
\end{figure*}
%%%%%%%%%%%%%%%%%%%%%%%%%%%%%%%%%%%%%%%%
%Spectra of ch3oh, hcn, hnc, h2co and hnco

%\begin{figure*}
    %\centering
    %\includegraphics[width=\textwidth]{chapters/images/both_maps_16_arsec.pdf}
    %\addtolength{\leftskip} {-4.5cm}
    %\addtolength{\rightskip}{-1.cm}
     %\includegraphics[height=25cm,width=15cm,]{chapters/images/remaining_spectra.pdf}
    %\vspace*{-21mm}
    %\caption{Spectra of CH$_3$OH, H$_2$CO, HN$^{13}$C, HCN and HNCO at all positions. The dashed lines indicates the narrower range of velocity 77-82 km.s$^{-1}$ which has been selected based on the spectra of N$_2$H$^+$ and H$^{13}$CO$^+$}. 
    %\label{fig:remaining_spectra}
%\end{figure*}

%%%%%%%%%%%%%%%%%%%%%%%%%%%%%%%%%%%%%%%%

%To obtain the velocity range of the detected signal, first the line peak velocity is identified. Then, the line is integrated at blue (lower) and red-shifted (higher) velocities until the lowest positive temperature is reached.
%%%%%%%%%%%%%%%%%%%%%%%%%%%%%%%%%%%%%%%%%%%%%%%%%%

\section{Results}
\label{S:results}

\subsection{Abundances, including density dependence}

Here we summarise the abundances of the various species obtained in the IRDC positions (see also Table~\ref{tab:landscape-tab}). For CO, we find a typical value of ${\rm [CO]} \equiv N_{\rm CO}/N_{\rm H} \sim 2 \times 10^{-5}$, which corresponds to a CO depletion factor of $f_D\sim7$. In Fig.~\ref{fig:fd_n}a we plot $f_D$ versus $n_{\rm H}$, which reveals a trend of increasing depletion factor (i.e. smaller gas phase CO abundance) with increasing density. While correlated uncertainties are present in the measurements of $f_D$ and $n_{\rm H}$ (estimated via $N_{\rm H}$), the trend is significant, given a Pearson correlation coefficient of $r=0.93$. For completeness, the same information about the trend of [CO] with $n_{\rm H}$ is also shown in Fig.~\ref{fig:species_nH}a, with this figure showing the abundances versus density for all eight species measured in our study.

Theoretically, for a given set of conditions, including cloud age, one expects higher CO depletion factors at higher densities, that is due to higher rates of freeze-out leading to shorter freeze-out timescales. To illustrate this quantitatively, Fig.~\ref{fig:fd_n}a shows some example models of the time evolution of $f_D$ versus $n_{\rm H}$, with other parameters set to values of $T=15\:$K, $A_V=20\:$mag, and three values of $\zeta=10^{-18}$, $10^{-17}$ and $10^{-16}\:{\rm s}^{-1}$ explored. The observed trend of $f_D$ versus $n_{\rm H}$ can be explained by a model in which the ten positions have fairly similar ages of $2\times 10^5\:$yr. We note that this result is quite insensitive to certain choices of the astrochemical models, such as $\zeta$, $A_V$ (for values $>5\:$mag) and $T$ (for values $<20\:$K). However, as we discuss below, certain choices in the models about how nonthermal desorption processes are implemented can have impact on the inferred chemical age from these data.

%(see Fig.~\ref{fig:fd_n}b {\bf NE:dotted line}) and $A_V$ (see Fig.~\ref{fig:fd_n}c {\bf NE:dashed-dotted line}).

Figure~\ref{fig:fd_n}b plots $f_D$ versus $\bar{A}_{V}$.
%({\bf which we show it with ${A}_{V}$ symbol}). 
While we see a trend of increasing observed $f_D$ versus $\bar{A}_{V}$, we note that $\bar{A}_{V}\propto N_{\rm H} \propto n_{\rm H}$. At a given age and volume density, the astrochemical models do not predict a significant variation of $f_D$ with $A_V$ (at least for $A_V>5\:$mag, which is the relevant range). Thus, we consider that trend we observe here is more likely to be caused by variations in density in the regions.

\begin{figure*}[h]
\centering
\includegraphics[width=\textwidth]{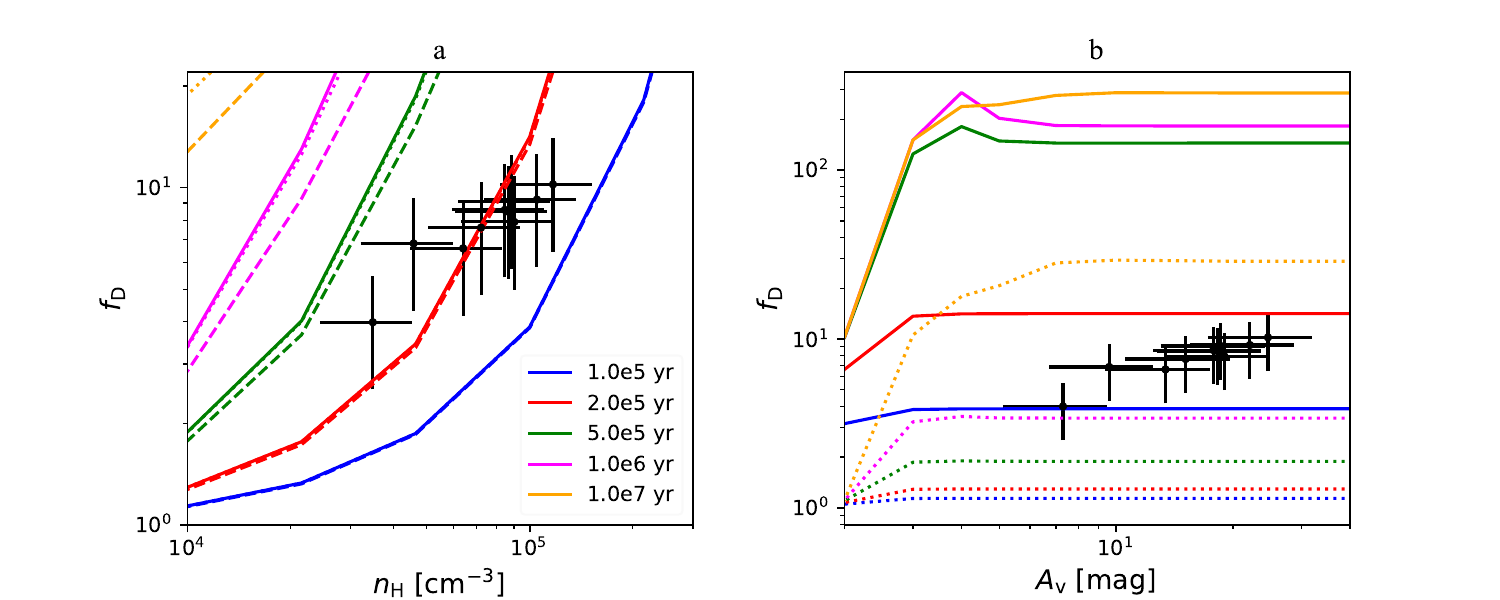}
\vspace*{-5mm}
\caption{{\it (a) Left:} CO depletion factor ($f_D$) versus number density of hydrogen nuclei ($n_{\rm H}$). The data for each position are shown, along with their uncertainties. Astrochemical models are shown for ages ranging from $10^5\:$yr to $10^7\:$yr and for $\zeta= 10^{-18}\:{\rm s}^{-1}$ (dotted lines) and $10^{-17}\:{\rm s}^{-1}$ (solid lines) and $10^{-16}\:{\rm s}^{-1}$ (dashed lines) with other parameters set to be $T=15\:$K and $A_V=20\:$mag.
{\it (b) Right:} CO depletion factor ($f_D$) versus visual extinction ($A_V$). The data for each position are shown, along with their uncertainties. Astrochemical models are shown here for $n_{\rm H}=10^{5}\: {\rm cm}^{-3}$, $T=15\:$K and $\zeta=10^{-17}\:{\rm s}^{-1}$ with solid lines (colors represent different times, as in legend of (a)). An equivalent model for $n_{\rm H}=10^{4}\: {\rm cm}^{-3}$ is shown with dotted lines.}
     \label{fig:fd_n}
\end{figure*}

%%%%%%%%%%%%%%%%%%%%%%%%%%%%%%%%%%%%%%%%%

\begin{figure*}[!ht]
%\centering
%\addtolength{\leftskip} {-8cm}
%\addtolength{\rightskip}{-8.5cm}
\hspace{-2cm}
\includegraphics[width=1.2\textwidth]{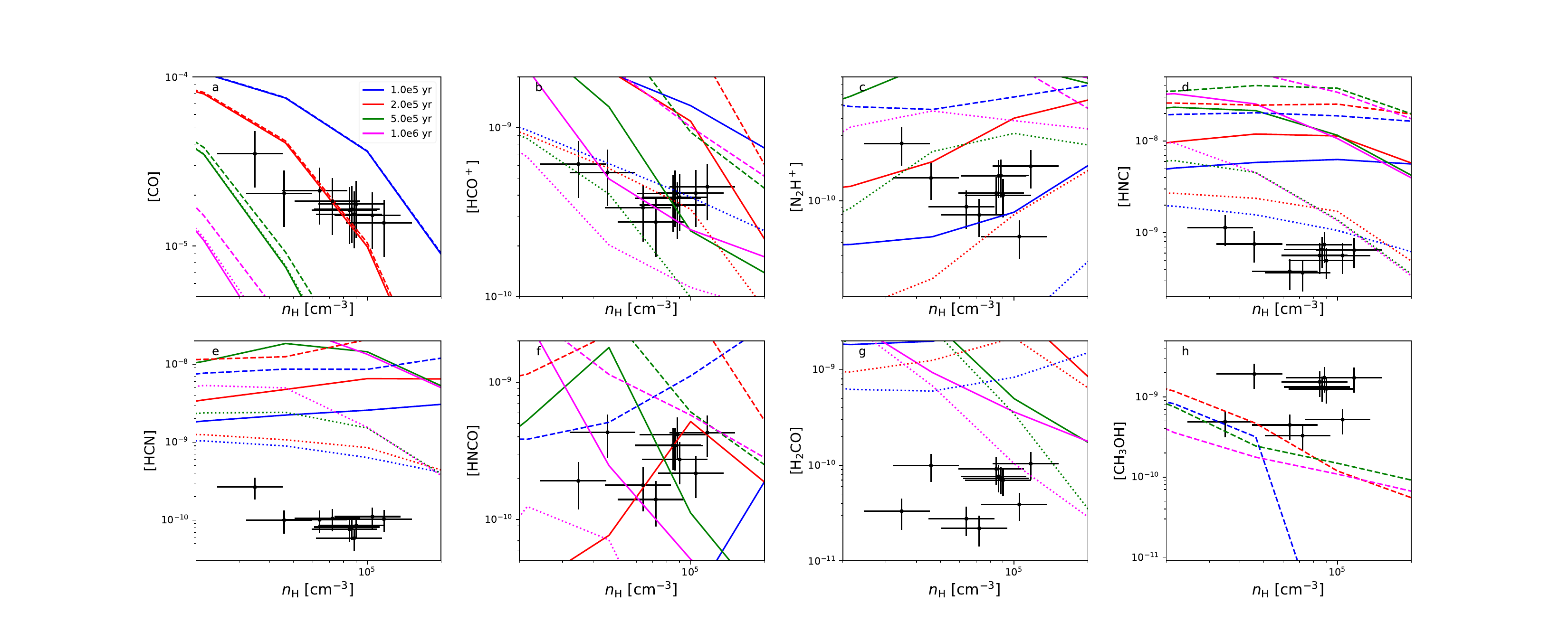}
\vspace*{-10mm}
\caption{Observed abundances of (a) CO, (b) HCO$^+$, (c) $\rm N_2H^+$, (d) HNC, (e) HCN, (f) HNCO, (g) H$_2$CO and (h) $\rm CH_3OH$ versus $n_{\rm H}$. Astrochemical model results are also shown for $\zeta= 10^{-18}\:{\rm s}^{-1}$ (dotted lines), $10^{-17}\:{\rm s}^{-1}$ (solid lines) and $10^{-16}\:{\rm s}^{-1}$ (dashed lines)  with other parameters set to be $T=15\:$K and $A_V=20\:$mag.}
\label{fig:species_nH}
\end{figure*}

%with 37$\%$ uncertainty in all positions except P6, with relatively higher abundance at starless positions (p7-p10). As it can be seen from Figure\ref{fig:sigma_T_map}, and estimated in table~\ref{tab:physical_parameters_IRDC} as well, P2 is located in the center of mass condensation and it contains the highest Hydrogen volume density. Estimated CO abundance at P2 is 1.37$\times$ 10$^{-5}$, which is the lowest value with respect to the other regions which consequently implies that at P2, we see the higher value of CO depletion factor ($\simeq$ 10). Contrariwise, at P6 that we see the lowest Hydrogen density, the highest CO abundance is estimated. It can be said that as hydrogen density reduces, CO abundance increases.

%... discussion of the estimated abundances and the implied CO depletion factors... is there any trend of CO depletion factor with density, T, spatial location in the IRDC?

For HCO$^+$, using emission from H$^{13}$CO$^+$ we infer typical abundances of [HCO$^+$]~$\sim 4 \times 10^{-10}$. The abundances inferred from  HC$^{18}$O$^+$ are systematically larger by about 30$\%$. However, such a difference is within the isotopologue ratio uncertainties. While the two lowest density regions, P6 and P10, are seen to have the highest abundances, $\sim 6 \times 10^{-10}$, the overall correlation with density is not as strong as in the case of [CO] (see Fig.~\ref{fig:species_nH}b).

%As examples, we estimated [HCO$^+$]~$\sim 4 \times 10^{-10}$ 
%with $\sim$ 37$\%$ uncertainty obtained 
%from H$^{13}$CO$^+$, with the highest abundance at P6 and the lowest one at P8.

%and 6~$\times$ 10$^{-10}$ with $\sim$ 40$\%$ uncertainty obtained from HC$^{18}$O$^+$. 

We estimate [N$_2$H$^+$] $\sim 10^{-10}$ on average (see Fig.~\ref{fig:species_nH}c). Again, the lowest density region has the highest abundance, but, as with HCO$^+$, there is no significant trend with density. Similar results are seen for HNC, with typical abundance of $6\times 10^{-10}$ (Fig.~\ref{fig:species_nH}d) and HCN, with typical abundance of [HCN]~$\sim 10^{-10}$ (Fig.~\ref{fig:species_nH}e).

Finally, we find [HNCO]$\sim 3\times 10^{-10}$, [H$_2$CO]$\sim 5\times 10^{-11}$ and [CH$_3$OH]$\sim 10^{-9}$ as typical values (see Fig.~\ref{fig:species_nH}f, g and h). Here there are hints of a bifurcation in the regions, with P1, P6, P7 and P8 having relatively low values, and the other six regions having abundances that are several times larger. 

We note that P7 and P8 were specifically chosen to be in regions that are relatively quiescent with respect to star formation activity as traced by CO(2-1) outflows \citep{kong2019widespread}, while still containing relatively dense structures. P1 also includes the candidate C1-N and S massive pre-stellar cores \citep{2013ApJ...779...96T}, that is significant quantities of dense, quiescent gas, although there are also two protostellar sources (C1-Sa and C1-Sb) in the same region. P6 also appears to be quiescent, that is not containing a $70\:{\rm \mu m}$ detected protostar \citep{moser2020high}, but is outside the region mapped for CO(2-1) outflows. Thus, we tentatively conclude that regions with relatively low star formation activity also have relatively low abundances of the relatively complex species of HNCO, H$_2$CO and CH$_3$OH. Such a result could be due to requirement of protostellar warming or outflow shock activity to liberate these species efficiently from dust grain ice mantles.

%We estimated [N$_2$H$^+$] approximately 2~$\times$ 10 $^{-10}$, with approximately 30$\%$ uncertainty. The highest [N$_2$H$^+$] has been estimated at P6, the less dense region and the lowest abundance has been estimated at P1, the coldest region.

%For other species, the abundances with~30-35$\%$ uncertainties, are varied from position to positions. for HNC and HCN, the highest abundance has been estimated at P6. Although for HNCO, CH$_3$OH and H$_2$CO, relatively lower abundances has been estimated at P6, the highest abundance of HNCO and and H$_2$CO has been estimated at P2.

%It is clear that, at p2 with the highest volume density and P6 with the lowest hydrogen density, we see outstanding values of abundances which are related to the the properties of these positions.

%... typical value and dispersion? Which region has highest/lowest abundance.

%Follow similar pattern for other species... does any region stand out?... 

\subsection{Astrochemical implications for cosmic ray ionisation rate and cloud age}

In this section we use the abundance measurements of the IRDC positions to constrain conditions in the clouds, especially $\zeta$ and $t$, via the astrochemical models described in \S\ref{astrochemical_model}. In general, we compare a set of the observed abundances, $[X_i]_{\rm obs}$, in a given position with those of theoretical model abundances, $[X_i]_{\rm theory}$. We search the model grid parameter space $(n_{\rm H}, T, A_V, \zeta, t)$ to minimise a $\chi^2$ metric of the following form
\begin{equation}
    \chi^2=\sum_i W_i [({\rm log}_{10} [X_i]_{\rm obs} - {\rm log}_{10} [X_i]_{\rm theory}) / {\rm log}_{10}(\sigma_i)]^2,
    \label{rel:chi-square}
\end{equation}
where $\sigma_i$ is an estimate of the uncertainty in the abundance, including the basic observational uncertainty and any allowed theoretical model uncertainty, and $W_i$ is a weighting factor, which allows certain species to be given greater or lesser importance in the analysis.

%Below we describe the method used to identify the model that best reproduces the observed molecular abundances and that provides us with an estimate of the cosmic ionization rate across the ten positions. For each positions, the best model is considered the one that minimizes the reduced $\chi^2$ in logarithmic scale, defined as:\\

%where \textit{X[obs]} and \textit{X[model]} refers to the abundances of the observed and astrochemical model data respectively and \textit{X[obs]$_{error}$} is the uncertainty of the observed abundances.\\

\subsubsection{Constraints using [CO], [HCO$^+$] and [$\rm N_2H^+$] (Case 1)}

As first step (Case 1), we only consider the abundances of CO, HCO$^+$ (obtained from H$^{13}$CO$^+$) and N$_2$H$^+$, that is CO and the two molecular ions in our set of species. 
We have first searched the entire model grid parameter space to find best fitting models and map out the $\chi^2$ landscape.
However, unrealistic solutions can be found with low values of $\chi^2$, that is at high temperatures ($\gtrsim 500\:$K), where low values of [CO] are caused by gas-phase chemical processing of the CO, rather than by CO freeze out onto dust.
Thus, we focus on a restricted (R) search over the ranges: $n_{\rm H} = 10^4 - 10^6\: {\rm cm}^{-3}$; $T= 10 - 50\:$K and $A_V = 5 - 100\:$mag. We also consider ultra-restricted (UR) searches in which $n_{\rm H}$ and $A_V$ are set to the closest values in the model grid to those of the observed region and with only adjacent values also considered. For this UR case we limit the search to $T=10 - 20\:$K. Finally, we have an extreme-restricted (XR) case where we only search the $\zeta$-$t$ parameter space in the best physically matched model for $n_{\rm H}$, $A_V$ and $T$ (we note that in this case we always set $T=15\:$K: this is always the closest match, except for P1, where $T=11.8\:$K, but even here, for simplicity, we consider 15~K for the XR case).

The $\chi^2$ landscape in the $\zeta$ versus $t$ plane for the XR search at the P2 position (i.e. with $n_{\rm H}= 1.0\times 10^5\:{\rm cm}^{-3}$, $A_V=20\:$mag and $T=15\:$K) with Case 1 ([CO], [HCO$^+$] and [N$_2$H$^+$]) fitting is shown in the central panel of Fig.~\ref{fig:p2_n_T}. The surrounding eight panels show the effect on the $\chi^2$ landscape of stepping in the model grid to the next higher and lower values of density and temperature, while keeping $A_V$ held fixed at 20~mag.

We notice several main features in Fig.~\ref{fig:p2_n_T}. First, the best result in the central panel that is matched to the physical conditions of P2 has $\zeta=2.2\times 10^{-18}\:{\rm s}^{-1}$ and $t=2\times10^5\:$yr and yields a value of $\chi^2=0.69$. These values are listed in Table~\ref{tab:gen_table}, along with those obtained for the UR and R searches. 

Second, we see that this position of lowest $\chi^2$ sits in an `island' that extends to higher values of $\zeta$ at earlier times and lower values of $\zeta$ at later times. There is also a separate, secondary island of relatively low values of $\chi^2$ at later times and higher $\zeta$, with local minimum at $t=2.82\times10^6\:$yr and $\zeta=2.2\times 10^{-16}\:{\rm s}^{-1}$.

As we vary the density of the model, the above features are broadly preserved, with a shift in the best position to earlier times and higher $\zeta$ if the density is raised, and later times and lower $\zeta$ if it is lowered. We have also inspected the equivalent figures for the cases with $A_V=10$ and 50~mag and find only modest differences compared to the $A_V=20\:$mag cases. Considering temperature variation, the models at $T=10\:$K have a similar structure, with the best position now at moderately higher values of $\zeta$, but with worse values of $\chi^2$. However, at $T=20\:$K the results are quite different, with the previous best $\chi^2$ island now erased and the new best solutions being at much later times, with the best value unrealistically old.

To understand the above behaviour, we next consider the detailed time evolution of some example models from the XR conditions: we examine the overall best model and the best model in the secondary, late-time island (see Fig.~\ref{fig:p2_n_T_evolution}). With a temperature of 15~K, for both models the evolution of gas-phase CO abundance is initially one of monotonic decline due to freeze-out onto dust grains. At the XR density of $n_{\rm H}=1.0\times 10^5\:{\rm cm}^{-3}$, this reaches the observed depletion factor of $\sim10$ by about $2\times10^{5}\:$yr, independent of $\zeta$. The influence of the CRIR is seen on the (quasi-)equilibrium level of gas phase CO abundance that is reached, after about 1~Myr. However, even in the case with $\zeta=2.2\times 10^{-16}\:{\rm s}^{-1}$, this abundance of CO is about 10 times smaller than the observed level. Thus the process of CO freeze-out essentially sets the timescale of the best model. 

Next, considering the abundance of HCO$^+$, it is seen to have a fairly constant evolution in time. In the overall best model it is gradually declining in the first few hundred thousand years. Finally, the evolution of [$\rm N_2H^+$] shows an early-time rapid rise from very low values associated with our assumed initial conditions, that is all N in atomic form. The rate of this rise is more rapid when $\zeta$ is higher. This is followed by a plateau phase and subsequent decline in abundance. The overall best model is achieved on the first time rise of [$\rm N_2H^+$], while the higher CRIR later-time model is achieved on the decline from the plateau phase. It is [HCO$^+$] and [$\rm N_2H^+$] that mostly set the CRIR of the best model, in particular favouring a low value of $\zeta=2.2\times 10^{-18}\:{\rm s}^{-1}$. This is illustrated in Fig.~\ref{fig:p2_n_T_evolution_high_low_CRIR}, where models with ten times higher and lower values of CRIR compared to this value are shown. 

%%%%%%%%%%%%%%%%%%%%%%%%%%%%%%%%%%%%%%%%%%%%%%%%%%%%%%
\begin{figure*}[htb!]
    \centering
    \addtolength{\leftskip} {-1.2cm}
    \addtolength{\rightskip}{-1cm}
     \includegraphics[width=20cm, height=16cm]
     {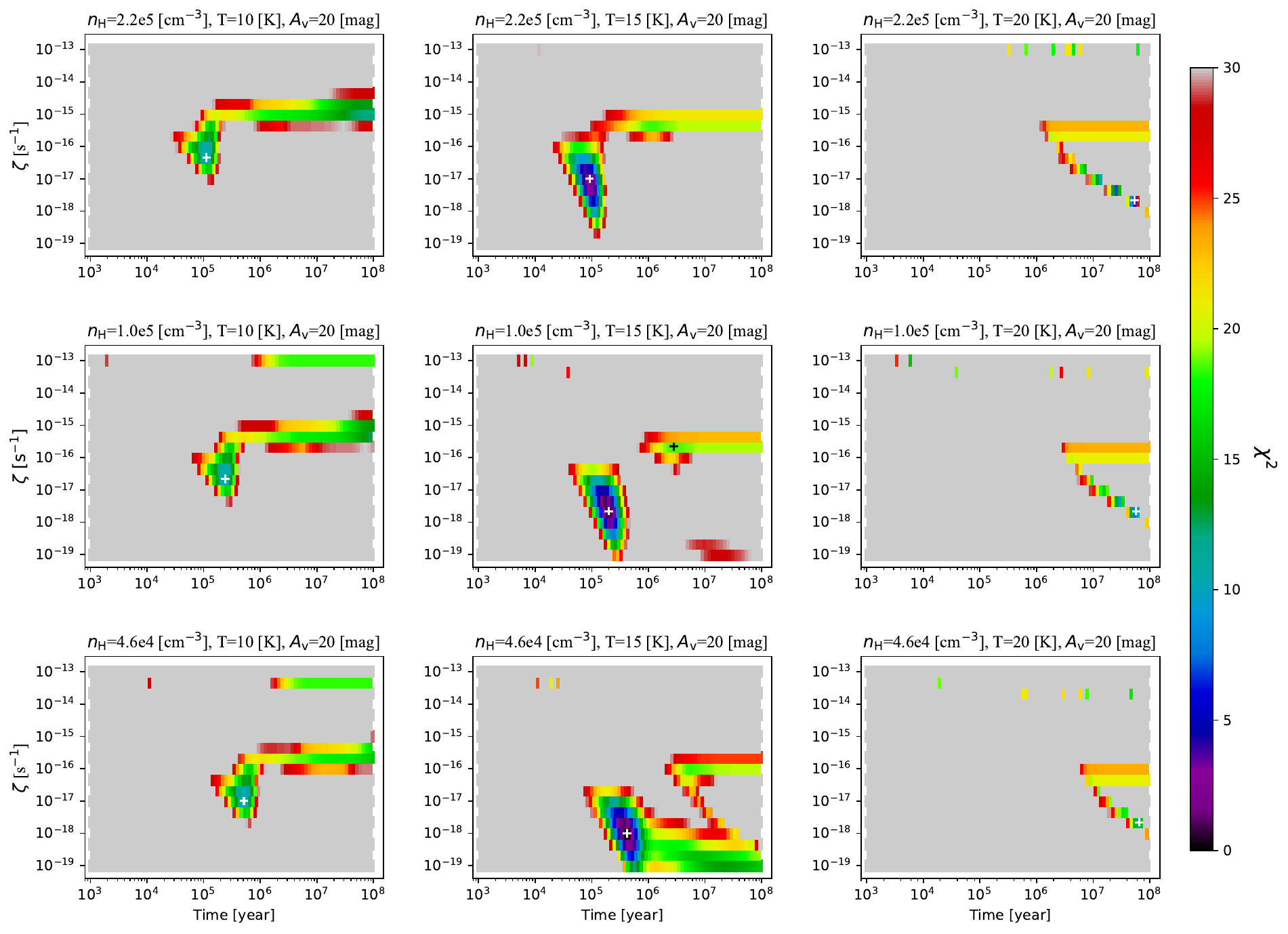}
     \vspace*{-3mm}
     \caption{The central panel shows the $\chi^2$ landscape (up to values of 30) in the $\zeta$ versus $t$ plane for the P2 position for the XR search (i.e. with $n_{\rm H}= 1.0\times 10^5\:{\rm cm}^{-3}$, $A_V=20\:$mag and $T=15\:$K) Case 1 (fitting to abundances of CO, HCO$^+$ and N$_2$H$^+$). The location of the minimum $\chi^2$ is marked with a white cross. The surrounding eight panels show the effect on the $\chi^2$ landscape of stepping in the model grid to the next higher and lower values of density and temperature (as labelled). Note, $A_V$ is held fixed at 20~mag in all these panels (so these are just a subset of the UR search).}
     \label{fig:p2_n_T}
 \end{figure*}
 
\begin{figure*}[htb!]
    \centering
    %\addtolength{\leftskip} {-2.2cm}
    %\addtolength{\rightskip}{-1cm}
     \includegraphics[width=1.0\textwidth]
     {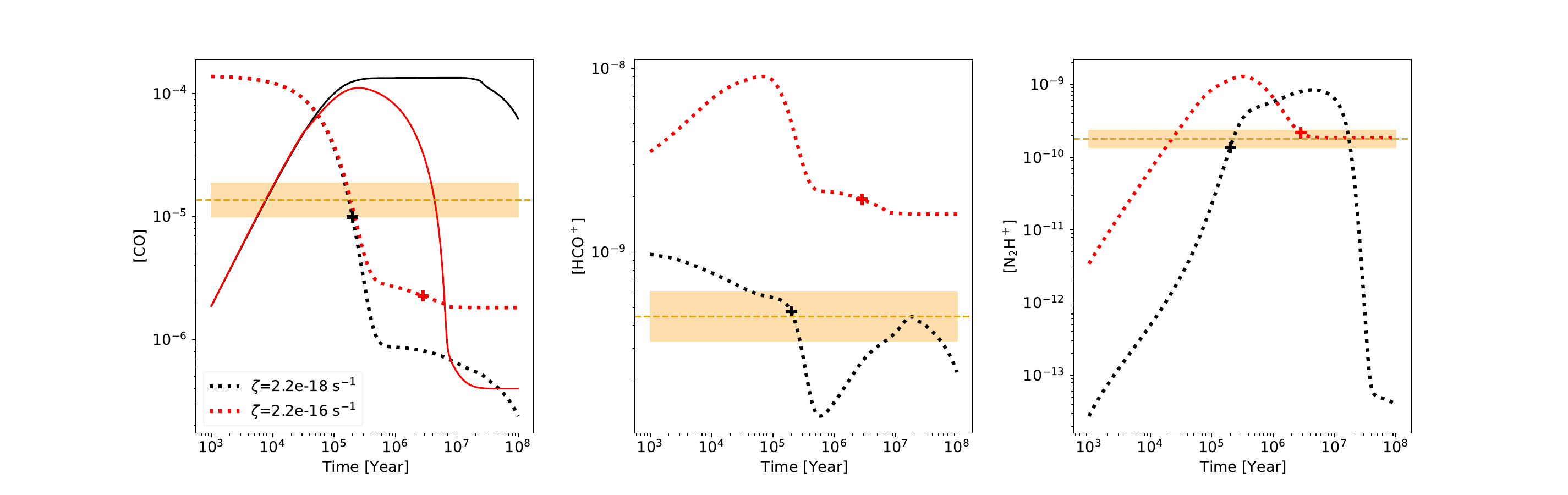}
     \vspace*{-6mm}
     \caption{Time evolution of [CO] (gas [dotted] and ice [solid] phases) (left), [$\rm HCO^+$] (middle) and [$\rm N_2H^+$] (right) for the overall best fitting model (black lines, $\zeta=2.2\times10^{-18}\:{\rm s}^{-1}$) and best `late-time' model (red lines, $\zeta=2.2\times 10^{-16}\:{\rm s}^{-1}$) for the XR search of the P2 clump, i.e. with $n_{\rm H}=1.0\times 10^5\:{\rm cm}^{-3}$, $A_V=20\:$mag and $T=15\:$K. The time of the best models along each of these tracks is marked with a `+' symbol. Horizontal dotted lines show the observed values of abundances, with the uncertainties shown by the shaded bands.
     }
     \label{fig:p2_n_T_evolution}
 \end{figure*} 

%jct - please delete the legend box. Please increase the size of the axis labels and numbers.

%jct - I do not understand the GCO and GCH4 results... which CR values are these for? All red lines should be for the high CR model, black for low CR model. Maybe we will only show GCO here... Also, the y-axis range should be reduced, i.e., to be like that used in Fig. 15.

%jct - I don't think the legend should have GCO twice - best remove both

\begin{figure*}[htb!]
    \centering
    %\addtolength{\leftskip} {-2.2cm}
    %\addtolength{\rightskip}{-1cm}
     \includegraphics[width=\textwidth]
     {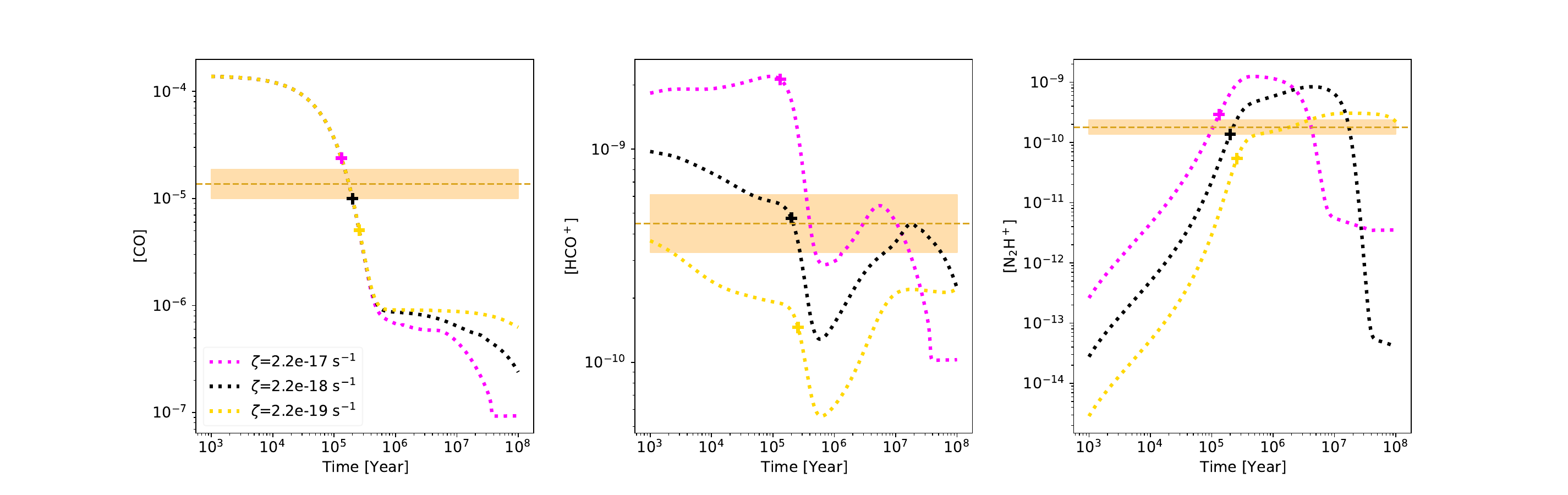}
     \vspace*{-6mm}
     \caption{As Fig.~\ref{fig:p2_n_T_evolution}, but now showing the effect of varying the CRIR by a factor of 10 to higher (magenta) and lower (yellow) values compared to the XR best model value of $\zeta=2.2\times10^{-18}\:{\rm s}^{-1}$. The best timescales for each of the models are shown with corresponding `+' symbols, given the observed abundances of the P2 location.
     %     Evolution of the main species, in the best model shown in the central panel of Figure~\ref{fig:p2_n_T}, as well as two other models with higher and lower CRIR. The density of the models is $n_{\rm H}$=1.0e5 cm$^{-3}$, $A_{\rm v}$ = 20 mag and T=15~K. $\zeta$ of the best model has been marked with a white cross in the location of minimum $\chi^2$ in central panel of Figure~\ref{fig:p2_n_T}. The same locations have been shown with colored crosses at each panel of the evolution of the species. The horizontal band indicates observed species with associated uncertainties at P2.
}
     \label{fig:p2_n_T_evolution_high_low_CRIR}
 \end{figure*}

%jct - same comment as before - remove title legend - put into panel - increase size of labels.

Next we consider the results of the UR and R search results for P2. As mentioned, the best fitting models for these search ranges are also listed in Table~\ref{tab:gen_table}. Figure~\ref{fig:p2_global_cr} shows the projected $\chi^2$ landscapes in the $\zeta$ versus $t$ plane for XR, UR and R searches. We see that the regions of low values of $\chi^2$ expand, as expected. The early-time, low-$\zeta$ island expands to cover a broader range of times, with the later time solutions being those of lower density and vice versa. In particular, the overall best models are now moved to these later time, lower density, lower $\zeta$ regions. However, it should be noted that there are quite widespread regions of relatively low $\chi^2$, so the relevance of the particular overall best model is less significant.

\begin{figure*}[!ht]
   \centering
     %\addtolength{\leftskip} {-8cm}
    %\addtolength{\rightskip}{-8.5cm}
     \includegraphics[width=\textwidth]{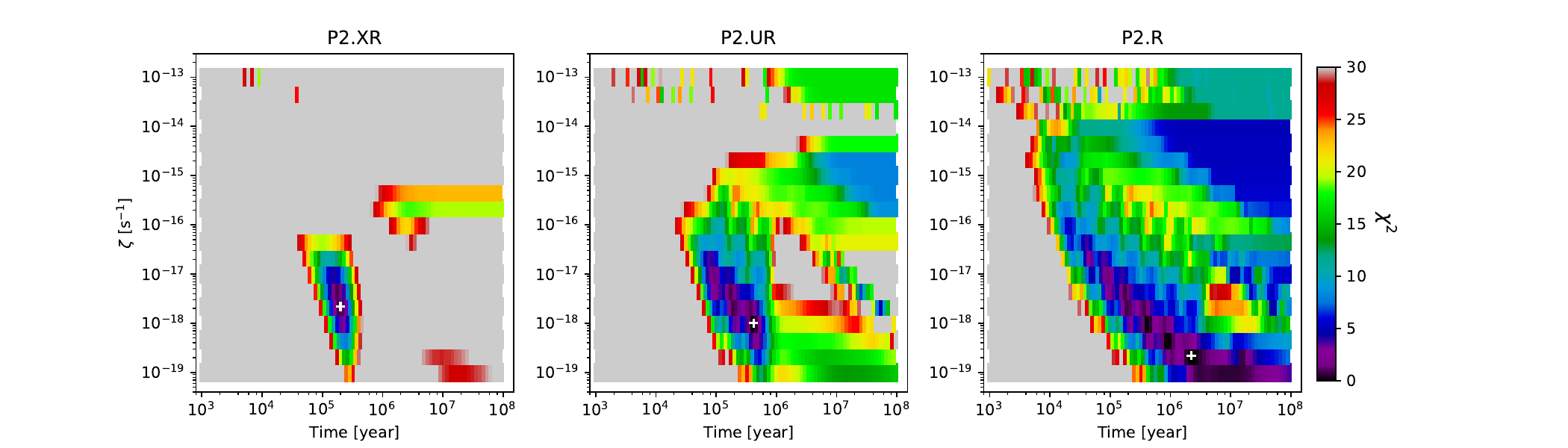}
     %\vspace{0.01cm}
     \caption{Projected best $\chi^2$ values in the $\zeta$ versus $t$ plane for the P2 position for the XR (left), UR (middle) and R (right) search methods with Case 1 (fitting to abundances of CO, HCO$^+$ and N$_2$H$^+$). The location of the minimum $\chi^2$ is marked with a white cross in each panel.
    % Distribution of $\zeta$ associated with $\chi^2$ using abundances of three main species CO, HCO$^+$ and N$_2$H$^+$ at P2, XR) Extra-Restricted area at which n$_H$ = 1.0e5 cm$^{-3}$, T=15 K and A$_v$=20 mag are held to be fixes and $\zeta$ has been changed, UR) Ultra-Restricted area at which n$_H$=4.6e4,1.0e5,2.2e5 cm$^{-3}$, T=10,15,20 K and A$_v$=10,20,50 mag, R) Restricted area at which n$_H$=1.0e4e4-1.0e6 cm$^{-3}$, T=10-50 K and A$_v$=5-50 mag.
    }
     \label{fig:p2_global_cr}
 \end{figure*}

Next we consider the Case 1 results for the rest of the IRDC positions. In particular, we next consider the P6 position (Fig.~\ref{fig:p6_global_cr}), which has been selected to be starless. We see that the overall feature of the $\chi^2$ distributions for P6 are quite similar to those of P2. The equivalent figures for the other regions follow the same general patterns and we do not present them here.
%are shown in Appendix~\ref{appendixA}. 
In general, all these regions also exhibit quite similar $\chi^2$ landscapes for their XR, UR and R Case 1 search results.

Thus, in summary, the main features of the Case 1 fitting are that the best solutions have relatively low CRIRs ($\sim 10^{-18}\:{\rm s}^{-1}$) and yield ages in the range $\sim 10^5$ to $\sim 10^6\:$yr. There is an extended `valley' of low $\chi^2$ values extending to higher CRIR solutions (e.g. $\sim 10^{-17}\:{\rm s}^{-1}$), but requiring smaller ages (e.g. $10^5\:$yr). This type of solution is based on the gas being in a phase of early CO freeze-out, that is with the age set by the time needed to freeze-out from the adopted initial conditions of all CO in gas phase.

However, as we discuss later in \S\ref{S:discussion}, we consider that there are caveats with this interpretation. In particular, it seems to be unlikely that all the ten selected positions would be in this initial phase of CO freeze out and the implied values of CRIR are relatively low compared to previous estimates in the more diffuse ISM. Thus, we also discuss possible ways in which later-time solutions ($t\gtrsim 1\:$Myr) can be achieved.

\begin{figure*}[!ht]
   \centering
     %\addtolength{\leftskip} {-8cm}
    %\addtolength{\rightskip}{-8.5cm}
     \includegraphics[width=\textwidth]{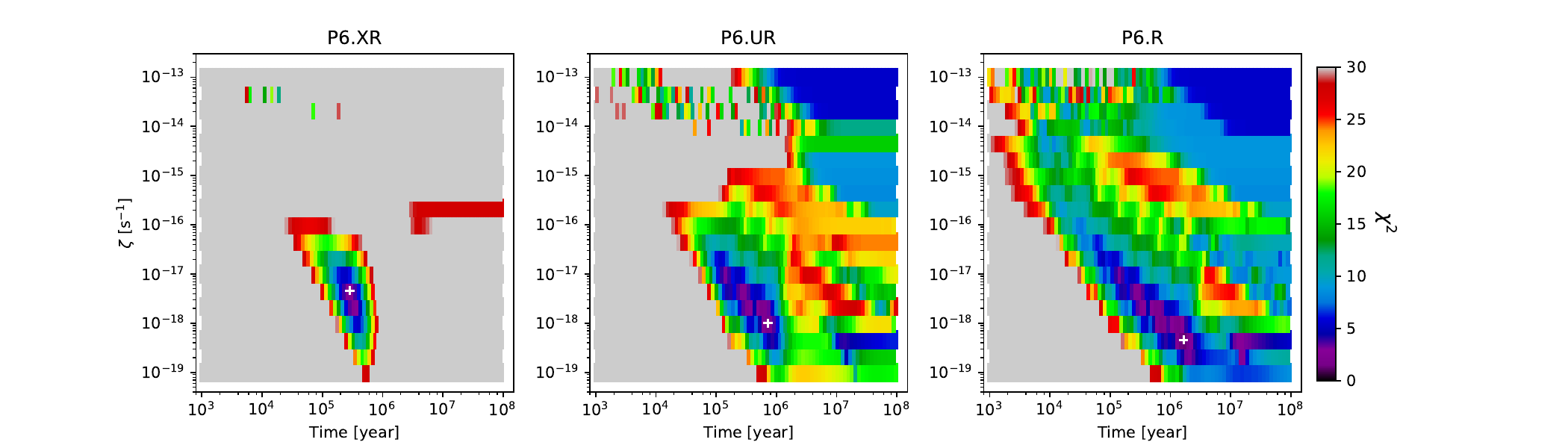}
     %\vspace{0.01cm}
     \caption{As Fig.~\ref{fig:p2_global_cr}, but now for P6.}
     \label{fig:p6_global_cr}
 \end{figure*}

\subsubsection{Constraints using all species (Case 2)}

We next show the equivalent results for constraining the astrochemical conditions when the abundances of all the eight species observed at the P2 position are used, which we refer to as Case 2. However, for these additional five species we set their weights to only contribute 25\% of the total weighting of $\chi^2$, that is 5\% each. This choice is made given that we consider that these species have somewhat more uncertain astrochemistry and that, being neutral species, they have a less direct connection to the CRIR than HCO$^+$ and N$_2$H$^+$.

%({\bf NE: Again, due to unrealistic solution obtained by searching through the entire model grid parameter space, 
%As for Case 1, we focus on the restricted (R) search over the ranges: $n_{\rm H} = 10^4 - 10^6\: {\rm cm}^{-3}$; $T= 10 - 50\:$K and $A_V = 5 - 100\:$mag. As we have done before for Case 1, we keep working on considering ultra-restricted (UR) searches, as well as extreme-restricted (XR) case.

The Case 2 $\chi^2$ landscape in the $\zeta$ versus $t$ plane for the XR search at the P2 position (i.e. with $n_{\rm H}= 1.0\times 10^5\:{\rm cm}^{-3}$, $A_V=20\:$mag and $T=15\:$K) is shown in the central panel of Fig.~\ref{fig:p2_n_T_case2}. The surrounding eight panels show the effect on the $\chi^2$ landscape of stepping in the model grid to the next higher and lower values of density and temperature, while keeping $A_V$ fixed at 20~mag.

We notice several main features in Fig.~\ref{fig:p2_n_T_case2}. First, the best result in the central panel that is matched to the physical conditions of P2 has $\zeta=2.2\times 10^{-17}\:{\rm s}^{-1}$ and $t=4.47\times10^6\:$yr and yields a value of $\chi^2=60.82$. These values are listed in Table~\ref{tab:gen_table}, along with those obtained for the UR and R searches. We see that this position of lowest $\chi^2$ sits close to the later-time, higher CRIR region that was found as a secondary solution area in the Case 1 XR search. The previous best $\chi^2$ region from Case 1 that is at earlier times and lower CRIRs is still seen in Case 2 as a local minimum of $\chi^2$, but now does not contain the global minimum. We note also that for Case 2 the overall values of $\chi^2$ are much greater than for Case 1. We examine below the reason for the poorer level of fitting.

%an ``island'' that extends to higher values of $\zeta$ at earlier times and lower values of $\zeta$ at later times.

As we vary the density of the model, the general, overall shape of the $\chi^2$ landscape does not change too much, however, we note that going to higher densities leads to the best fit model being at earlier times ($\sim 10^5\:$yr) and with relatively high CRIR ($\zeta =2.2\times 10^{-16}\:{\rm s}^{-1}$). This solution region is also found when the temperature is lowered to 10~K. When raising the temperature to 20~K we see that later time solutions are preferred, with there being some overlap of these with the best region found in the XR case. Similar to Case 1, having inspected the equivalent results for $A_V=10$ and 50~mag, we find only modest differences compared to the $A_V=20\:$mag cases.

%above features are broadly preserved, with a shift in the best position to earlier times and higher $\zeta$ if the density is raised, and later times and lower $\zeta$ if it is lowered. Also we see that in the panel of higher density with $T=15\:$K, an earlier time solution island with higher $\zeta=2.2\times 10^{-16}\:{\rm s}^{-1}$ and slightly smaller $\chi^2=59.57$ appears as well.

%Considering temperature variation, the models at $T=10\:$K have a similar structure, with the best position now at moderately higher values of $\zeta$, but with better values of $\chi^2$. Also in the models with $T=10\:$K, the solution island appears in earlier time (<1Myrs), with respect to the models with $T=15\:$K. However, at $T=20\:$K the results are quite different but similar to the results obtained for Case 1, with the previous best $\chi^2$ island now erased and the new best solutions being at much later times, with the best value unrealistically old.

To understand the above behaviour, we next consider the detailed time evolution of some example models from the XR conditions. First, we examine the overall best model (see Fig.~\ref{fig:p2_n_T_evolution_case2}). As seen before in Case 1, with a temperature of 15~K, the evolution of gas-phase CO abundance is initially one of monotonic decline due to freeze-out onto dust grains. At the XR density of $n_{\rm H}=1.0\times 10^5\:{\rm cm}^{-3}$, this reaches the observed depletion factor of $\sim10$ by about $2\times10^{5}\:$yr. However, because of the abundance of other species, especially $\rm CH_3OH$, later time solutions are preferred, even though the gas-phase abundance of CO is now in worse agreement compared to the observed value. We also see why the overall value of $\chi^2$ is worse, since the abundances of several of the species, such as CO, HNC, HCN, $\rm H_2CO$ and $\rm CH_3OH$, are not very well fit even for the best model, with differences greater than a factor of 10.

%Next, considering the abundance of HCO$^+$, it is seen to have a fairly constant evolution up to $\sim 1\times10^{5}\:$yr. Then, it is gradually declining and following a small valley phase, it keeps to be declining more.

%The evolution of [$\rm N_2H^+$] shows an early-time rapid rise from very low values associated with our assumed initial conditions, i.e., all N in atomic form. This is followed by a plateau phase and subsequent decline in abundance. Although the general time evolution of [$\rm HNC$] and [$\rm HCN$] is very similar to [$\rm N_2H^+$], but these models are far above the observed values of abundances. 

%The evolution of [$\rm HNCO$] and [$\rm H_2CO$] follows the same pattern, although their general appearance are different. Both show an early-time rise, followed by a decline, raise and a declining phase subsequently. Observed value of [$\rm HNCO$] can be reproduce by the model at some points, but the model again is far from the observed value of [$\rm H_2CO$] abundance.
 
%Finally, although the time evolution of [$\rm CH_3OH$] follows the same pattern of [$\rm HNCO$] and [$\rm H_2CO$], but this model is not able to reproduce [$\rm CH_3OH$] abundance and is evolving far below the observed value of [$\rm CH_3OH$] abundance.

A comparison of models with different CRIRs is shown in Fig.~\ref{fig:p2_n_T_evolution_case2_2}, where models with ten times higher and lower values of $\zeta$ compared to that of the best XR model are shown. In general higher values of $\zeta$ lead to higher abundances of the considered species and vice versa.

Next we consider the Case 2 UR and R search results for P2. As mentioned, the best fitting models for these search ranges are also listed in Table~\ref{tab:gen_table}. Figure~\ref{fig:p2_global_cr_all} shows the projected $\chi^2$ landscapes in the $\zeta$ versus $t$ plane for XR, UR and R searches. We see that the regions of low values of $\chi^2$ expand, as expected. In the UR search, a solution at earlier time and higher CRIR is preferred, which is in a region already seen in Fig.~\ref{fig:p2_n_T_case2}. In the less constrained R search, a solution with an unrealistically old age ($\gg 10^7\:$yr) is found, although earlier times with only moderately worse $\chi^2$ values are apparent. In general, the estimated values of $\zeta$ via Case 2 at P2 are higher than those found in Case 1. Later time solutions are also more favoured. However, the overall fits are worse, with larger values of $\chi^2$.

%The results of Case 2 differ from those obtained for Case 1, although the values of $\chi^2$ of the best models are high ($>20$) implying that the models fail to reproduce all observed abundances. 

%The early-time, low-$\zeta$ island in UR panel of Fig.~\ref{fig:p2_global_cr_all} expands to cover a broader range of times, with the later time solutions being those of lower density and vice versa. But in XR and R panels, the best solution island appears after 1.0e7 years with $\zeta=1.0\times 10^{-17}\:{\rm s}^{-1}$. 
 
Next we consider the Case 2 results for the rest of the IRDC positions. In particular, we next consider the P6 position (Fig.~\ref{fig:p6_global_cr_all}). There are some noticeable differences in the Case 2 $\chi^2$ landscapes of P6 compared to P2, especially for the XR and UR searches. For XR in P6 there is a greater preference for earlier time solutions, while for UR a later time solution is favoured. The equivalent figures for the other regions are quite similar and we do not present them here.
%shown in Appendix~\ref{appendixA}. 

%These regions also exhibit quite similar $\chi^2$ landscapes for their XR, UR and R Case 2 search results.

Thus, in summary, the main features of the Case 2 fitting are that, compared to Case 1, the best solutions have relatively larger values of CRIRs ($\sim 10^{-17}\:{\rm s}^{-1}$) and a more extended range of ages, from $\sim 10^5$ to $\sim 10^7\:$yr. However, the overall goodness of the fits is worse.

%However, as we mentioned before, we will discuss later in \S\ref{S:discussion} to search for the models being able to reproduce observed values of abundances in later time with relatively higher $\zeta$.

\begin{figure*}[htb!]
    \centering
    \addtolength{\leftskip} {-1.2cm}
    \addtolength{\rightskip}{-1cm}
     \includegraphics[width=20cm, height=16cm]
     {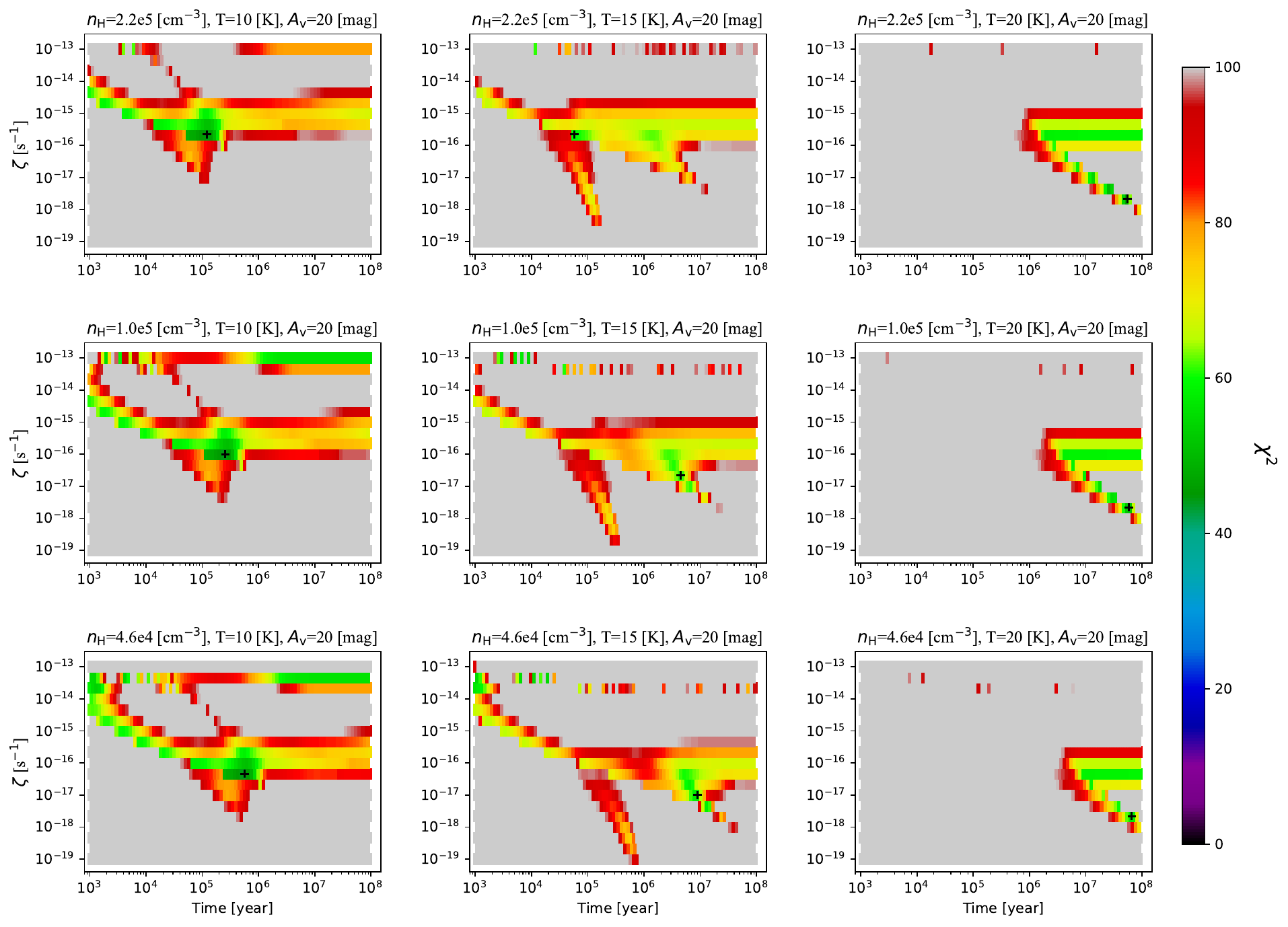}
     \vspace*{-3mm}
     \caption{The central panel shows the $\chi^2$ landscape (up to values of 100) in the $\zeta$ versus $t$ plane for the P2 position for the XR search (i.e. with $n_{\rm H}= 1.0\times 10^5\:{\rm cm}^{-3}$, $A_V=20\:$mag and $T=15\:$K) Case 2 (fitting to abundances of all species). The location of the minimum $\chi^2$ is marked with a black cross. The surrounding eight panels show the effect on the $\chi^2$ landscape of stepping in the model grid to the next higher and lower values of density and temperature (as labelled). We note that $A_V$ is held fixed at 20~mag in all these panels (so these are just a subset of the UR search).}
     \label{fig:p2_n_T_case2}
 \end{figure*}

\begin{figure*}[htb!]
    \centering
    %\addtolength{\leftskip} {-1.2cm}
    %\addtolength{\rightskip}{-1cm}
     \includegraphics[width=\textwidth]
     {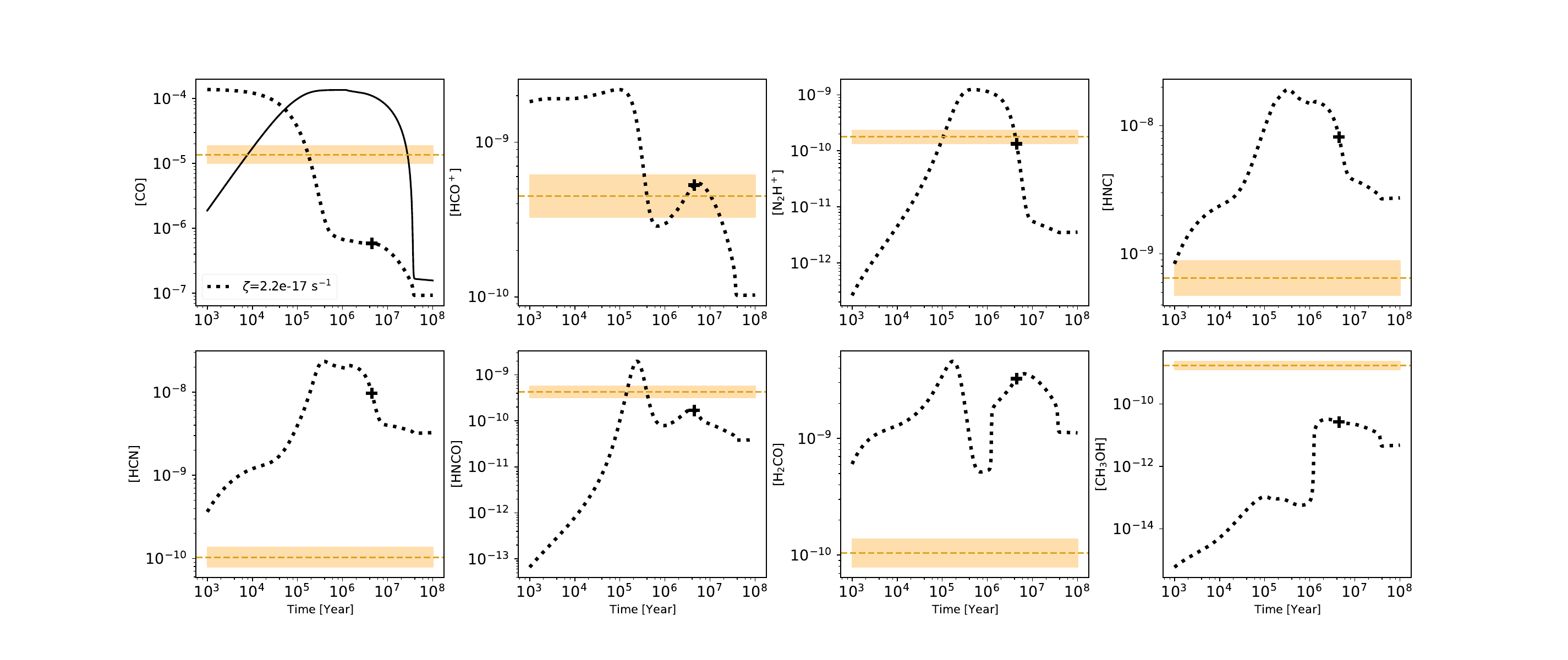}
     \vspace*{-10mm}
     \caption{Time evolution of [CO] (gas [dotted] and ice [solid] phases) (top left). The remaining panels show the equivalent panels for the gas phase abundances of the rest of the species for the overall best fitting model ($\zeta=2.2\times10^{-17}\:{\rm s}^{-1}$), for the XR search of the P2 clump, that is with $n_{\rm H}=1.0\times 10^5\:{\rm cm}^{-3}$, $A_V=20\:$mag and $T=15\:$K. The time of the best overall model along each of these tracks is marked with a `+' symbol. Horizontal dotted lines show the observed values of abundances, with the uncertainties shown by the shaded bands.}
     \label{fig:p2_n_T_evolution_case2}
 \end{figure*}

 \begin{figure*}[htb!]
    \centering
    %\addtolength{\leftskip} {-1.2cm}
    %\addtolength{\rightskip}{-1cm}
     \includegraphics[width=\textwidth]
     {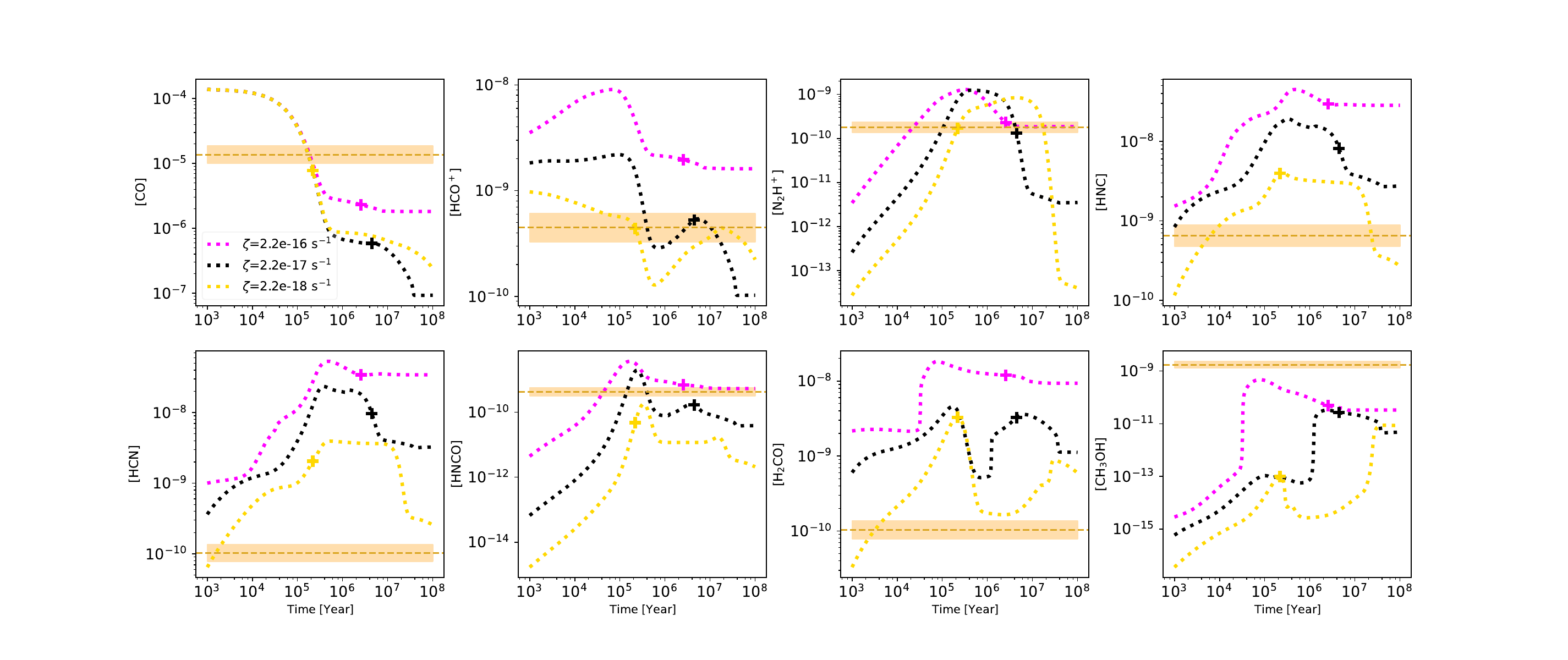}
     \vspace*{-10mm}
     \caption{As Fig.~\ref{fig:p2_n_T_evolution_case2}, but now showing the effect of varying the CRIR by a factor of 10 to higher (magenta) and lower (yellow) values compared to the XR best model value of $\zeta=2.2\times10^{-17}\:{\rm s}^{-1}$. The overall best timescales for each of the models are shown with corresponding `+' symbols, given the observed abundances of the P2 location.}
     \label{fig:p2_n_T_evolution_case2_2}
 \end{figure*}

 %%%%%%%%%%%%%%%%%%%%%%%%%%%%%%%%%%%%%%%%%%%%%
\begin{figure*}[!ht]
   \centering
     %\addtolength{\leftskip} {-8cm}
    %\addtolength{\rightskip}{-8.5cm}
     \includegraphics[width=\textwidth]{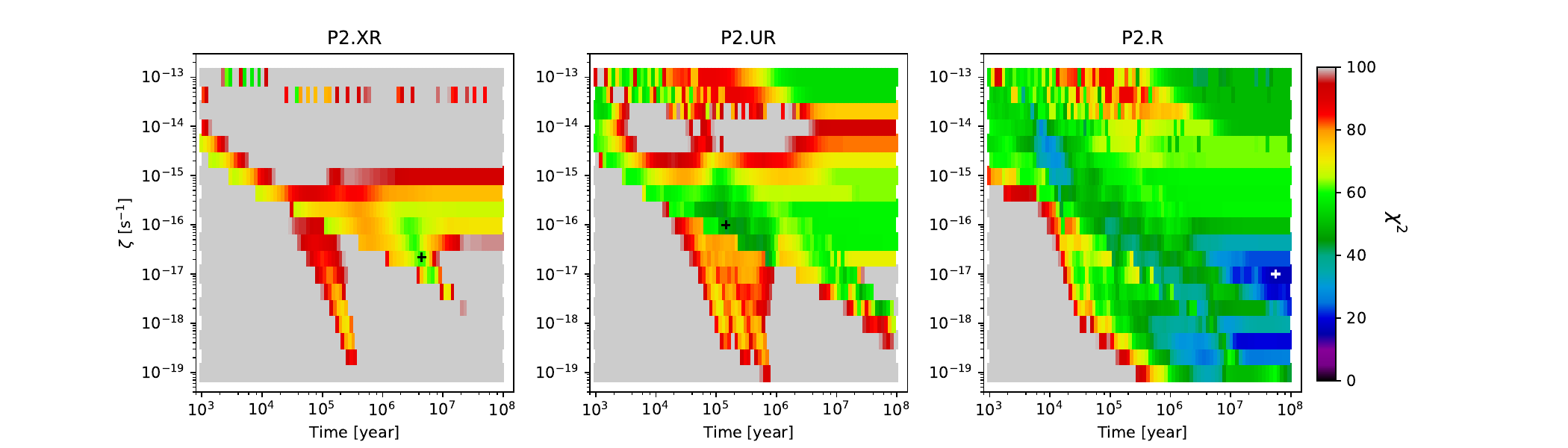}
     %\vspace{0.01cm}
     \caption{Projected best $\chi^2$ values in the $\zeta$ versus $t$ plane for the P2 position for the XR (left), UR (middle) and R (right) search methods with Case 2 (fitting to abundances of all species). The location of the minimum $\chi^2$ is marked with a black or white cross in each panel.}
     \label{fig:p2_global_cr_all}
 \end{figure*}

%%%%%%%%%%%%%%%%%%%%%%%%%%%%%%%%%%%%%%%%%%%%%
%%%%%%%%%%%%%%%%%%%%%%%%%%%%%%%%%%%%%%%%%%%%%
\begin{figure*}[!ht]
   \centering
     %\addtolength{\leftskip} {-8cm}
    %\addtolength{\rightskip}{-8.5cm}
     \includegraphics[width=\textwidth]{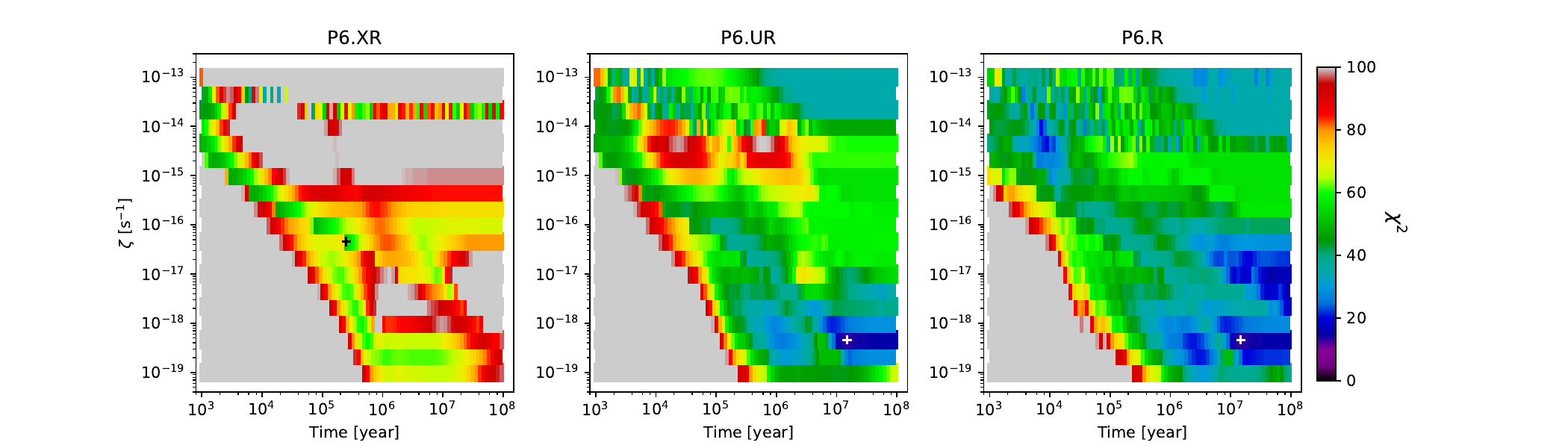}
     %\vspace{0.01cm}
     \caption{As Fig.~\ref{fig:p2_global_cr_all}, but now for P6.}
     \label{fig:p6_global_cr_all}
 \end{figure*}
%%%%%%%%%%%%%%%%%%%%%%%%%%%%%%%%%%%%%%%%%%%%%%%%

\begin{figure*}[!ht]
   \centering
     %\addtolength{\leftskip} {-8cm}
    %\addtolength{\rightskip}{-8.5cm}
     \includegraphics[width=\textwidth]{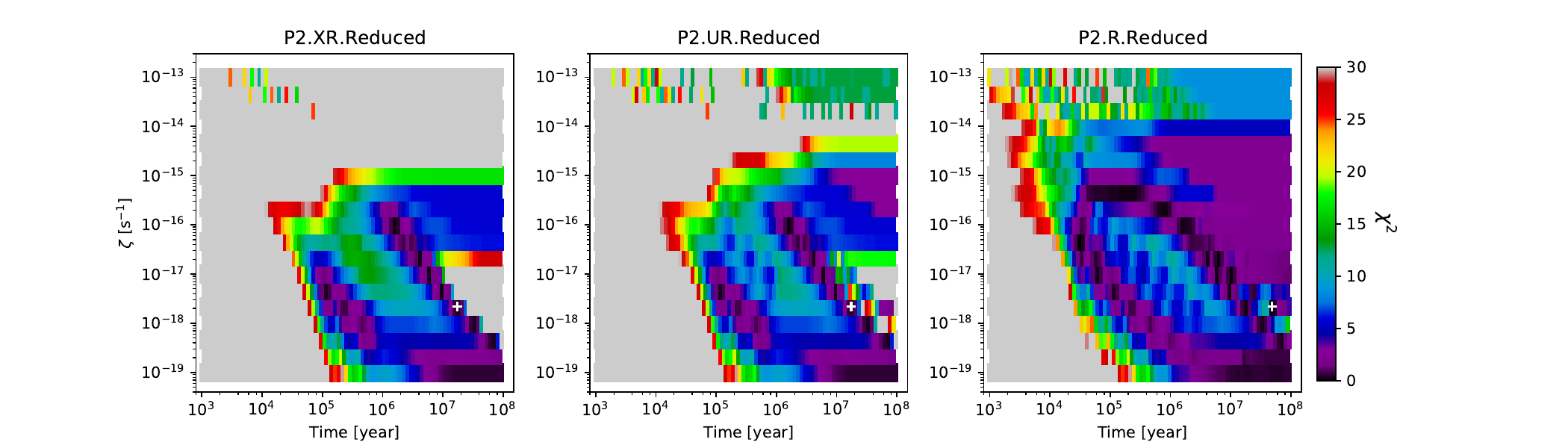}
     %\vspace{0.01cm}
     \caption{Investigation of effect of CO envelope contamination showing the projected best $\chi^2$ values in the $\zeta$ versus $t$ plane for the P2 position for the XR (left), UR (middle) and R (right) search methods with Case 1 (fitting to abundances of CO, HCO$^+$ and N$_2$H$^+$). For this fitting, the observed abundance of CO has been reduced by a factor of 10 and the associated uncertainty is set to be a factor of 3. The location of the minimum $\chi^2$ is marked with a white cross in each panel.}
     \label{fig:p2_global_cr_by10F}
 \end{figure*}

%%%%%%%%%%%%%%%%%%%%%%%%%%%%%%%%%%%%%%%
\begin{figure*}[htb!]
    \centering
    %\addtolength{\leftskip} {-2.2cm}
    %\addtolength{\rightskip}{-1cm}
     \includegraphics[width=\textwidth]
     {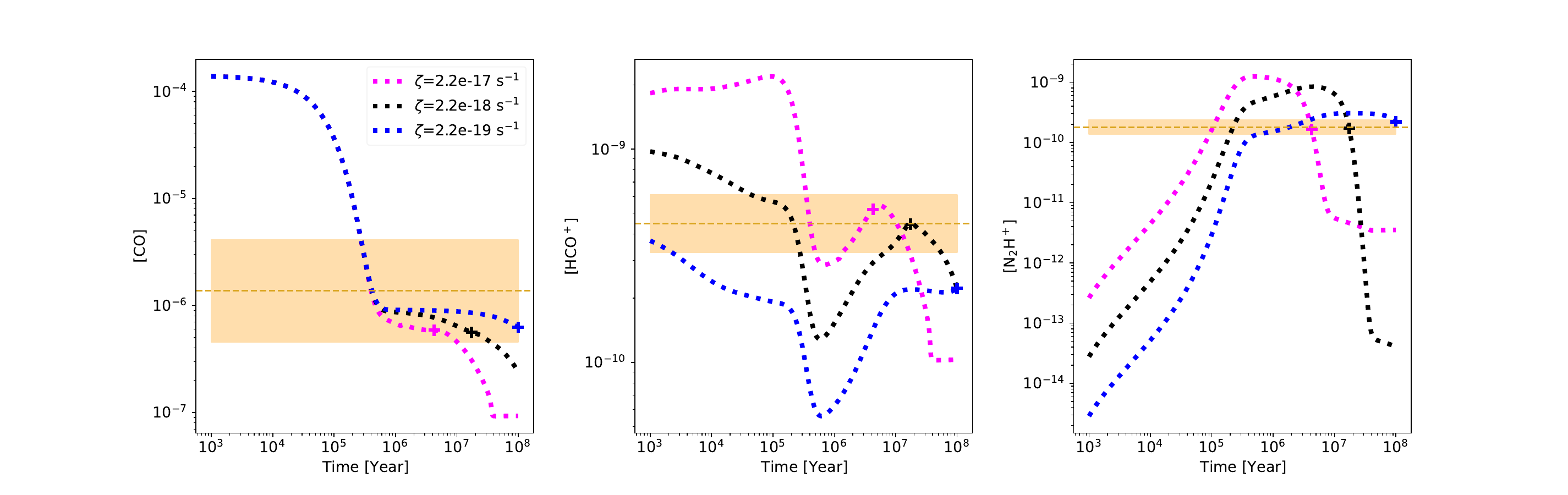}
     \vspace*{-6mm}
     \caption{Investigation of effect of CO envelope contamination showing time evolution of [CO] (left), [$\rm HCO^+$] (middle) and [$\rm N_2H^+$] (right) for the overall best fitting model (black line, $\zeta=2.2\times10^{-18}\:{\rm s}^{-1}$). Here the observed CO abundance has been reduced by a factor of 10 and the total uncertainty set to be a factor of 3. The effects of varying the CRIR by a factor of 10 to higher (magenta) and lower (blue) values compared to the XR best model value are shown. The best timescales for each of the models are marked with corresponding `+' symbols.}
     \label{fig:p2_XR_10F}
 \end{figure*}

%%%%%%%%%%%%%%%%%%%%%%%%%%%%%

\section{Discussion and extended analysis}\label{S:discussion}

Here we investigate the effects of several aspects of the observational analysis and astrochemical modelling that may have a major influence on the fiducial Case 1 results of a preference for relatively early time ($\sim 10^5\:$yr) and low $\zeta$ ($\sim 10^{-18}\:{\rm s}^{-1}$) solutions. We consider that such young ages are questionable because it is unlikely that all ten regions, P1 to P10, are in such early evolutionary stages with ages similar to their current local free-fall times, $t_{\rm ff}=(3\pi/[32G\rho])^{1/2}\sim 10^5\:{\rm yr}$ (see Table~\ref{tab:landscape-tab}). Furthermore, the derived values of CRIR are significantly lower than those derived from previous studies (see \S\ref{S:introduction}).

\subsection{Contamination by envelope CO emission}\label{S:contamination}

The CO(1-0) has a much lower critical density than HCO$^+$(1-0) and $\rm N_2H^+$(1-0). While we have made efforts to restrict the velocity range used to estimate the CO column density to closely match that of the higher density tracers, it remains possible that our measurement suffers from contamination from lower density, envelope gas that is surrounding the region of interest, but overlapping in velocity space. To account for this possibility and to investigate its potential effects, we simply assume a contamination level of 90\%, and reduce the observed CO column density by a factor of 10 (and we set the uncertainty in CO abundance to then be a factor of 3).

Figure~\ref{fig:p2_global_cr_by10F} presents XR, UR and R $\chi^2$ projections in the $\zeta$ versus $t$ plane for P2 with this reduced gas phase CO abundance. We find that later-time solutions are now favoured. For example, the best XR solution has $t\simeq 2 \times 10^7\:$yr and $\zeta\simeq 2\times 10^{-18}\:{\rm s}^{-1}$. However, there is also a degeneracy with higher CRIR and earlier times that yield similarly low $\chi^2$ values. Figure~\ref{fig:p2_XR_10F} shows time histories of abundances for several example XR models that illustrate these kinds of solutions. The observed HCO$^+$ abundance is seen to give a constraint on the CRIR, while that of $\rm N_2H^+$ helps select models at relatively later times.

\subsection{Influence of temperature near the CO sublimation limit}

The next aspect we investigate is that of the precise choice of temperature in the range 15~K to 20~K\footnote{This effect is also similar to uncertainties in the precise value of the CO-ice binding energy, which we discuss below in \S\ref{S:combined}.}. Within this range there is a dramatic difference in the evolution of gas phase CO abundance due to freeze-out on dust grains. Recall also that the early-time solutions found in Case 1 are driven largely by the evolution of CO freeze-out from an initial condition of $\rm [CO]$ $\sim 10^{-4}$ to observed values of $\sim 10^{-5}$ (i.e. CO depletion factor of $f_D\sim 10$).

Figure~\ref{fig:TH_T_var_p2} shows the time evolution of the abundances of CO, HCO$^+$ and N$_2$H$^+$ in the fiducial Case 1 modelling of P2 with finer temperature variation from 15 to 20~K. We see that a small difference in adopted temperature, that is 16~K compared to 15~K, makes a large difference in the gas phase CO abundance during timescales up to $\sim 10^7\:$yr. In particular, the model with $T=16\:$K reaches a quasi equilibrium gas phase CO abundance of $\sim 10^{-5}$ after about 300,000~yr, which then persists at about this level for several Myr. Given the sensitivity of this aspect of the astrochemical model to the choice of $T$ and recognising the systematic uncertainties in both the CO ice binding energy and the measurement of $T$ in the IRDC, it is conceivable that the IRDC clumps are in this quasi-equilibrium phase. 

In Fig.~\ref{fig:TH_CR_var_p2} we show examples of the time evolution of [CO], [HCO$^+$] and [N$_2$H$^+$] at $T=16$~K for various CRIRs. Later time solutions are possible, although the preferred value of $\zeta$ remains relatively low, $\sim 10^{-18}\:{\rm s}^{-1}$, set by [HCO$^+$]. The full projected Case 1 $\chi^2$ landscapes for P2 with XR, UR and R, but with the temperature always fixed to be $T=16$~K, are shown in Fig.~\ref{fig:p2_glob_16K}. Here we see the full extent of these later-time, lower-CRIR solutions.

%and \ref{fig:p6_glob_16K}, 
%for Case 1.
%and \ref{fig:p2_all_glob_16K} 
%and \ref{fig:p6_all_glob_16K} 
%for Case 2.

\begin{figure*}[htb!]
    \centering
    %\addtolength{\leftskip} {-1.2cm}
    %\addtolength{\rightskip}{-1cm}
     \includegraphics[width=\textwidth]
     {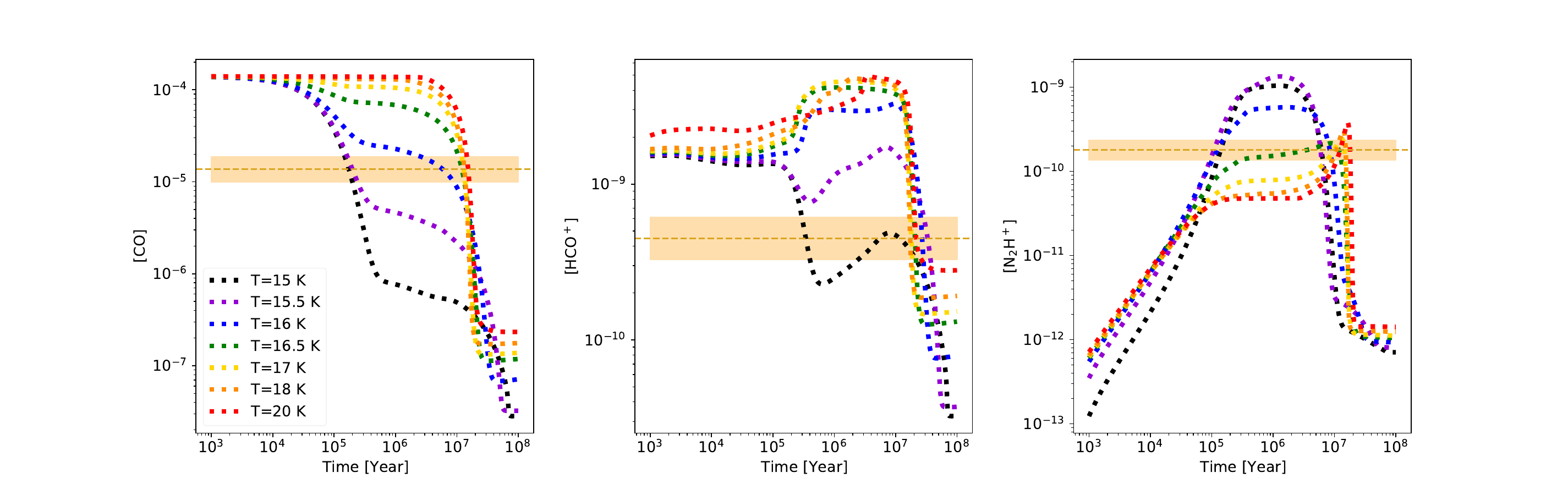}
     \vspace*{-6mm}
     \caption{Time evolution of [CO], [HCO$^+$] and [N$_2$H$^+$] for XR conditions for P2, that is $n_{\rm H}=1\times 10^5\:$cm$^{-3}$, $A_{V}$ =20 mag, but with various temperatures from 15 to 20~K. Here, $\zeta=1.0\times 10^{-17}\:$s$^{-1}$. The horizontal band shows the observed abundances, together with corresponding uncertainties, at P2.}
     \label{fig:TH_T_var_p2}
 \end{figure*}

\begin{figure*}[htb!]
    \centering
    %\addtolength{\leftskip} {-1.2cm}
    %\addtolength{\rightskip}{-1cm}
     \includegraphics[width=\textwidth]
     {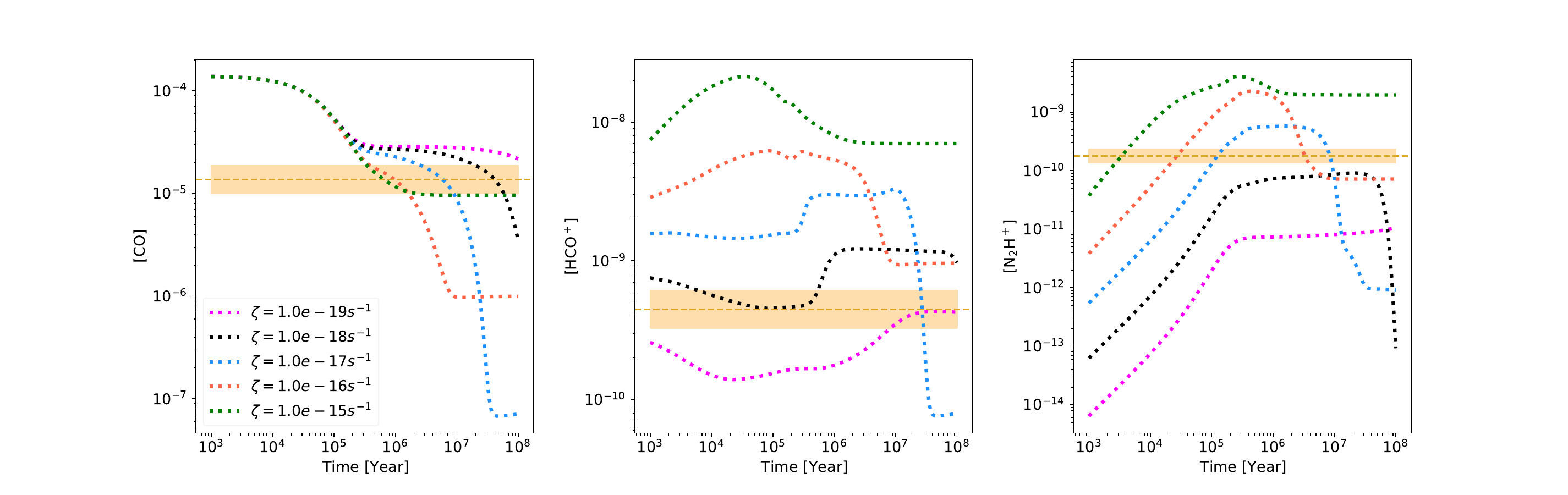}
     \vspace*{-6mm}
     \caption{Time evolution of CO, HCO$^+$ and N$_2$H$^+$ abundances in the models with $n_{\rm H}=1\times 10^5\:{\rm  cm}^{-3}$, $T=16\:$K , $A_{V}=100\:$mag and for different values of $\zeta$. The horizontal band shows the observed abundances together with corresponding uncertainties at P2.}
     \label{fig:TH_CR_var_p2}
 \end{figure*}

\begin{figure*}[htb!]
    \centering
     \includegraphics[width=\textwidth]
     {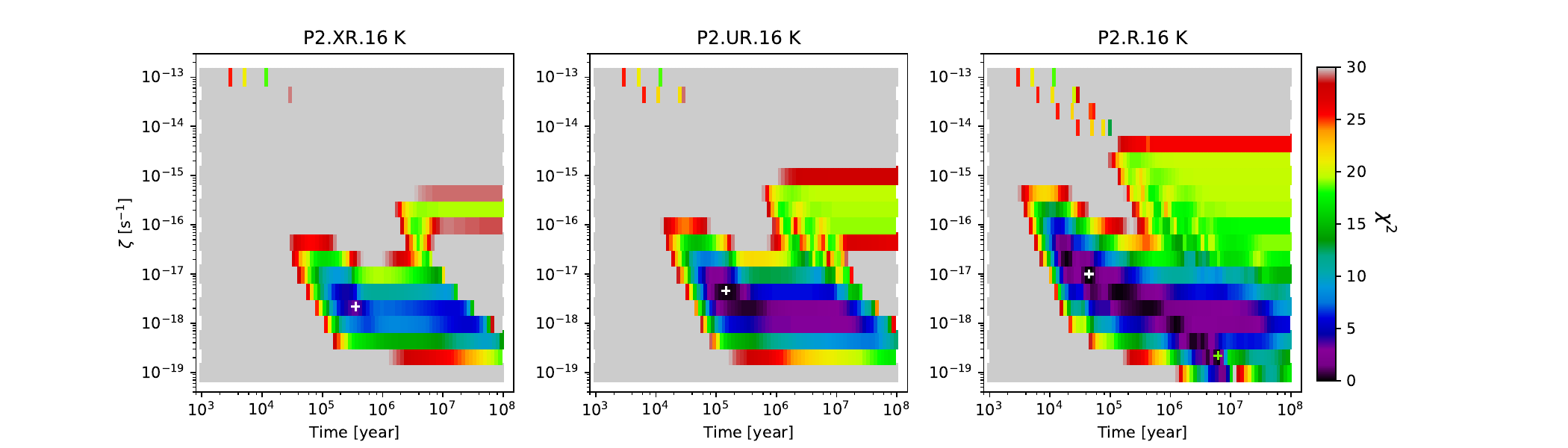}
     \vspace*{-5mm}
     \caption{Investigation of effect of temperature variation between 15 and 20~K. The panels show the projected best $\chi^2$ values in the $\zeta$ versus $t$ plane for the P2 position for the XR (left), UR (middle) and R (right) search methods, but with a further constraint of $T=16\:$K, for Case 1 (fitting to abundances of CO, HCO$^+$ and N$_2$H$^+$). The location of the minimum $\chi^2$ is marked with a white cross in each panel. 
%     Distribution of $\zeta$ associated with $\chi^2$, using observed abundances at P2;XR)$n_{\rm H}$ =1.0e5 cm$^{-3}$ and $A_{\rm v}$ =20 mag, UR)$n_{\rm H}$ =2.2e5, 1.0e5, 4.6e4 cm$^{-3}$, $A_{\rm v}$ =10, 20, 50 mag, R)$n_{\rm H}$ =1.0e4-1.0e6 cm$^{-3}$, $A_{\rm v}$ =5-100 mag. In all panels T=16 K and white crosses show the location of minimum $\chi^2$ at each panel.
     }
     \label{fig:p2_glob_16K}
 \end{figure*}

\subsection{Influence of cosmic ray induced desorption (CRID)}

\begin{figure*}[htb!]
    \centering
     \includegraphics[width=\textwidth]
     {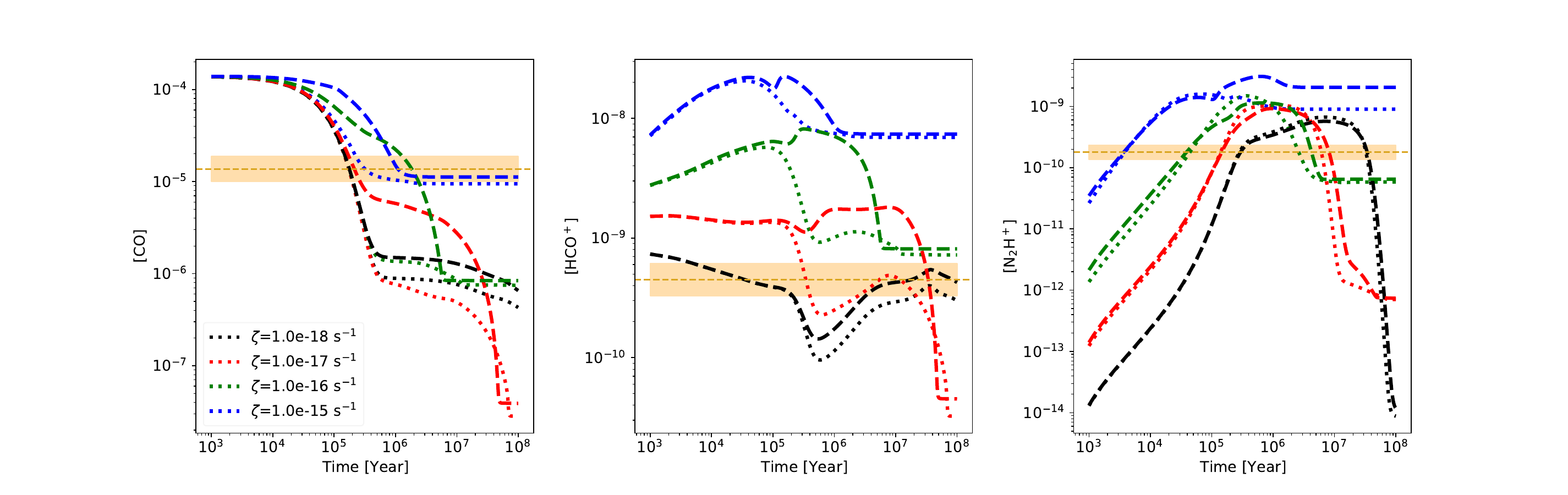}
     \vspace*{-7mm}
     \caption{Effects of CR induced desorption (CRID). Time evolution of the abundances of the Case 1 species CO (left), $\rm HCO^+$ (middle) and $\rm N_2H^+$ (right) in the models with $n_{\rm H} =1.0\times 10^5\:{\rm cm}^{-3}$, $T=15\:$K and $A_{V}=20\:$mag for several different CRIRs. The dotted lines show the fiducial case in which CRID is not included. The dashed lines show the case in which CRID is included with a standard rate (see text). The horizontal bands show the observed abundances together with their uncertainties at P2.}
     \label{fig:CRIDR-on}
 \end{figure*}

%(\textbf{NE: I tried to explain time history of the species when Cosmic Ray direct desorption mechanism is ON, in section 4.3.8})

\begin{figure*}[htb!]
    \centering
     \includegraphics[width=\textwidth]
     {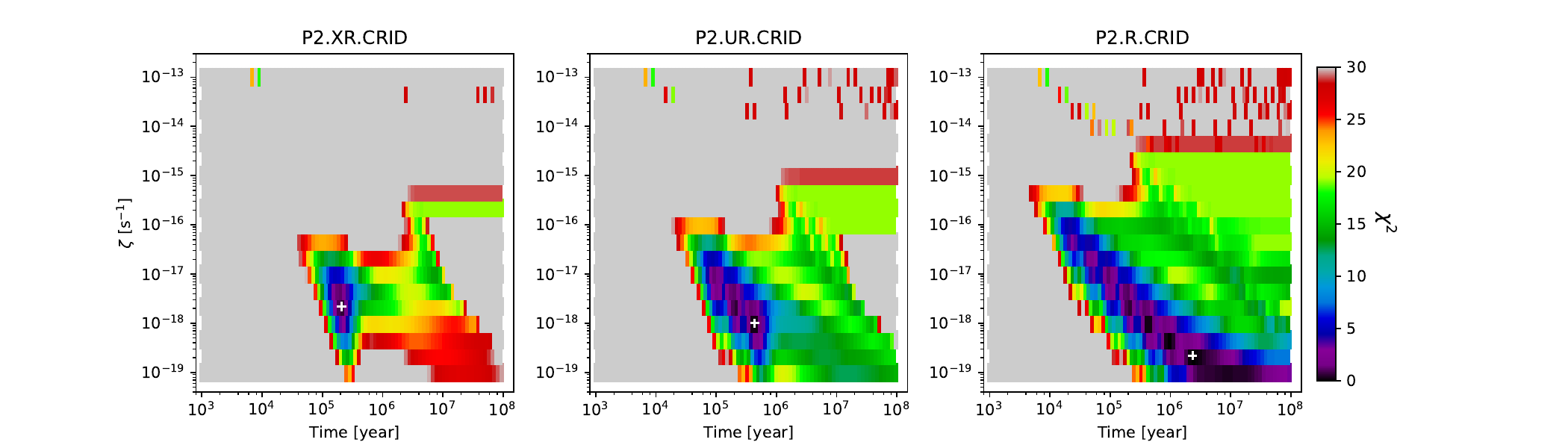}
     \vspace*{-6mm}
     \caption{Effects of CRID. The panels show the projected best $\chi^2$ values in the $\zeta$ versus $t$ plane for the P2 position for the XR (left), UR (middle) and R (right) search methods using models with CRID included for Case 1 (fitting to abundances of CO, HCO$^+$ and N$_2$H$^+$). The location of the minimum $\chi^2$ is marked with a white cross in each panel. Temperature is set to be 15 K for all panels.
%     Distribution of $\zeta$ associated with $\chi^2$, using abundances of CO, HCO$^+$ and N$_2$H$^+$ at P2. In this panel $n_{\rm H}$ =1.0e5 cm$^{-3}$, T=15~K and $A_{\rm v}$ =20 mag and CRIDR of the models are turned ON.
     }
     \label{fig:dotplot_CRIDR-on}
\end{figure*}

%jct - this figure should be updated to show XR, UR, R
%jct - title in each panel can be, e.g., "P2.XR.CRID"

When cosmic rays impact dust grains they cause local heating of the material, which can lead to enhanced thermal desorption of species from the ice mantles. This process has been described by \cite{walsh2010chemical} and incorporated into the astrochemical network. However, so far, in the fiducial modelling described above, we have not included this process due to its inherent large uncertainties. Here we investigate its effects with the standard rates described by \cite{walsh2010chemical} and \cite{sipila2021revised}. In particular, the maximum temperature of the transiently heated grains is assumed to be $T_{\rm CRID}=70\:$K (see \S\ref{astrochemical_model}).

Figure~\ref{fig:CRIDR-on} shows the results of modelling of [CO], [$\rm HCO^+$] and [$\rm N_2H^+$] for XR conditions for P2 with several different values of $\zeta$. For each $\zeta$ we show a model with and without CRID, with the latter being the fiducial models already presented. We see that for [CO] the effect of CRID is, as expected, to maintain a higher gas phase abundance, especially from about 0.5 to 10~Myr. As a result of this higher CO abundance, the level of [$\rm HCO^+$] is also increased in the CRID models.

Figure~\ref{fig:dotplot_CRIDR-on} shows the projected $\chi^2$ landscapes in the $\zeta$ versus $t$ plane for the Case 1 XR, UR and R search methods for the P2 abundances using models in which CRID is included. While the best overall XR model is similar to that found before without CRID, in general we see that inclusion of CRID raises up the values of $\chi^2$ at later times, that is around $t=1\:$Myr.

We conclude that CRID, along with the already investigated effects of CO envelope contamination and intermediate temperatures in the range 15 to 20~K, can make later time and higher CRIR solutions more consistent with the observational constraints of the P2 position (and other IRDC positions). In the next subsection we use the Case 1 data from all the positions, including the observed trends with density, to give some constraints on combinations of these effects that can give reasonable later time solutions.

%(\textbf{NE: Some explanation in section 4.4.6})

\subsection{Combined effects and application to the full sample}\label{S:combined}

%%%%%%%%%%%%%%%%%%%%%%%%%%%%%%%%%%%%
\begin{figure*}[h]
\centering
\includegraphics[width=\textwidth]{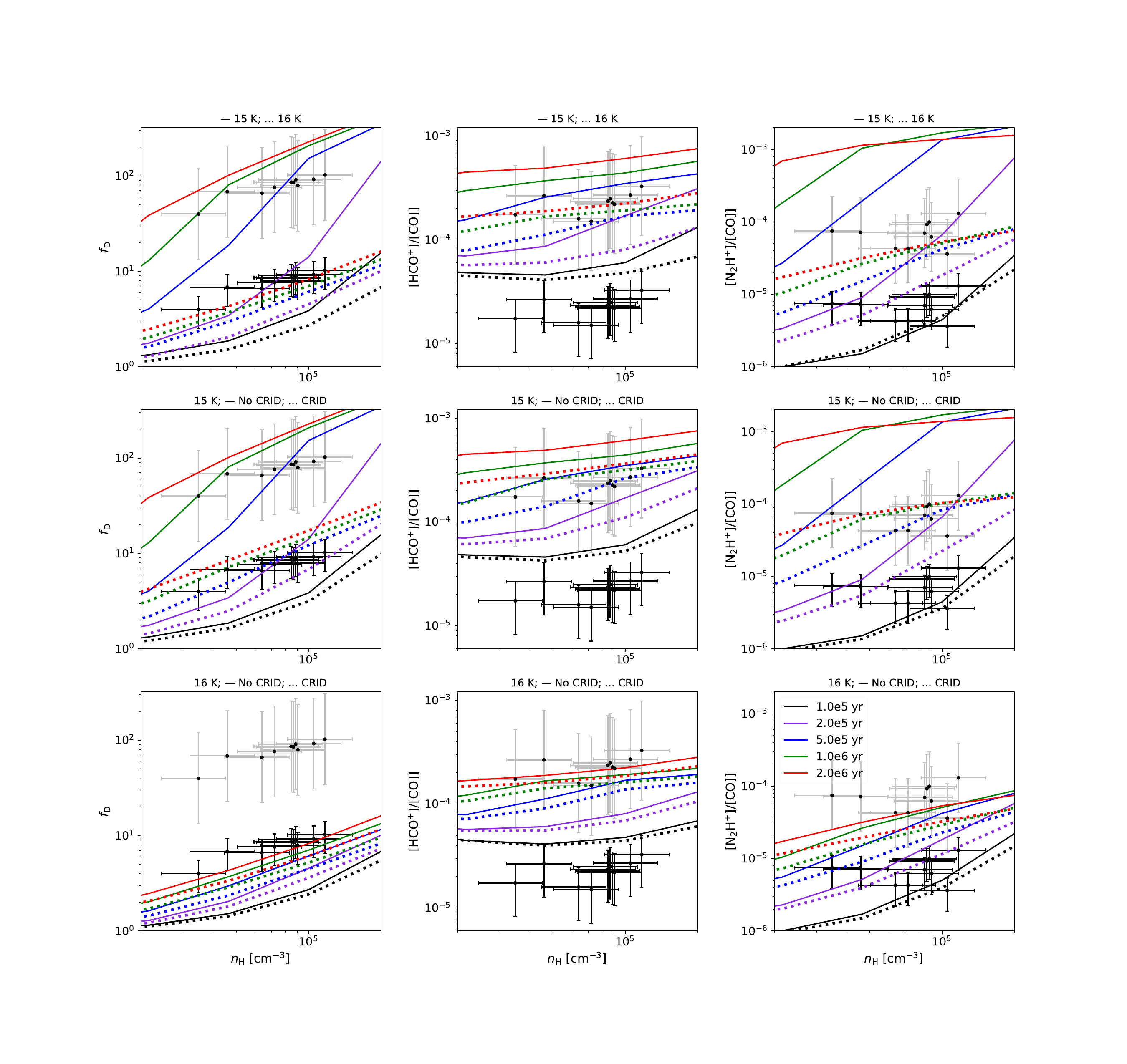}
\vspace*{-20mm}
\caption{Exploration of Case 1 fitting of the whole IRDC clump sample, including trends with density. Left column panels show $f_D$ versus $n_{\rm H}$; middle column panels show [$\rm HCO^+$]/[CO] versus $n_{\rm H}$; right column panels show [$\rm N_2H^+$]/[CO] versus $n_{\rm H}$. In all panels, the black points are the observed data, while grey points assume CO envelope contamination of 90\% (see text).
{\it (a) First row:} Temperature comparison; models have $\zeta= 2.2 \times 10^{-17}\:{\rm s}^{-1}$, $A_V=20\:$mag. Solid lines are from the models with $T=15$~K and dotted lines from those of $T=16$~K. 
{\it (b) Second row:} CRID comparison at 15~K; models have $\zeta= 2.2 \times 10^{-17}\:{\rm s}^{-1}$, $A_V=20\:$mag and $T=15$~K. Solid lines are from the models without CRID; dotted lines are with CRID. 
{\it (c) Third row:} CRID comparison at 16~K; results as (b), but with $T=16$~K.}
     \label{fig:fd_n_16K_2.2e-17}
\end{figure*}

%%%%%%%%%%%%%%%%%%%%%%%%%%%%%%%%%%%
\begin{figure*}[h]
\centering
\includegraphics[width=\textwidth]{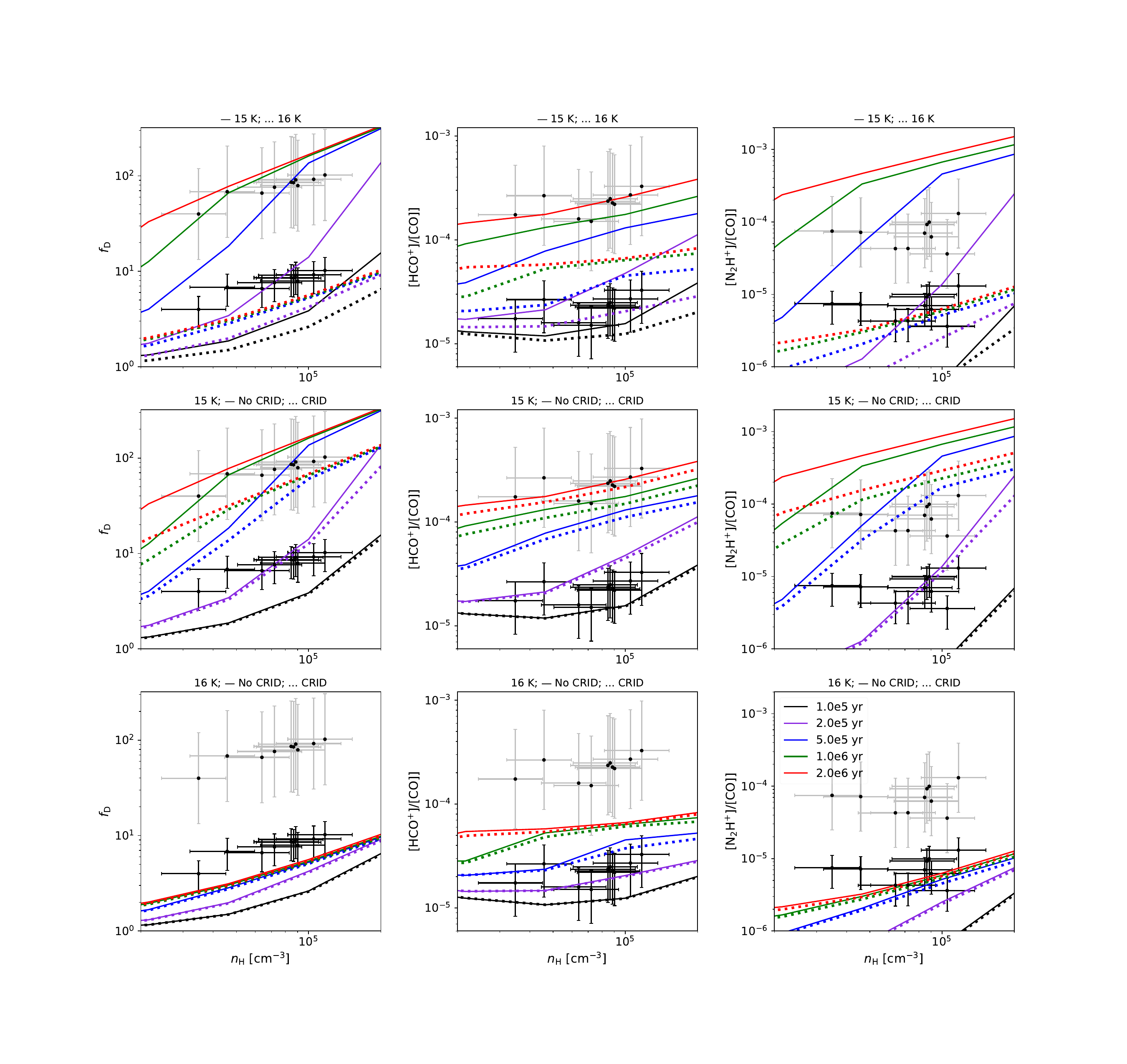}
\vspace*{-20mm}
\caption{Same as Fig.~\ref{fig:fd_n_16K_2.2e-17}, but with all models having $\zeta= 2.2 \times 10^{-18}\:{\rm s}^{-1}$.{\it (a) Top row; (b) middle row; (c) bottom row}
%{\it First row panels}: \textbf{Temperature Comparison}; The models have $\zeta= 2.2 \times 10^{-18}\:{\rm s}^{-1}$, $A_V=20\:$mag. Solid lines are from the models with T=15~K and dotted lines from those of T=16~K. {\it Second row panels: }\textbf{CRID ON and OFF Comparison}; The models have $\zeta= 2.2 \times 10^{-18}\:{\rm s}^{-1}$, $A_V=20\:$mag and T=15~K. Solid lines are from the models with OFF Cosmic Ray Induced Desorption and dotted lines are from those at which it is set to be ON. {\it Third row panels: }\textbf{CRID ON and OFF Comparison}; The models have $\zeta= 2.2 \times 10^{-18}\:{\rm s}^{-1}$, $A_V=20\:$mag and T=16~K. Solid lines are from the models with OFF Cosmic Ray Induced Desorption and dotted lines are from those at which it is set to be ON. In all panels, the black data points are actual observed data, and the gray data points have been obtained by setting CO observed abundance to be lower by the factor of 10.}
}
     \label{fig:fd_n_16K}
\end{figure*}

Here we examine the [CO], [$\rm HCO^+$] and [$\rm N_2H^+$] data for all positions P1 to P10 and search for later time (quasi-equilibrium) models, that is in the range from $t\sim0.2$ to 1~Myr, that have CRIRs that are closer to `standard' values, that is $\zeta\sim 10^{-17}\:{\rm s}^{-1}$, to see if they are able to explain the observed trends. Figure~\ref{fig:fd_n_16K_2.2e-17} shows examples of this analysis. In each row the left panel shows $f_D$ versus $n_{\rm H}$, the middle panel shows [$\rm HCO^+$]/[CO] versus $n_{\rm H}$, and the right panel shows [$\rm N_2H^+$]/[CO] versus $n_{\rm H}$. In each panel the black points show the observed data of positions P1 to P10. The grey points show the measurements assuming that 90\% of the inferred [CO] is contamination from a lower density envelope (i.e. see \S\ref{S:contamination}). In general we show models for conditions with $A_{V}=20\:$mag, that is a relatively high level of extinction, which should also be applicable for higher values of $A_{V}$, and their results at times of 0.1, 0.2, 0.5, 1 and 2~Myr. 

The first row of Fig.~\ref{fig:fd_n_16K_2.2e-17} shows models with $\zeta=2.2 \times 10^{-17}\:{\rm s}^{-1}$ and with $T=15\:$K (solid lines) and $16\:$K (dotted lines). The high sensitivity of CO depletion to this choice is clearly seen. Examining the three panels, if the goal is to match the black points, then we see that this is best done at earlier times ($\sim 0.1$ to 0.2~Myr), although the [HCO$^+$]/[CO] metric is hard to match for any of the models. If the goal is to match the grey points, then this can be done with 15~K models at ages of about 0.2 to 0.5~Myr, although the $\rm N_2H^+$ abundance can be difficult to match in the highest density sources. If the goal is to find the best later time ($\gtrsim 0.5~$Myr) solution, then it is easiest to do when allowing for some CO envelope contamination, that is fitting to grey points, and with the 15~K models. However, overall, in all of these three cases, there is no self-consistent single solution.

The second row of Fig.~\ref{fig:fd_n_16K_2.2e-17} shows models with and without CRID included, with other parameters as in the first row for the 15~K models. If CO envelope contamination is negligible, that is attempting to fit black data points, then solutions with relatively young ages are preferred, although there is no global model that gives a good match to all the data. Similarly, if fitting to the grey data points, the models without CRID are preferred, that is the same as those found in the analysis of the first row. These are also the best later-time solutions. In summary, these $\zeta=2.2\times 10^{-17}\:{\rm s}^{-1}$ models with CRID at 15~K can only match the data if there is little CO envelope contamination and only at relatively young ages.

The third row of Fig.~\ref{fig:fd_n_16K_2.2e-17} also shows these $\zeta=2.2\times 10^{-17}\:{\rm s}^{-1}$ models with and without CRID, but now for 16~K conditions. We see from the $f_D$ results that there is little scope for CO envelope contamination in these cases and that relatively late time solutions are needed to reach even the minimum CO depletion factor estimates. However, such models tend to overproduce [$\rm HCO^+$] and [$\rm N_2H^+$] abundances.

Figure~\ref{fig:fd_n_16K} presents the same type of results as Fig.~\ref{fig:fd_n_16K_2.2e-17}, but now for models with CRIR of $\zeta=2.2\times 10^{-18}\:{\rm s}^{-1}$. It can be seen that the effect of lowering the CRIR is to lower the abundance of $\rm HCO^+$ and $\rm N_2H^+$ for a given model as a function of density and at a given age. This generally gives better agreement with the observed values of [$\rm HCO^+$]/[CO], and is needed for the case of no CO envelope contamination (black data points). 

From a global consideration of the results of Figs.~\ref{fig:fd_n_16K_2.2e-17} and \ref{fig:fd_n_16K}, we draw the following conclusions. First, we note that the observational data for $f_D$ versus density form a very tight relation, especially in comparison to what is seen in the time evolution of many of the models. We consider that this indicates that quasi-equilibrium solutions that converge to this $f_D$ versus $n_{\rm H}$ relation are favoured. Furthermore, this argues against there being a significant effect of CO envelope contamination, since this would be expected to introduce scatter in the $f_D$ - $n_{\rm H}$ relation at about the level of the contamination factor. Late time, quasi-equilibrium solutions that match the observed $f_D$ - $n_{\rm H}$ relation can be found, for example with 16~K models or with 15~K models with CRID, but to better match [$\rm HCO^+$] and [$\rm N_2H^+$] constraints, then relatively low values of CRIR ($\lesssim 2.2 \times 10^{-18}\:{\rm s}^{-1}$) are preferred. There is a strong sensitivity of the normalisation of the late-time, quasi-equilibrium $f_D$ - $n_{\rm H}$ relation to temperature occurring at values of temperature close to 15~K. While this means that agreement between the models and the data can be achieved by fine tuning the choice of temperature, again the lack of dispersion in the observational data argues against this solution. Indeed, it would favour a quasi equilibrium solution that is insensitive to temperature, thus pointing to cosmic ray induced desorption mechanisms for liberating CO to the gas phase as being more important than simple thermal desorption. In the next subsection we consider a chemical network with revised CO ice binding energy and CRID parameters that can yield a reasonable match to the observed constraints, that is with later time solutions that converge on the observed $f_D$ - $n_{\rm H}$ relation and that are insensitive to temperature for values below $\sim 15\:$K.

\subsection{Models with revised astrochemical parameters of CO ice binding energy and $T_{\rm CRID}$}\label{S:grid2}

%which are close to the observed values in the IRDC positions (their average is 13.5~K). The precise value of temperature where the variation of the normalization of the $f_D$ - $n_{\rm H}$ relation becomes large is also expected to depend sensitively on the choice of CO-ice binding energy (set to be 855~K in the models explored so far).

\begin{figure*}[h]
\centering
\includegraphics[width=\textwidth]{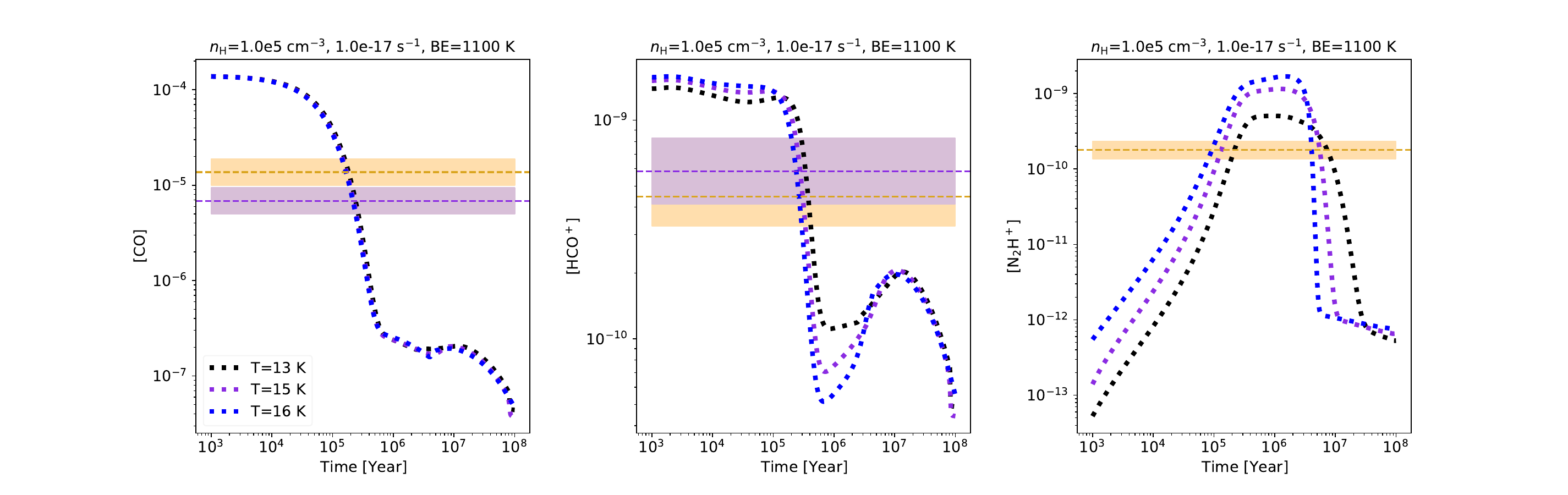}
\vspace*{-6mm}
\caption{Evolution of [CO], [HCO$^+$] and [N$_2$H$^+$] in models with CO ice binding energy of 1100~K, $n_{\rm H}=1.0\times 10^{5}\:{\rm cm}^{-3}$, $A_V = 20\:$mag, $\zeta= 1.0\times 10^{-17}\:{\rm s}^{-1}$ and no CRID, exploring effects of different temperatures in the range relevant to IRDCs. The golden bands at each panel represent observed abundances at P2 ([HCO$^+$] from [H$^{13}$CO$^+$]) and the purple bands show [CO] reduced by a factor of 2 (allowing for potential CO envelope contamination) and [HCO$^+$] estimated from [HC$^{18}$O$^+$].}
     \label{fig:BE-1}
\end{figure*}

\begin{figure*}[h]
\centering
\includegraphics[width=\textwidth]{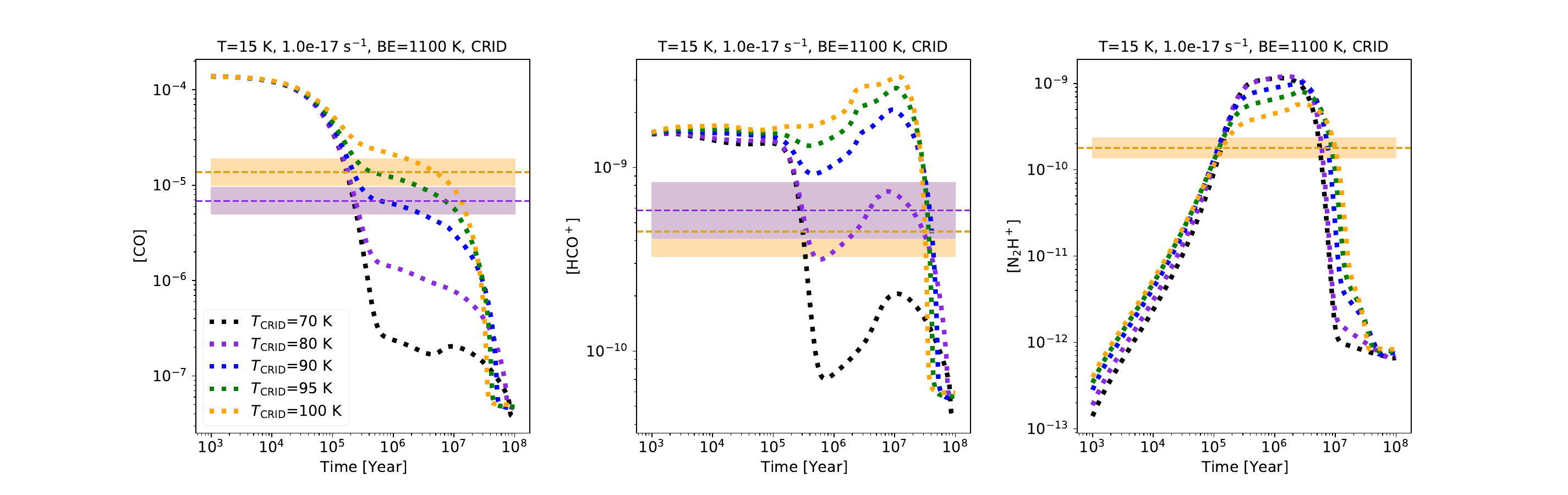}
\vspace*{-6mm}
\caption{Evolution of [CO], [HCO$^+$] and [N$_2$H$^+$] in models with CO ice binding energy of 1100~K, $n_{\rm H}=1.0\times 10^{5}\:{\rm cm}^{-3}$, $A_V = 20\:$mag, $\zeta= 1.0\times 10^{-17}\:{\rm s}^{-1}$ and CRID, exploring values of $T_{\rm CRID}$ from 70 to 100~K. The golden bands in each panel represent observed abundances at P2 ([HCO$^+$] from [H$^{13}$CO$^+$]) and the purple bands show [CO] reduced by a factor of 2 (allowing for potential CO envelope contamination) and [HCO$^+$] estimated from [HC$^{18}$O$^+$].}
     \label{fig:CRT-variation}
\end{figure*}

\begin{figure*}[h]
\centering
\includegraphics[width=\textwidth]{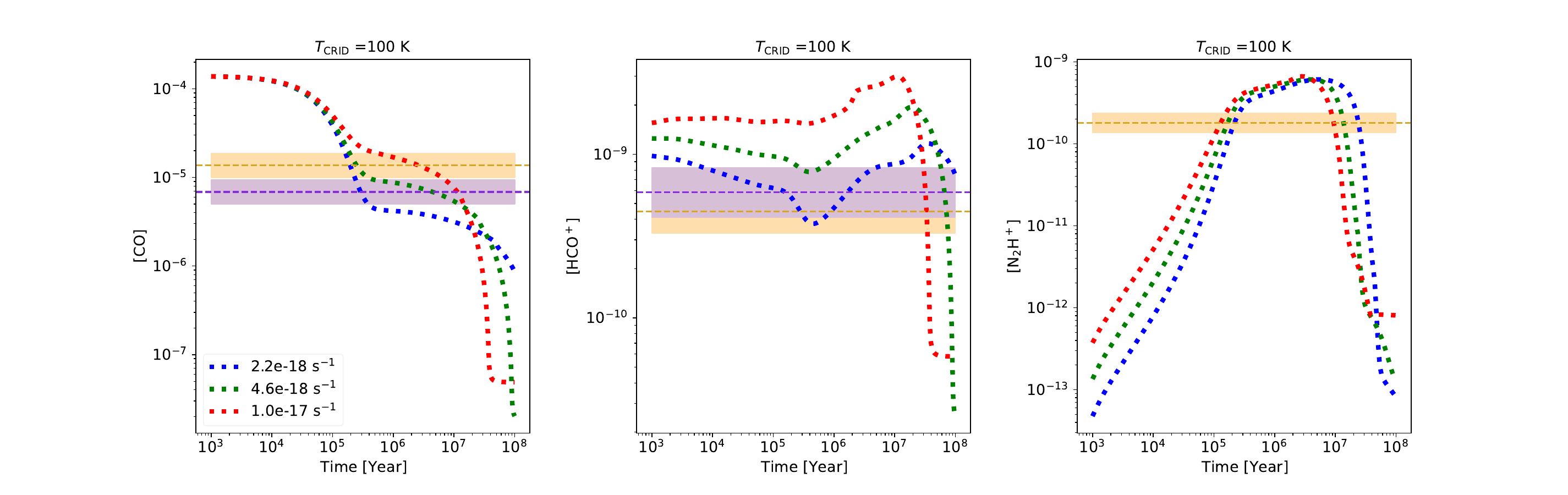}
\vspace*{-6mm}
\caption{Evolution of [CO], [HCO$^+$] and [N$_2$H$^+$] in models with CO ice binding energy of 1100~K, $n_{\rm H}=1.0\times 10^{5}\:{\rm cm}^{-3}$, $A_V = 20\:$mag, $T_{\rm CRID}=100\:$K and $\zeta= (0.22, 0.46, 1.0)\times 10^{-17}\:{\rm s}^{-1}$. The golden bands at each panel represent observed abundances at P2 ([HCO$^+$] from [H$^{13}$CO$^+$]) and the purple bands show [CO] reduced by a factor of 2 (allowing for potential CO envelope contamination) and [HCO$^+$] estimated from [HC$^{18}$O$^+$].}
     \label{fig:XR-best-models-n}
\end{figure*}

In the models we have investigated so far, CO ice binding energy was set to the fiducial value of 855~K, leading to a strong sensitivity of CO depletion to temperature variation between 15 and 16 K. While this value of binding energy is suitable for pure CO ice \citep{oberg2005competition}, in more realistic situations in which CO is mixed with water ice, its binding energy is expected to increase, with values of $\sim$1100~K preferred  \citep{oberg2005competition,cuppen2017grain}. As shown in Fig.~\ref{fig:BE-1}, by increasing CO ice binding energy to 1100~K, CO abundance variation between 13 and 16~K becomes negligible, that is the gas phase abundance of CO decreases rapidly towards very low values. 
%which means CO is still remained on the grain surfaces. 
Thus, to release CO and raise its later-time gas phase abundance closer to observed levels for temperatures in the observed range $\lesssim15\:$K, we need to rely on CRID.

In the astrochemical network, the main parameter which affects significantly the final CRID rate is $T_{\rm CRID}$, which is the maximum temperature reached by grains after suffering a CR impact. Within the network utilised so far, the fiducial value of $T_{\rm CRID}=70\:$K. Figure~\ref{fig:CRT-variation} shows the effect of varying $T_{\rm CRID}$ from 70 to 100~K. We see that values of $T_{\rm CRID}\sim 90\:$ to 100~K are able to sustain CO depletion factors of about 10 for relatively long timescales, up to $\sim 10\:$Myr. Furthermore, as illustrated in Fig.~\ref{fig:XR-best-models-n}, in these models the level of CO abundance around the timescales from 0.5 to 10~Myr also depends on the level of CRIR. 

\begin{figure*}[h]
\centering
\includegraphics[width=\textwidth]{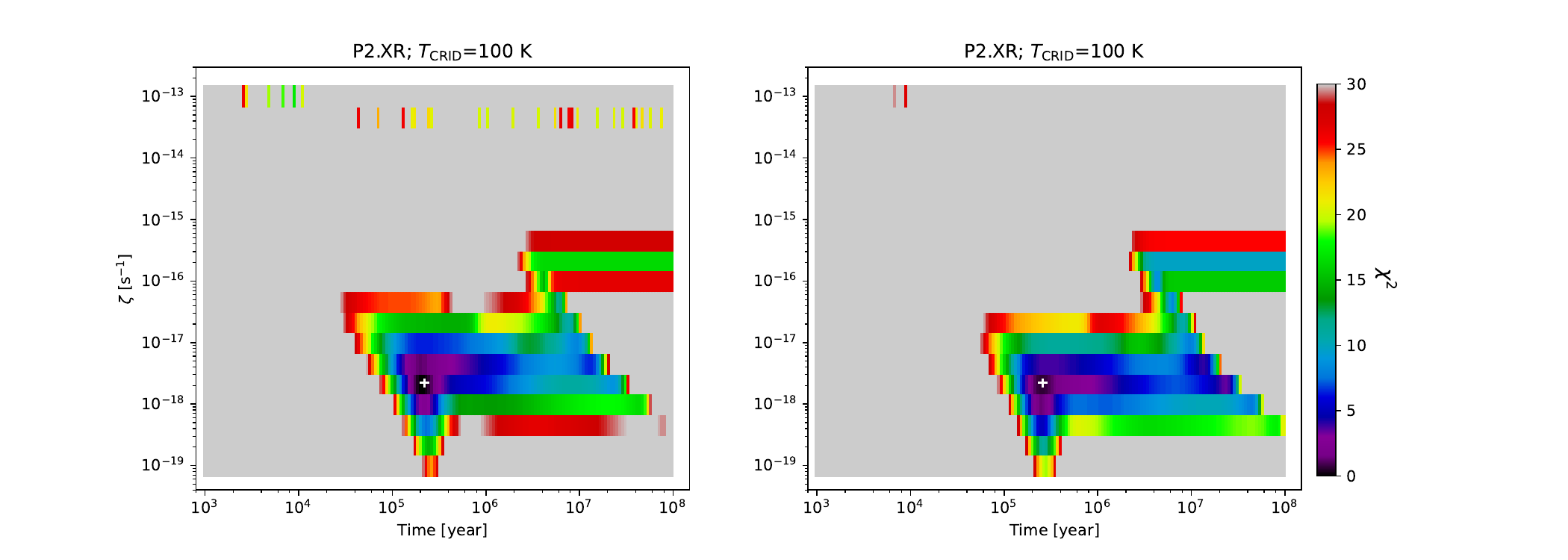}
\vspace*{-6mm}
\caption{{\it (a) Left:} Example of XR Case 1 fitting of P2 abundances (HCO$^+$ from HC$^{18}$O$^+$) using astrochemical model Grid 2 (i.e. CO ice binding energy of 1100~K and $T_{\rm CRID}=100\:$~K) showing the projected best $\chi^2$ values in the $\zeta$ versus $t$ plane. The location of minimum $\chi^2$ is marked with a white cross. {\it (b) Right:} As (a), but now for [CO] reduced by a factor of two (allowing for potential CO envelope contamination).
%Investigation of effect of $T_{\rm CRID}=100\:$~K, showing the projected bext $\chi^2$ values in the $\zeta$ versus $t$ plane for the P2 position for the XR range and using the search method with Case1, fitting to abundances HCO$^+$ (obtained from HC$^{18}$O$^+$) and N$_2$H$^+$ and CO in the left panel and reduced CO (by the factor of 2) in the right panel. The location of minimum $\chi^2$ is marked with a white cross in each panel.
}
     \label{fig:P2-XR-CRT100}
\end{figure*}

Figure~\ref{fig:P2-XR-CRT100} shows examples of XR fitting of P2 for a grid of models with CO ice binding energy of 1100~K and $T_{\rm CRID}=100\:$K, that is `Grid 2'. Relatively early time and low CRIR solutions are again found as the best overall models, although good solutions with low $\chi^2$ values are now found extending to later times at similar values of CRIR. The main difference compared to the equivalent Grid 1 results is that these later time solutions do not require much lower values of CRIR.

\begin{figure*}[h]
\centering
\includegraphics[width=\textwidth]{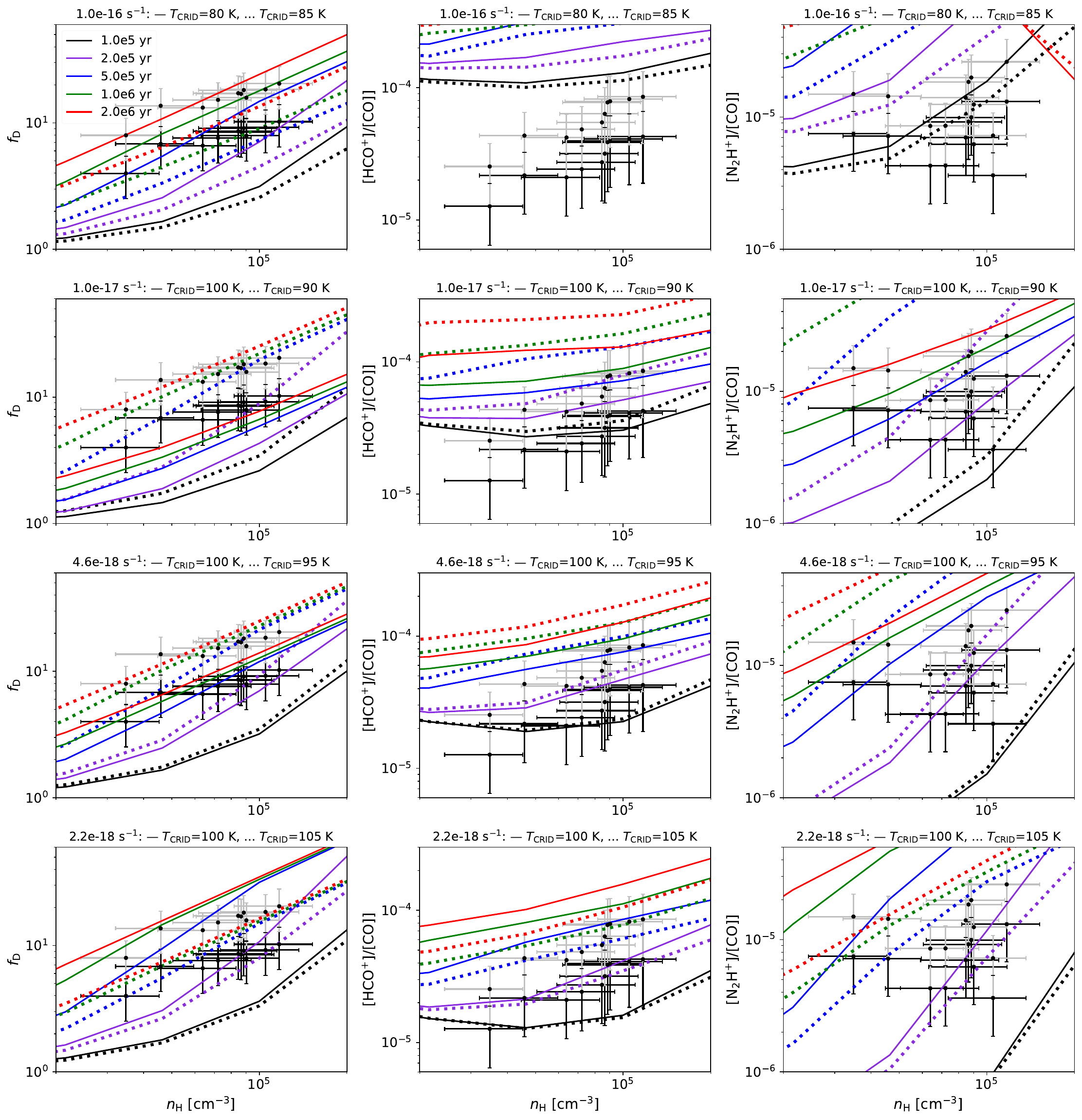}

\vspace*{-1mm}
\caption{
Exploration of Case 1 fitting of the whole IRDC clump sample, including trends with density, using models with gas temperature of $T=15$~K, $A_V=20\:$~mag, CO ice binding energy of 1100~K and exploring effects of cosmic ray induced desorption parameter $T_{\rm CRID}$. Left column panels show $f_D$ versus $n_{\rm H}$; middle column panels show [$\rm HCO^+$]/[CO] versus $n_{\rm H}$; right column panels show [$\rm N_2H^+$]/[CO] versus $n_{\rm H}$. In all panels, black points are the actual observed data (with HCO$^+$ obtained from HC$^{18}$O$^+$), while the grey points assume CO abundance is reduced by a factor of 2 (due to CO envelope contamination).
{\it (a) First row:} High CRIR case with $\zeta= 1.0 \times 10^{-16}\:{\rm s}^{-1}$. Solid lines are models with $T_{\rm CRID}=80\:$K and dotted lines are those with $T_{\rm CRID}=85\:$K. As shown in the legend, the various colours correspond to models from 0.1 to 2~Myr.
{\it (b) Second row:} Case with $\zeta= 1.0 \times 10^{-17}\:{\rm s}^{-1}$. Solid lines are models with $T_{\rm CRID}=100\:$K and dotted lines are those with $T_{\rm CRID}=90\:$K.
{\it (c) Third row:} Case with $\zeta= 4.6 \times 10^{-18}\:{\rm s}^{-1}$. Solid lines are models with $T_{\rm CRID}=100\:$K and dotted lines are those with $T_{\rm CRID}=95\:$~K. 
{\it (d) Fourth row:} Case with $\zeta= 2.2 \times 10^{-18}\:{\rm s}^{-1}$. Solid lines are models with $T_{\rm CRID}=100\:$~K and dotted lines are those with $T_{\rm CRID}=105\:$~K. 
}
     \label{fig:1100BE-2}
\end{figure*}

Now we examine again the [CO], [$\rm HCO^+$] and [$\rm N_2H^+$] data for all positions P1 to P10 and search for later-time (quasi-equilibrium) models with CO ice binding energy of 1100~K and different values of $T_{\rm CRID}$. Following the format of Figs.~\ref{fig:fd_n_16K_2.2e-17} and \ref{fig:fd_n_16K}, Fig.~\ref{fig:1100BE-2} shows examples of this type of analysis. In each row the left panel shows $f_D$ versus $n_{\rm H}$, the middle panel shows [$\rm HCO^+$]/[CO] versus $n_{\rm H}$, and the right panel shows [$\rm N_2H^+$]/[CO] versus $n_{\rm H}$. In each panel the black points show the observed data of positions P1 to P10. The grey points show the measurements assuming that 50\% of the inferred [CO] is contaminated (i.e. see \S\ref{S:contamination}). We show models for conditions with $T=15\:$K (although results are relatively insensitive to temperatures near this value) and $A_{V}=20\:$mag and their results at times of 0.1, 0.2, 0.5, 1 and 2~Myr. 

The first row of Fig.~\ref{fig:1100BE-2} presents a high CRIR case with $\zeta=1.0 \times 10^{-16}\:{\rm s}^{-1}$. Matching the observed $f_D$-$n_{\rm H}$ relation with later-time solutions requires $T_{\rm CRID}\simeq 80$ to 85~K. However, we see that these models are not good fits to the [$\rm HCO^+$]/[CO] and [$\rm N_2H^+$]/[CO] data. The second row shows $\zeta=1.0 \times 10^{-17}\:{\rm s}^{-1}$ models with $T_{\rm CRID}=90$ and 100~K (the latter being the fiducial Grid 2 value). Again, later-time solutions can match the $f_D$-$n_{\rm H}$ relation and are now seen to be closer to, but still above, the [$\rm HCO^+$]/[CO] and [$\rm N_2H^+$]/[CO] constraints. The third and fourth rows bring the CRIR down to $\zeta=4.6 \times 10^{-18}\:{\rm s}^{-1}$ and $\zeta=2.2 \times 10^{-18}\:{\rm s}^{-1}$, respectively. Values of $T_{\rm CRID}$ within $100\pm5\:$K are displayed. In the third row, models with an age of about 0.5 Myr with $T_{\rm CRID}=95\:$K are a good match to the $f_D$-$n_{\rm H}$ relation with 50\% CO envelope contamination and the [$\rm HCO^+$]/[CO] versus $n_{\rm H}$ data. They are also within about a factor of two of the [$\rm N_2H^+$]/[CO] data. Finally, in the fourth row the 0.5~Myr models with $T_{\rm CRID}=105\:$K are a good match to all the constraints provided by the $f_D$, [$\rm HCO^+$]/[CO] and [$\rm N_2H^+$]/[CO] versus $n_{\rm H}$ data (assuming 50\% CO envelope contamination).

Thus the conclusion of this analysis is that relatively late-time ($\sim 0.5\:$Myr) solutions (i.e. much older than a local free-fall time) exist that are insensitive to the gas temperature (in the relevant range $\lesssim 15\:$K) requiring only modest levels of CO envelope contamination (as implied by the relatively low scatter in the $f_D$-$n_{\rm H}$ relation). However, these solutions require thermal heating of dust grain ice mantles up to $T_{\rm CRID}\sim 100\:$K. The models also imply that the cosmic ray ionisation rate is a factor a few times smaller than the canonical value of $\zeta=10^{-17}\:{\rm s}^{-1}$.

\subsection{Relation to previous works and implications for massive star and star cluster formation}

Summarising our results, we have found individual astrochemical models that can be a good match to observed abundances of CO, HCO$^+$ and $\rm N_2H^+$, that is focusing on relatively simple species that also have high sensitivity to the CRIR, implying chemical ages as short as $\sim 10^5\:$yr, based on the timescale to freeze-out CO onto dust grains at densities of $\sim 3$ to $10\times 10^{4}\:{\rm cm}^{-3}$ at cold temperatures of $\lesssim 15\:$K. However, we have argued against such models given the relatively narrow dispersion in the observed CO depletion factor versus density relation among ten independent positions. Instead, we have concentrated on what conditions would be needed for relatively late time solutions, that is $\gtrsim 3 t_{\rm ff} \sim 0.5\:$Myr to explain [CO], [$\rm HCO^+$] and [$\rm N_2H^+$]. Here we have seen that such solutions are possible, but require an adjustment of one the main parameters involved in cosmic ray induced thermal desorption, which is $T_{\rm CRID}\sim 100\:$K, as well as favouring values of CRIR of $\lesssim 5\times 10^{-18}\:{\rm s}^{-1}$.

One may ask how the timescale of $\sim 0.5\:$Myr compares with expected massive star and star cluster formation times. For massive star formation from prestellar cores (PSCs) of mass $M_c$ that are in approximate pressure equilibrium with their local clump environment that has a mass surface density of $\Sigma_{\rm cl}\sim 0.3\:{\rm g\:cm}^{-2}$, \cite{mckee2003formation} [MT03] estimated star formation times of $t_{\rm *f} \simeq 3.2\times 10^5 (M_c/60\:M_\odot)^{1/4} (\Sigma_{\rm cl} / 0.3\:{\rm g\:cm}^{-2})^{-3/4}\:{\rm yr}$. Thus, we see that a chemical age of $\sim 0.5\:$Myr would be quite similar to the timescale expected for massive star formation in these environments, especially as the MT03 estimate is for the protostellar phase, rather than the earlier prestellar phase. For star cluster formation, the overall timescale of the process remains debated between the extremes of relatively short formation in $\sim 1 t_{\rm ff}$ 
\citep[e.g.][]{2007ApJ...668.1064E,2019MNRAS.490.3061V} to longer timescales of $\gtrsim 10 t_{\rm ff}$ \citep[e.g.][]{2006ApJ...641L.121T,2014ApJ...795...55D}. We suggest that the lack of dispersion in the observed $f_D$ versus $n_{\rm H}$ relation is evidence in favour of longer duration star formation models.
%, however, a more quantitative comparison would require 

\citet{2012ApJ...751..105V} presented astrochemical models of two IRDC positions, one estimated from $\rm NH_3$ observations to be cold (IRDC 013.90-1 with $T\sim 13\:$K) and another estimated to be warm (IRDC 321.73-1 with $T\sim 22\:$K). They included an investigation on the role of surface chemistry. They considered densities in the range $n_{\rm H2}=10^5$ to $10^6\:{\rm cm}^{-3}$, temperatures in the range $T=10$ to $40\:$K and CRIR for $\rm H_2$ of $1.3\times 10^{-17}\:{\rm s}^{-1}$. For their most complete `surface' network applied to their cold IRDC, they found a most likely chemical age (based on matching to abundances of seven species) of $\sim 10^5\:$yr, but with older ages up to about 1~Myr still having confidence parameter values within a factor of about 5 of the best value. We note that they did not have observational constraints on the abundance of CO. We consider that these results of \citet{2012ApJ...751..105V} are broadly consistent with results that we have found, that is identifying a potential early-time solution at $\sim10^5\:$yr, but with later time solutions not completely excluded. However, given the significant differences in the astrochemical modelling and analysis methods, it is difficult to draw more general and definitive conclusions from this comparison.

\citet{2014A&A...563A..97G} have estimated a chemical age of IRDC regions to be very short, that is $\sim10^4$ years. However, this is an order of magnitude shorter than the free-fall time of the regions we have considered, so seems unlikely to be valid for our case. The method of \citet{2014A&A...563A..97G} is based on the time to evolve a chemical model, assuming $\zeta=5\times 10^{-17}\:{\rm s}^{-1}$ and $A_V=10\:$mag, from its initial conditions to an observed set of abundances, but with agreement a factor of 10, and with equal weighting given to relatively complicated species whose chemistry is more uncertain). Furthermore, the possibility that the system is older is not excluded by this analysis.

\citet{2021A&A...652A..71S} studied the chemical timescales of the different phases of massive star formation, including an early $70\:{\rm \mu m}$-weak phase that was inferred to have a duration of only $\sim 5\times 10^4\:$yr. However, their analysis involves several significant differences compared to our work. For example, they only considered one value of CRIR ($5 \times 10^{-17} {\rm s}^{-1}$), one value of temperature (15~K) and one starting density ($n_{\rm H} = 10^4 {\rm cm}^{-3}$) for their early phase models (though they consider evolution of the density up to various values in the range $10^5 - 10^8 {\rm cm}^{-3}$). They also focused on the relatively complicated species of $\rm CH_3CCH$, $\rm CH_3CN$, $\rm H_2CO$ and $\rm CH_3OH$. \citet{2021A&A...652A..71S} described how changing the ice binding energies by $\sim 10\%$ leads to changes in abundances of more than one order of magnitude by the end of the evolution. Furthermore, their comparison to another chemical network \citep[e.g.][]{2006A&A...457..927G}, led to order of magnitude differences in the abundances of their considered species. Given these difficulties, it appears that it is very difficult to use this type of modelling and comparison to these relatively complex species to constrain the chemical timescale of the pre-stellar phase with any great precision. Such considerations have led us to focus on the simpler species of CO, HCO$^+$ and $\rm N_2H^+$ as the primary species to be modelled.

Another general difference between our work and studies such as those of \citet{2014A&A...563A..97G} and \citet{2021A&A...652A..71S} is our use of a simple one-zone model, that is of constant density, temperature, etc., rather than a more complicated model with a structured density and temperature profile. We consider that the single zone model, while approximate, is a reasonable first estimate for the average conditions in the regions traced by our molecular line observations, especially since we are focusing on the pre-stellar phase, rather than trying to model protostellar systems, which would be better described by concentrated structures with well-defined centres. While our IRDC regions likely do include higher density substructures, our modelling does allow for a range of densities to be considered above and below the average value inferred from our column density maps.

\subsection{Caveats}

One potential caveat is the relative simplicity of the astrochemical modelling. In regard to the ability of the models to match [$\rm HCO^+$] and [$\rm N_2H^+$] as the main constraint on CRIR, we note that in these presented models $\rm HCO^+$ is one of the main carriers of the positive charge, that is with [$\rm HCO^+$]/[$e^-$]~$\sim 0.5$. One may consider whether the lack of deuterated species in the astrochemical network may influence these results, especially if the deuterated species, such as $\rm H_2D^+$ and $\rm DCO^+$ were to make a significant contribution to the charge budget. To examine this issue, we have considered the gas phase only deuteration modelling results of \citet{hsu2021deuterium} with $f_D=10$ and similar conditions as XR for P2. We find that deuteration only makes $\sim 20\%$ differences to the abundance of $\rm HCO^+$ and so is not likely to be a major influence on our analysis.

Another caveat is the relative simplicity of the treatment of cosmic ray induced desorption. In particular, the modelling assumes a single value for the maximum temperature reached by the grains when they are hit by a cosmic ray. A more realistic treatment would allow for a distribution of such temperatures. In addition, variation of heating within an individual grain \citep[e.g.][]{2015ApJ...805...59I} is a feature that is also likely to be relevant and worthy of consideration in future analysis.

Finally, as discussed previously, there are a number of observational limitations. We have measured physical properties and chemical abundances averaged on scales of about 1~pc. In reality, these regions will contain a distribution of densities, temperatures, abundances, etc. In addition, a number of the regions considered are known to have sites of star formation, which will have very localised and strong variations in physical and chemical properties. While IRDCs do contain large amounts of relatively quiescent molecular gas and our study has included an effort to select regions that are free of known star formation, future studies making use of higher angular resolution data are warranted.

%In summary, we see that the observed $f_D$ - $n_{\rm H}$ relation can be matched with late-time limit results of models that involve either a fine-tuned temperature of 15.8~K or a lower temperature of $\lesssim 15\:$K, but with a slightly smaller value of the CO-ice binding energy than the fiducial one assumed in the astrochemical model. However, such fine-tuning of temperature seems unrealistic, especially as it would introduce scatter in the $f_D$ - $n_{\rm H}$ relation among the 10 positions. Thus, we consider that non-thermal desorption processes, i.e., due to CRs, are more likely to be setting the normalization of this relation, so that it is insensitive to temperatures in the range of about 10 to 15~K. Such nonthermal desorption processes may not yet be included adequately in our astrochemical network. Further work is needed to understand if they can be modeled, while at the same time giving consistent results for [$\rm HCO^+$] and [$\rm N_2H^+$].

%{\bf Some discussion of uncertainties in the CO-ice binding energy - is this CO on water or CO on CO? Help from astrochemists needed here. As well as any thoughts on the last paragraph.}

%{\bf \subsection{CO Binding Energy}

\section{Conclusions}\label{S:conclusions}

We have studied the astrochemical conditions in ten positions in a massive IRDC. The abundances of eight species were measured via line emission observed from single dish IRAM-30m telescope observations, in combination with total mass surface density measurements from both {\it Herschel} dust emission and NIR+MIR extinction maps. The dust emission maps also enabled a measurement of temperature of the positions. To interpret these results an extensive grid of gas-grain astrochemical models was computed, exploring the effects of density, temperature, visual extinction of the Galactic FUV radiation field and cosmic ray ionisation rate on the time evolution of the abundances of the observed species. Furthermore, additional parameters and features of the models, which are CO ice binding energy and effect of cosmic ray induced desorption, including an exploration of the transient heating temperature, were explored.

With various restrictions on the density, temperature, and visual extinction, the standard astrochemical models (Grid 1) that best match CO, $\rm HCO^+$ and $\rm N_2H^+$ (Case 1) and all species (Case 2) have been found. For Case 1 fitting, these generally favour early time solutions (i.e. $t\sim 10^5\:$yr), driven by CO freeze-out timescales, and with relatively low CRIRs. However, there are islands of low $\chi^2$ at later times and higher CRIRs. Furthermore, later times can be preferable for Case 2 fittings, although with worse values of $\chi^2$.

We have discussed several aspects of the methodology that may be producing systematic errors, including the possibility of CO envelope contamination, high sensitivity to precise values of temperature near 15~K (or equivalently CO-ice binding energy assumptions) and cosmic ray induced desorption.

We have argued that the observed tightness of the CO depletion factor ($f_D$) versus density ($n_{\rm H}$) relation argues against significant CO envelope contamination, since this would introduce significant scatter that is not observed. We have also proposed that the lack of scatter in the $f_D$-$n_{\rm H}$ relation does not favour solutions that rely on a fine tuning of temperature, but rather rely on the process of cosmic ray induced desorption to achieve a quasi-equilibrium later-time solution. Comparing to free-fall times, one would then favour models that show relatively tight, convergent behaviour after a few $\times t_{\rm ff} \sim 0.3\:$Myr. Such a conclusion would argue against rapid evolution of the IRDC structures on timescales that are as short as their local free-fall times.

%We have also argued that the lack of dispersion in the $f_D$-$n_{\rm H}$ relation among the 10 independent positions favors astrochemical models that have late-time ($\gtrsim$ few $\times t_{\rm ff} \sim 0.3\:$Myr) limiting behaviour that matches the observed relation. 

We have found that such models exist that well match the shape and normalisation of the $f_D$-$n_{\rm H}$ relation: which are the models with $\zeta \sim 2$ to $4\times 10^{-18}\:{\rm s}^{-1}$, $T_{\rm CRID}\sim 100\:$K, CO ice binding energy of 1100~K and ages of $\sim 0.5\:$Myr (see Fig.~\ref{fig:1100BE-2}c and d). Furthermore, these models also provide good matches to the observed [$\rm HCO^+$]/[CO] and [$\rm N_2H^+$]/[CO] versus $n_{\rm H}$ data. Within the framework of the presented models, we are not able to find good solutions with cosmic ray ionisation rates $\gtrsim {\rm few} \times 10^{-17}\:{\rm s}^{-1}$, since these overpredict the abundances of HCO$^+$ and $\rm N_2H^+$. Such low rates of cosmic ray ionisation are in contrast to measurements in the diffuse ISM \citep[e.g.][]{neufeld2017cosmic}, which may point to shielding of IRDC regions via their high column densities and/or magnetic field gradients.

Future work to extend this type of study to other IRDCs, including in different Galactic environments, and utilising higher angular resolution molecular line observations is desirable.

\newpage

\begin{acknowledgements}
We thank an anonymous referee for helpful comments that improved the paper. This work is based on that submitted in partial requirements for Ms. Negar Entekhabi's Masters thesis in Physics at Chalmers University of Technology, supervised by Prof. Jonathan C. Tan. We acknowledge Mr. Arturo Cevallos Soto for serving as the student opponent for the Masters defense. JCT acknowldeges support from VR grant 2017-04522 and ERC project MSTAR. I.J.-S. has received partial support from the Spanish State Research Agency (AEI; project number PID2019-105552RB-C41).
\end{acknowledgements}

 \bibliographystyle{aa}
 \bibliography{aa}

\end{document}